\documentclass[fleqn,usenatbib]{mnras}

\usepackage{newtxtext,newtxmath}

\usepackage[T1]{fontenc}

\DeclareRobustCommand{\VAN}[3]{#2}
\let\VANthebibliography\thebibliography
\def\thebibliography{\DeclareRobustCommand{\VAN}[3]{##3}\VANthebibliography}

\usepackage[dvipdfmx]{graphicx}	
\usepackage{amsmath}	
\usepackage{bm}  
\usepackage{color}
\usepackage{ulem}

\definecolor{ao_en}{rgb}{0.0, 0.5, 0.0}

\defcitealias{2020MNRAS.497.3830F}{Paper I}

\title[YMC formation under radiative feedback]{
Radiation hydrodynamics simulations of massive star cluster formation in giant molecular clouds
}

\author[Fukushima et al.]{
Hajime Fukushima$^{1}$\thanks{E-mail:fukushima@ccs.tsukuba.ac.jp},
Hidenobu Yajima$^{1}$
\\
$^{1}$Center for Computational Sciences, University of Tsukuba, Ten-nodai, 1-1-1 Tsukuba, Ibaraki 305-8577, Japan\\
}
\date{Accepted XXX. Received YYY; in original form ZZZ}

\pubyear{2020}

\begin{document}
\label{firstpage}
\pagerange{\pageref{firstpage}--\pageref{lastpage}}
\maketitle

\begin{abstract}
By performing three-dimensional radiation hydrodynamics simulations, we study the formation of young massive star clusters (YMCs, $M_{*}>10^4~M_{\odot}$) in clouds with the surface density ranging from  $\Sigma_{\rm cl} = 80$  to $3200~M_{\odot}\;{\rm pc^{-2}}$.
We find that photoionization feedback suppresses star formation significantly in clouds with low surface density. Once the initial surface density exceeds $\sim 100~M_{\odot}\;{\rm pc^{-2}}$ for clouds with $M_{\rm cl}=10^{6}~M_{\odot}$ and $Z= Z_{\odot}$, most of the gas is converted into stars because the photoionization feedback is inefficient in deep gravitational potential.
In this case, the star clusters are massive and gravitationally bounded as YMCs.
The transition surface density increases as metallicity decreases, and it is $\sim 350~M_{\odot}\;{\rm pc^{-2}}$ for $Z=10^{-2}~Z_{\odot}$. We show that more than 10 percent of star-formation efficiency (SFE) is needed to keep a star cluster gravitationally bounded even after the disruption of a cloud.
Also, we develop a semi-analytical model reproducing the SFEs obtained in our simulations. We find that the SFEs are fit with a power-law function with the dependency  $\propto \Sigma_{\rm cl}^{1/2}$ for low-surface density and rapidly increases at the transition surface densities.
The conditions of the surface density and the metallicity match recent observations of giant molecular clouds forming YMCs in nearby galaxies.
\end{abstract}

\begin{keywords}
stars: formation - stars: massive - stars: Population II - H II regions - galaxies: star clusters: general - galaxies: star formation
\end{keywords}


\section{Introduction}\label{introduction} 
Young massive clusters (YMCs) form in giant molecular clouds (GMCs).
Its mass $(\gtrsim 10^4~M_{\odot})$ and density $(\gtrsim 10^3~M_{\odot}{\rm pc^{-3}})$  are larger than typical open clusters in the Milky Way \citep{2010ARA&A..48..431P}. 
YMCs have been frequently observed in local starburst or merger galaxies \citep{2014prpl.conf..291L}. 
Also, YMCs are likely to form in the early Universe, and some of
them remain as globular clusters (GCs) in the present-day galaxies \citep[e.g.,][]{2014CQGra..31x4006K}.
These massive dense clusters can be 
main formation sites of massive stars which regulate the star formation in galaxies via stellar feedback such as photoionization, stellar wind, and supernovae \citep[e.g.,][]{2010MNRAS.402.1536S, 2012ApJ...745...50W, 2013MNRAS.428..154H, 2017ApJ...846...30Y, 2020arXiv201111663Y}.
Besides, YMCs are gravitationally bounded and can produce black hole binaries via multi-body gravitational interaction
\citep[e.g.,][]{2000ApJ...528L..17P, 2017PASJ...69...94F, 2019MNRAS.486.3942K}, likely resulting in gravitational wave events \citep{2016PhRvL.116f1102A}.
Therefore, understanding the formation of YMCs is crucial in astronomy.
However, their formation processes are still debated, especially for the physical conditions of GMCs forming YMCs. 
\citep[e.g.,][]{2014prpl.conf..291L}.

As a cloud collapses gravitationally, gas converts into stars.
Once massive stars form, they give energy and momentum into the surrounding medium via radiation, outflow, and stellar wind referred to as "feedback" \citep[e.g.,][]{2019ARA&A..57..227K}.  
These feedback processes hamper gravitational contraction and star formation. Hence the stellar feedback can be a key factor in the determination of star formation efficiency (SFE) that is defined by the ratio of final stellar mass to initial cloud one.
The SFE needs to be larger than $15-30$ percent to keep the star cluster gravitationally bounded 
\citep[e.g.,][]{1984ApJ...285..141L, 2001MNRAS.321..699K, 2007MNRAS.380.1589B, 2009Ap&SS.324..259G, 2017A&A...605A.119S}.
However, the SFEs are typically less than 10 percent in local galaxies \citep[e.g.,][]{2011ApJ...729..133M,2016ApJ...831...73V, 2019Natur.569..519K,2020MNRAS.493.2872C}.
In the Milky Way, most stars belong to star clusters initially, 
and then disperse gradually after disruption of gas clouds  \citep[e.g.,][]{2003ARA&A..41...57L}.
Therefore, special conditions are likely to be required for the formation of YMCs.

The observations of nearby galaxies showed that the lifetimes of the GMCs are less than $30~{\rm Myr}$ \citep{2009ApJS..184....1K, 2010ARA&A..48..547F}. 
\citet{2019Natur.569..519K} suggested that the duration time of star formation could be as short as the dynamical time of clouds, comparing the spatial correlation between star-forming molecular gas and H{\sc ii} regions.
This timescale is shorter than the lifetime of massive stars.
Therefore, star-forming clouds can be dispersed by pre-supernovae feedback, such as the radiative feedback or the stellar wind
\citep[e.g.,][]{2019Natur.569..519K,2020MNRAS.493.2872C, 2020arXiv201200019K, 2020arXiv201013788C}.

Extreme ultraviolet (EUV; $13.6~{\rm eV} \leqq h \nu \leqq 1~{\rm keV}$) photons ionize and heat up gas.
The high thermal pressure in H{\sc ii} bubbles pushes ambient neutral gas and allows rapid expansion of the bubbles \citep{1987degc.book.....S, 2005ApJ...623..917H, 2006ApJ...646..240H}.
Analytical studies showed that the photoionization feedback suppresses the star formation and makes the SFE lower 
\citep[e.g.,][]{1997ApJ...476..166W, 2002ApJ...566..302M, 2009ApJ...703.1352K, 2010ApJ...710L.142F, 2016ApJ...819..137K, 2020MNRAS.497.5061I}. 
If the density field is inhomogeneous, numerical simulations are required to evaluate the impacts of the photoionization feedback.
In particular, turbulence motion in GMCs induces the high-density contrast and the star formation in a stochastic manner \citep{1981MNRAS.194..809L}.
Recently, radiation hydrodynamics simulations have been performed to study star cluster formation and destruction of clouds 
\citep[e.g.,][]{2010ApJ...715.1302V, 2012MNRAS.427.2852D, 2013MNRAS.430..234D, 2017MNRAS.470.3346H,2017MNRAS.471.4844G, 2017MNRAS.472.4155G, 2019MNRAS.489.1880H, 2020MNRAS.497.4718D, 2018MNRAS.475.3511G, 2020arXiv200804453G, 2021MNRAS.501.4136A, 2021arXiv210302829F}.
\citet{2018ApJ...859...68K} showed that the photoionization feedback played a dominant role in the suppression of star formation. 
They indicated that the SFE increased with the surface density of clouds.
In their simulations, the SFE was less than 10 percent at $\Sigma \sim  100~M_{\odot} {\rm pc^{-2}}$, which is the typical surface density of GMCs in the Milly Way.
The photoionization feedback becomes ineffective if the escape velocity from a clouds exceeds the sound speed in H{\sc ii} regions $c_{\rm s} \sim 10~{\rm km s^{-1}}$ \citep{2012MNRAS.427.2852D, 2012ApJ...758L..28B}.
These massive star-forming clouds typically form in the starburst and merger galaxies \citep[e.g.,][]{2015ApJ...801...25L, 2018ApJ...869..126L, 2020PASJ..tmp..193T}.
Recently, the numerical simulations found that theses clouds could form in the high velocity colliding flow \citep[$\gtrsim 20~{\rm km/s}$, ][]{2020MNRAS.496L...1D, 2020MNRAS.499.1099L, 2020arXiv201111650M}.
Such GMCs with high surface density are likely to be needed for the formation of YMCs against the photoionization feedback.

GMCs are generally optically thick for UV radiation. Therefore, even if the photoionization feedback is ineffective, the radiation pressure can regulate the star formation as stellar luminosity increases \citep{2009ApJ...703.1352K, 2010ApJ...710L.142F, 2010ApJ...709..191M, 2016ApJ...819..137K}.
The energy absorbed by the dust is reprocessed as thermal emission at infrared (IR) wavelengths.
If a cloud is optically thick even for IR photons, the additional radiation pressure is exerted. This further suppresses the star formation \citep[e.g.,][]{2010ApJ...709..191M}.
\citet{2015ApJ...809..187S} performed RHD simulations with the radiation pressure from IR photons.
They suggested that the IR radiation pressure could regulate star formation significantly only if the opacity of a cloud is higher than $\kappa \sim 15~{\rm cm^2 \, g^{-1}}$ that corresponds to the super-solar environment.

The strength of radiative feedback sensitively depends on the metallicity of gas as shown in \citet[][hereafter \citetalias{2020MNRAS.497.3830F}]{2020MNRAS.497.3830F}.
The turbulent motion induces a filamentary structure in which the dust column density can become high enough to attenuate EUV radiation. Therefore, the expansion of H{\sc ii} bubbles is hampered because of the filaments in clouds enriched with metal and dust (see \citetalias{2020MNRAS.497.3830F}).
Whereas, in cases with the metallicity lower than $\sim 0.1~Z_{\odot}$, the filaments are easily disrupted due to the feedback, and the star formation is quenched rapidly.
Besides, the temperature of H{\sc ii} regions increases as the metallicity decreases because of the inefficient metal line cooling 
such as O{\sc ii} and O{\sc iii} \citep{2011piim.book.....D}.
This results in the faster propagation of ionizing front in the low-metallicity gas. 
Therefore, the photoionization feedback is reinforced at the lower-metallicity \citep{2019MNRAS.489.1880H, 2020MNRAS.497.3830F}.

The initial condition of turbulent motions can be another factor to control the SFE.
The turbulent motion is frequently characterized by the virial parameter $\alpha_{\rm vir}$ defined by the ratio of kinetic to gravitational energy \citep{1992ApJ...395..140B}.
Various values have been reported by observations of star-forming clouds in the Milky Way and nearby galaxies
\citep{2010ApJ...723..492R, 2016ApJ...831...16L, 2018ApJ...860..172S, 2020ApJ...901L...8S}.
Recently, \citet{2021ApJ...911..128K} showed that the SFEs decrease as the virial parameter increases.
However, they focused only on the clouds with the surface density similar to the typical one in the Milky Way. Therefore, 
the impacts of turbulent motions is still unclear in more massive compact clouds.

As stated above, the star formation in GMCs is related to various factors as the radiative feedback, the metallicity, the virial parameter in addition to the initial cloud mass and radius. Therefore, the formation of YMCs has not been understood yet because of the complicated processes of them.
In this paper, we study the conditions for the YMC formation, performing the simulations of clouds with various surface densities, metallicities, and turbulent strength.
We perform 3D RHD simulations with the SFUMATO-M1, the modified version of a self-gravitational magnetohydrodynamics code with an Eulerian adaptive mesh refinement (AMR), SFUMATO \citep{2007PASJ...59..905M, 2015ApJ...801...77M}.
We have newly developed the radiation transfer scheme based on the moment method with M1-closure.
In \citetalias{2020MNRAS.497.3830F}, we limited the parameters of clouds whose escape velocity is less than $10~{\rm km s^{-1}}$ to focus on the effects of the photoionization feedback. 
We here study star formation in more compact clouds in the range of surface densities, $ \Sigma = 32-3200~{M_{\odot} \, {\rm pc^{-2}}} $.
We also calculate non-equilibrium chemical reactions with $\rm H_2$ and CO molecules.
This allows us to follow the spatial distribution of the molecules in each evolutionary stage of a cloud.

We organize the rest of the paper as the following.
In Section \ref{numerical_method}, we describe the numerical method and the set-up of simulations.
We show the results of simulations in Section \ref{results}. 
Then, we develop a semi-analytical model with simulation results in Section \ref{analytical_arguments}.
Section \ref{summary} is summary and discussion.
In Appendix \ref{thermochemical_processes} and \ref{thermal_processes}, we show the detail of the chemical network and thermal processes
in our simulations.
We describe the numerical methods of the moment-based scheme of radiation transfer in Section \ref{RT_Moment_solver}.

\section{Numerical Method}\label{numerical_method}
\begin{table*}
    \caption{Models considered}
    \label{tab1}
    \centering

\begin{tabular}{|l|c|c|c|c|c|c|c|c|c|} \hline \hline
model & $M_{\rm cl} \, [ \, M_{\odot} \, ]$ & $R_{\rm cl}\, [ \, {\rm pc} \, ]$ & $Z  \, [ \, Z_{\odot} \, ] $ & $\alpha_{\rm vir}$   & $n_{\rm ini} \, [ \, {\rm cm^{-3}} \, ]$ & $\Sigma_{\rm cl} \, [\, M_{\odot} \, {\rm pc^{-2}} \, ]$ & $\sigma_0 \, [ \, {\rm km s^{-1}} \, ]$  & $t_{\rm ff} \, [ \, {\rm Myr} \, ]$ & $v_{\rm esc} \, [ \, {\rm km/s} \,] $ \\ \hline
M5R5Z0A1 & $10^{5}$ & $5$ & $1$ & 1 & $5600$ & $1300$ & $7.2$ & $0.59$ & $13$ \\
M5R5Z0A2 & $10^{5}$ & $5$ & $1$ & 1 & $5600$ & $1300$ & $10$ & $0.59$ & $13$ \\
M5R5Z-1A1 & $10^{5}$ & $5$ & $10^{-1}$ & 1 & $5600$ & $1300$ & $7.2$ & $0.59$ & $13$ \\
M5R5Z-2A1 & $10^{5}$ & $5$ & $10^{-2}$ & 1 & $5600$ & $1300$ & $7.2$ & $0.59$ & $13$ \\
M5R8Z0A1 & $10^{5}$ & $8$ & $1$ & 1 & $1400$ & $500$ & $5.7$ & $1.2$ & $10$ \\
M5R10Z0A1 & $10^{5}$ & $10$ & $1$ & 1  & $700$ & $320$ & $5.1$ & $1.7$ & $9.3$  \\
M5R10Z0A2 & $10^{5}$ & $10$ & $1$ & 2 & $700$ & $320$ & $7.2$ & $1.7$ &  $9.3$ \\
M5R10Z-1A1 & $10^{5}$ & $10$ & $10^{-1}$ & 1 & $700$ & $320$ & $5.1$ & $1.7$ &  $9.3$  \\
M5R10Z-2A1 & $10^{5}$ & $10$ & $10^{-2}$ & 1 & $700$ & $320$ & $5.1$ & $1.7$  &  $9.3$ \\
M5R12Z0A1 & $10^{5}$ & $12$ & $1$ & 1  & $410$ & $220$ & $4.6$ & $2.2$ & $8.5$  \\
M5R20Z0A1 & $10^{5}$ & $20$ & $1$ & 1  & $87$ & $80$ & $3.6$ & $4.7$ & $6.6$   \\
M5R20Z0A2 & $10^{5}$ & $20$ & $1$ & 2 & $87$ & $80$ & $5.1$ & $4.7$ & $6.6$  \\
M5R20Z-1A1 & $10^{5}$ & $20$ & $10^{-1}$ & 1 & $87$ & $80$ & $3.6$ & $4.7$ & $6.6$  \\
M5R20Z-2A1 & $10^{5}$ & $20$ & $10^{-2}$ & 1 & $87$ & $80$ & $3.6$ & $4.7$ & $6.6$  \\
M6R10Z0A1 & $10^{6}$ & $10$ & $1$ & 1  & $7000$ & $3200$ & $16$ & $0.52$ & $29$ \\
M6R10Z0A2 & $10^{6}$ & $10$ & $1$ & 2 & $7000$ & $3200$ & $23$ & $0.52$ & $29$ \\
M6R10Z-1A1 & $10^{6}$ & $10$ & $10^{-1}$ & 1 & $7000$ & $3200$ & $16$ & $0.52$ & $29$ \\
M6R10Z-2A1 & $10^{6}$ & $10$ & $10^{-2}$ & 1 & $7000$ & $3200$ & $16$ & $0.52$ & $29$ \\
M6R17.5Z-2A1 & $10^{6}$ & $17.5$ & $10^{-2}$ & 1 & $1300$ & $1000$ & $12$ & $1.2$ & $22$ \\
M6R20Z0A1 & $10^{6}$ & $20$ & $1$ & 1 & $870$ & $800$ & $11$ & $1.5$ & $20$  \\
M6R20Z0A2 & $10^{6}$ & $20$ & $1$ & 2 & $870$ & $800$ & $16$ & $1.5$ & $20$  \\
M6R20Z-1A1 & $10^{6}$ & $20$ & $10^{-1}$ & 1 & $870$ & $800$ & $11$ & $1.5$ & $20$  \\
M6R20Z-2A1 & $10^{6}$ & $20$ & $10^{-2}$ & 1 & $870$ & $800$ & $11$ & $1.5$  & $20$ \\
M6R25Z0A1 & $10^{6}$  & $25$ & $1$ & $1$ & $450$ & $510$ & $10$ & $2.1$ & $19$ \\  
M6R25Z-2A1 & $10^{6}$  & $25$ & $10^{-2}$ & $1$ & $450$ & $510$ & $10$ & $2.1$ & $19$ \\  
M6R30Z0A1 & $10^{6}$  & $30$ & $1$ & $1$ & $260$ & $350$ & $9.3$ & $2.7$ & $17$ \\  
M6R30Z-2A1 & $10^{6}$  & $30$ & $10^{-2}$ & $1$ & $260$ & $350$ & $9.3$ & $2.7$ & $17$ \\  
M6R32.5Z0A1 & $10^{6}$  & $32.5$ & $1$ & $1$ & $200$ & $300$ & $8.9$ & $3.1$ & $16$ \\  
M6R35Z0A1 & $10^{6}$  & $35$ & $1$ & $1$ & $160$ & $260$ & $8.6$ & $3.4$ & $16$ \\  
M6R35Z-2A1 & $10^{6}$  & $35$ & $10^{-2}$ & $1$ & $160$ & $260$ & $8.6$ & $3.4$ & $16$ \\  
M6R40Z0A1 & $10^{6}$ & $40$ & $1$ & 1 & $110$ & $200$ & $8.0$ & $4.2$ & $15$ \\
M6R40Z0A2 & $10^{6}$ & $40$ & $1$ & 2 & $110$ & $200$ & $11$ & $4.2$ & $15$ \\
M6R40Z-1A1 & $10^{6}$ & $40$ & $10^{-1}$ & 1 & $110$ & $200$ & $8.0$ & $4.2$ & $15$ \\
M6R40Z-2A1 & $10^{6}$ & $40$ & $10^{-2}$ & 1 & $110$ & $200$ & $8.0$ & $4.2$ & $15$ \\
M6R60Z0A1 & $10^{6}$ & $60$ & $1$ & 1 & $32$ & $88$ & $6.6$ & $7.7$ & $12$ \\
M6R60Z0A2 & $10^{6}$ & $60$ & $1$ & 2 & $32$ & $88$ & $9.3$ & $7.7$ & $12$ \\
M6R60Z-1A1 & $10^{6}$ & $60$ & $10^{-1}$ & 1 & $32$ & $88$ & $6.6$ & $7.7$ & $12$ \\
M6R60Z-2A1 & $10^{6}$ & $60$ & $10^{-2}$ & 1 & $32$ & $88$ & $6.6$ & $7.7$ & $12$ \\
\hline 
\end{tabular}
    \begin{minipage}{1 \hsize}
     Notes. Column 1: model names, Column 2: cloud masses, Column 3: cloud radii, Column 4: metallicity, Column 5: virial parameters, Column 6: initial number densities, Column 7: surface densities defined as $\Sigma_{\rm cl} = M_{\rm cl} / (\pi R_{\rm cl}^2)$, Column 8: three-dimensional velocity dispersions, Column 9: free fall times, Column 10: escape velocity 
     
   \end{minipage}
\end{table*}


To perform RHD simulations, we introduce radiative transfer (RT) into an adaptive mesh refinement code, {\sc SFUMATO} \citep{2007PASJ...59..905M, 2015ApJ...801...77M}, which is dubbed {\sc SFUMATO-M1}.
We develop the RT module based on the moment equations with the M1 closure, which has been adopted in previous studies \citep[e.g.,][]{2013MNRAS.436.2188R, 2013ApJS..206...21S, 2013ApJ...772..127T, 2015MNRAS.449.4380R, 2019MNRAS.485..117K}.
In the moment-based scheme, the RT is calculated from the gradient of radiation energy densities between adjacent cells.
In \citetalias{2020MNRAS.497.3830F}, we used the RT module based on the adaptive ray-tracing method \citep{2002MNRAS.330L..53A} which was developed by \citet{2020ApJ...892L..14S}.
Its calculation amount is proportional to the number of sources. 
Therefore, if the number exceeds $\sim 1000$, the computational cost is  quite expensive even with current computational facilities. This regulates the parameter space with low-mass and low-density clouds. 
Thus, in this study, we adopt the moment-based scheme of which the calculation amount is independent of the number of sources.

In addition, in the case of massive compact clouds, the dust column density becomes high and can be opaque even to infrared radiation from dust. 
This gives the additional force on gas.
Therefore, we here take into account the RT of IR photons.
The detailed methodology of RT and test simulations are described in Appendix \ref{RT_Moment_solver}.
Also, {\sc{SFUMATO}} includes the non-equilibrium chemistry solver developed by \citet{2020ApJ...892L..14S}, allowing us to calculate the thermal evolution of low-temperature gas accurately.
The details of the thermal processes and chemical network in our simulations are described in Appendix \ref{thermochemical_processes} and \ref{thermal_processes}.

In this study, we set the size of calculation boxes as three times the cloud radius $(R_{\rm cl})$ on a side.
The maximum refinement level is fixed at $l_{\rm max} =4$ and
the minimum cell size is $\Delta x = 0.059~{\rm pc} (R_{\rm cl}/20 ~{\rm pc})$.
The simulations end at when $4 \times$ free-fall time $(t_{\rm ff})$ elapses.

\subsection{Basic Equations of hydrodynamics}

We perform the three-dimensional hydrodynamics simulations with Cartesian coordinate.
We solve the following basic equations of compressible hydrodynamics: the equation of continuity, 
\begin{align}
	\frac{\partial \rho}{\partial t} + \nabla \cdot \left( \rho \bm{v} \right) = 0, \label{renzoku_eq}
\end{align}
the equation of motion, 
\begin{align}
	\frac{\partial \left( \rho \bm{v} \right)}{\partial t} + \nabla \cdot \left( \rho \bm{v} \otimes \bm{v} \right) + \nabla P = \rho \left( \bm{g} + \bm{f} \right) , \label{undo_eq}
\end{align}
and the energy equation,
\begin{align}
	\frac{\partial \left( \rho E \right)}{\partial t} + \nabla \cdot \left[ \left( \rho E + P \right) \bf{v} \right] = \rho \left( {\bf{g} + \bf{f}} \right) \cdot {\bf{v}} + \Gamma - \Lambda, \label{energy_eq}
\end{align}
where $E$ is total energy defined as
\begin{align}
	E = \frac{| {\bf v} |^2}{2} + \left( \gamma - 1 \right)^{-1} \frac{P}{\rho}, \label{energy_eq2}
\end{align}
$\rho$, $P$, $\bf{v}$, $\bf{g}$, $\Gamma$ and $\Lambda$ are the density, pressure, velocity, gravitational force, the heating and cooling functions.
We estimate the adiabatic exponent $\gamma$ as in
\citet{1998ApJ...508..141O}.
In equations \eqref{undo_eq} and \eqref{energy_eq}, $\bf{f}$ represents the radiation pressure force.
We consider radiation pressure induced by absorption of EUV photons by H{\sc i} and dust grains, FUV and IR photons absorbed by dust grains.

We take into account the chemical networks of 11 species:
$\rm H$, $\rm H_2$, $\rm H^-$, $\rm H^+$, $\rm H_2^+$, $\rm e$, $\rm CO$, C{\sc ii}, O{\sc i}, O{\sc ii}, and O{\sc iii} \citep[see also,][]{2020MNRAS.497.3830F}.
The number density of the $i$-th specie is calculated as
\begin{align}
  \frac{\partial (y_{i} n_{\rm H})}{\partial t} + \nabla \cdot (y_{i} n_{\rm H} {\bf{v}}) = y_{i}  n_{\rm H} R_{i}, \label{chemical_net}
\end{align}
where $n_{\rm H}$ is the number density of hydrogen nuclei, $y_{i}=n_{i}/n_{\rm H}$ is the fractional abundance of each chemical specie, and $R_{i}$ is the reaction rate of the $i$-th specie.

We adopt the simple chemical network of \citet{1997ApJ...482..796N} for CO formation, which has been used for the RHD simulations \citep[e.g.,][]{2006ApJ...646..240H, 2018ApJ...857...57N}.
To obtain the relative abundances of O{\sc i}, O{\sc ii}, and O{\sc iii}, we adopt the following procedure \citep[see also,][]{2020MNRAS.497..829F}.
We assume that the ionization rate of O{\sc i} is equal to that of H{\sc i} because the ionization potential energies of O{\sc i} and H{\sc i} are similar.
The abundances of doubly ionized oxygen are determined as the chemical equilibrium between O{\sc ii} and O{\sc iii}.

As the heating/cooling processes, we include (1) the thermal processes related to the chemical reactions, (2) line cooling of $\rm H_2$, C{\sc ii}, CO, O{\sc i}, O{\sc ii}, and O{\sc iii}, and (3) energy transfer between gas and dust grains.
As the radiative processes, we include heating of H{\sc i} photoionization and $\rm H_2$ photodissociation.
The dust grain temperature is estimated from the energy balance between the absorption/emission of radiation and energy transfer with gas.
We set the temperature floor at $T=10~{\rm K}$ as \citet{2020MNRAS.497.3830F}.
All thermal processes are summarized in Appendix \ref{thermal_processes}.

\subsection{Sink particles and radiation sources}\label{sec_radiation_sources}

If gas cells reach the maximum refinement level, they form sink particles as star clusters. 
The model of the sink particles was developed in \citet{2015ApJ...801...77M}.
We summarize their procedures as follows.
More details of them were described in Appendix A of \citet{2015ApJ...801...77M}.

The production procedure of sink particles is the same as used in \citet{2020MNRAS.497.3830F}.
The conditions for the production are as follows
\citep{2010ApJ...713..269F}: (1) the gas density is higher than the threshold value $\rho_{\rm thr}$; (2) the birthplace is the local minimum of gravitational potential; (3) the velocity divergence $\nabla \cdot \bf{v}$ and all the eigenvalues of the symmetric part of the velocity gradient tensor $\nabla \bf{v}$ are negative; (4) the sum of the thermal, kinetic and gravitational energy is negative.
The density threshold is set as $\rho_{\rm thr} = 8.86 c_{\rm s}^2/(\pi G \Delta x^2) = 9.3 \times 10^{-19} \, (R_{\rm cl}/20\,{\rm pc})^{-2} \, {\rm g \, cm^{-3}}$ \citep{2013ApJS..204....8G, 2018ApJ...859...68K} where $c_{\rm s}$ is the sound speed, and $T=20~{\rm K}$ is used here.
We set a sink radius as $r_{\rm sink} = 2 \Delta x$.
We do not consider mergers of sink particles in this work.

In a hydrodynamics step, each sink particle accretes the excess of the gas mass over the threshold density $\rho_{\rm thr}$ in a sink radius.
The sink particles also obtain momentum from the accreted gas.
Therefore, the total momentum of gas and sink particles is conserved.

The gravitational force among the sink particles is considered for calculating their motions. 
Here, we set the softening radius as $r_{\rm soft}= r_{\rm sink}$.
The gravitational force from gas is also taken into account.
When a cell is inside the softening radius of a sink particle, we subdivided each cell into $8^3$ cells, and the gravitational forces from the subcells are summed \citep{2004ApJ...611..399K}.
The force from sink particles to gas is evaluated as the back reactions of them. 
Thus the conservation of the momentum is ensured through the interaction between sink particles and gas.

We estimate the luminosity of a sink particle from the stellar isochrone of \citet{2015MNRAS.452.1068C} and the Chabrier IMF \citep{2003PASP..115..763C} with the stellar mass range from $0.1~M_{\odot}$ to $150~M_{\odot}$.
We use their isochrone at $1~{\rm Myr}$, which roughly corresponds to star formation duration in the simulations.
In our model, the mass-luminosity relation to all sink particles is fixed.
However, we may overestimate the luminosity if the sink mass is not massive enough to form a massive star that dominates the luminosity. Therefore, as stellar radiation sources, we consider only sink particles with masses higher than $50~M_{\odot}$
where the expected number of massive stars ($>10\,M_{\odot}$) is larger than unity.
In our models, resolution dependence of radiative properties appears when the typical sink mass is comparable to the threshold value.
As shown in Appendix \ref{appendix_resolution_study} and \ref{appendix_sink_evolv}, it occurs in the diffuse and low-mass cloud models (M5R20Z0A1 in Table \ref{tab1}) because the typical sink mass is comparable to $\sim 50~M_{\odot}$. 
On the other hand, the peaks of the sink mass distributions appear at $\sim 100~M_{\odot}$ in the cases of $M_{\rm cl}=10^6~M_{\odot}$.
In such a case, total emissivity of sink particles does not depend on the numerical resolution significantly.
Note that the results of star cluster formation can also change with the models of radiation sources, such as ones with the IMF-averaged values or stochastic sampling \citep{2019MNRAS.488.2970G, 2021MNRAS.502.5417S}.
To reproduce the IMF smoothly, the total stellar mass of $\gtrsim 10^{4}~M_{\odot}$ is required \citep[e.g.,][]{2016ApJ...819..137K}.
Thus, our results for star clusters with the mass $\gtrsim 10^{4}~M_{\odot}$ are unlikely to be sensitive to the models of radiation sources.

\subsection{Initial conditions}

As initial conditions, we consider the clouds with the mass $M_{\rm cl} = 10^5$ or $10^6 \, M_{\odot}$, and the radius ranging from $R_{\rm cl} = 5$ to $60~\rm pc$ as summarized in Table \ref{tab1}.
The above parameters include the typical values of surface density of GMCs in the Milky Way,  $\Sigma \sim 100 ~{M_{\odot} \, {\rm pc^{-2}}}$ \citep[e.g.,][]{1987ApJ...319..730S, 2010ApJ...723..492R}, and $\sim 3 \times 10^3~M_{\odot} \, {\rm pc^{-2}}$ as in merger galaxies \citep[e.g.,][]{2016ApJ...831...16L, 2018ApJ...860..172S}.
Massive stars can form only in massive clouds.
\citet{2010ApJ...723L...7K} found that there is the lower limit for massive star formation as $M_{\rm cl} \gtrsim 870~M_{\odot}(r/{\rm pc})^{1.33}$ in the Milky Way \citep[see also][]{2018MNRAS.473.1059U}.
Our models fulfill this condition.
In paper I, we focused only on the clouds whose escape velocities are less than the sound speed of ionized gas $(v_{\rm esc} < 10 ~{\rm km/s})$ and investigated the cloud disruption process due to the photo-ionization feedback.
As the cloud becomes massive and compact, the deep gravitation potential well can host gas against the feedback, likely resulting in higher star formation efficiency.
Therefore we here extend the parameter range to take the clouds with $v_{\rm esc} > 20~{\rm km/s}$ into account.
Also, we change the metallicity from $Z=10^{-2} Z_{\odot}$ to $Z_{\odot}$ and investigate its impacts on the star-formation process.
The initial gas temperature is set as $T_{\rm g} = 10~{\rm K}$. 
The abundance of molecules sensitively depends on the metallicity.
We assume that the gas is fully molecular at $Z=10^{-1}Z_{\odot}$ and $Z_{\odot}$.
In the case with $Z=10^{-2}~Z_{\odot}$, 
we set $y_{\rm H_2} = 3 \times 10^{-3}$ as the initial abundance because the formation time scale of $\rm H_{2}$ is longer than the dynamical time of the cloud.

The simulations finish when the elapsed time reaches four times the free-fall time of the clouds.

As in paper I, we take into account the turbulent velocity field. 
We assume that the velocity power spectrum of $P(k) \propto k ^{-4}$ where $k$ is the wave number.
The velocity field is generated with the random seeds.
We re-scale the same velocity fields in each simulation to investigate feedback dependence on the cloud mass and compactness.
We note that the results, such as the SFEs, are affected by choice of this seed \citep{2020arXiv200804453G, 2021ApJ...911..128K}.
The amplitude of the turbulent motion is defined by the virial parameter as 
\begin{align}
    \alpha_0 = \frac{E_{\rm kin}}{|2E_{\rm grav}|} = \frac{5 \sigma_0^2 R_{\rm cl}}{3 G M_{\rm cl}}, \label{virial_param}
\end{align}
where $\sigma_0$, $E_{\rm kin}$, and $E_{\rm grav}$ are the 3D velocity dispersion, kinetic, and gravitational energy.
We adopt the value of virialized clouds as $\alpha_0 = 1$ as the fiducial value. 
Recent observations showed that the virial parameters somewhat change depending on the formation sites of the clouds.
\citep[e.g.,][]{2010ApJ...723..492R, 2016ApJ...831...16L, 2018ApJ...860..172S}.
Thus, we also perform the simulations with $\alpha = 2$ only with $Z=Z_{\odot}$ to clarify the effects of turbulent motions.
Here, the turbulence decays freely from the start of the simulations, and it leads to the rapid collapse of the cloud compared with the cases with the external turbulent sources \citep[e.g.,][]{2011ApJ...740...74K}.

\section{Results}\label{results}

We first study the effects of the cloud compactness by comparing M6R20Z0A1 and M6R40Z0A1 in Section \ref{star_cluster_form}.
In Section \ref{star_formation_each_cloud}, we show the SFEs and properties of star clusters in all models.
In Section \ref{rapid_increase_of_SFE}, we discuss the threshold surface density for the formation of gravitationally bounded clusters.
In Section \ref{virial_parameter} and \ref{dpdce_on_esc_velocity}, we describe the dependence of SFEs on the virial parameters and the escape velocities of the clouds.
In Table \ref{tab_results}, we summarize the results obtained in our simulations.
In Appendix \ref{appendix_resolution_study}, we compare the results with the different maximum refinement levels and their impacts on our study. 

\subsection{Star cluster formation in massive cloud}\label{star_cluster_form}
\subsubsection{compact cloud model}

\begin{figure*}
    \begin{center}
    \includegraphics[width=170mm]{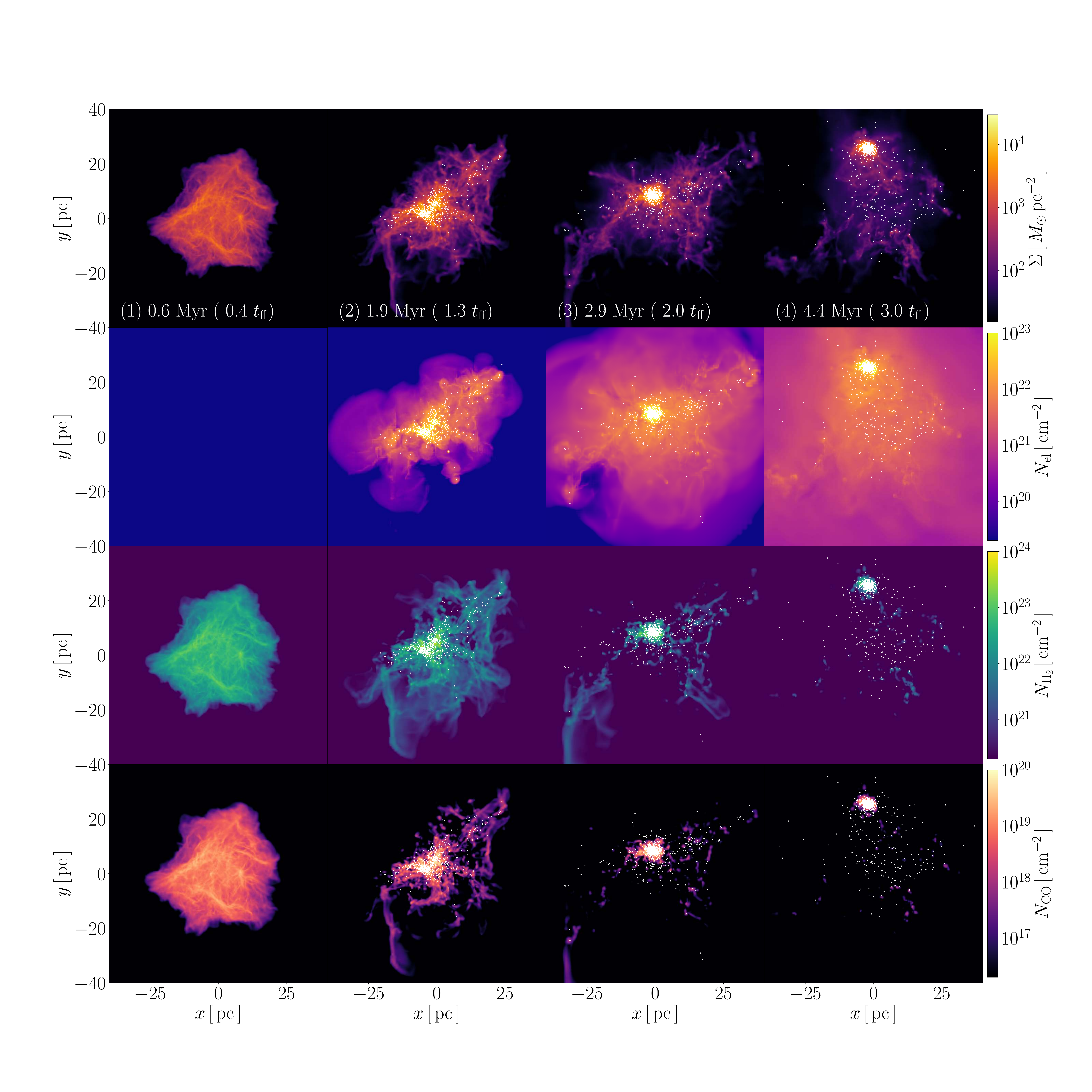}
    \end{center}
    \caption{
    The structure of a cloud with $(M_{\rm cl}, R_{\rm cl}, Z) = (10^{6}~M_{\odot}, 20~{\rm pc}, Z_{\odot})$ at $t=0.6 (0.4~t_{\rm ff}), 1.9 (1.3~t_{\rm ff}), 2.9 (2 ~t_{\rm ff})$ and $4.4 ~{\rm Myr} (3~t_{\rm ff})$.
    Each panel shows the surface density of gas, the number column densities of electron, $\rm H_2$ and CO molecules from top to bottom.
    Stellar particles are shown as white dots. 
   }
    \label{fig_sigma_m6r20z1}
\end{figure*}

\begin{figure*}
    \begin{minipage}{0.45\hsize}
    	\begin{center}
    		\includegraphics[width=80mm]{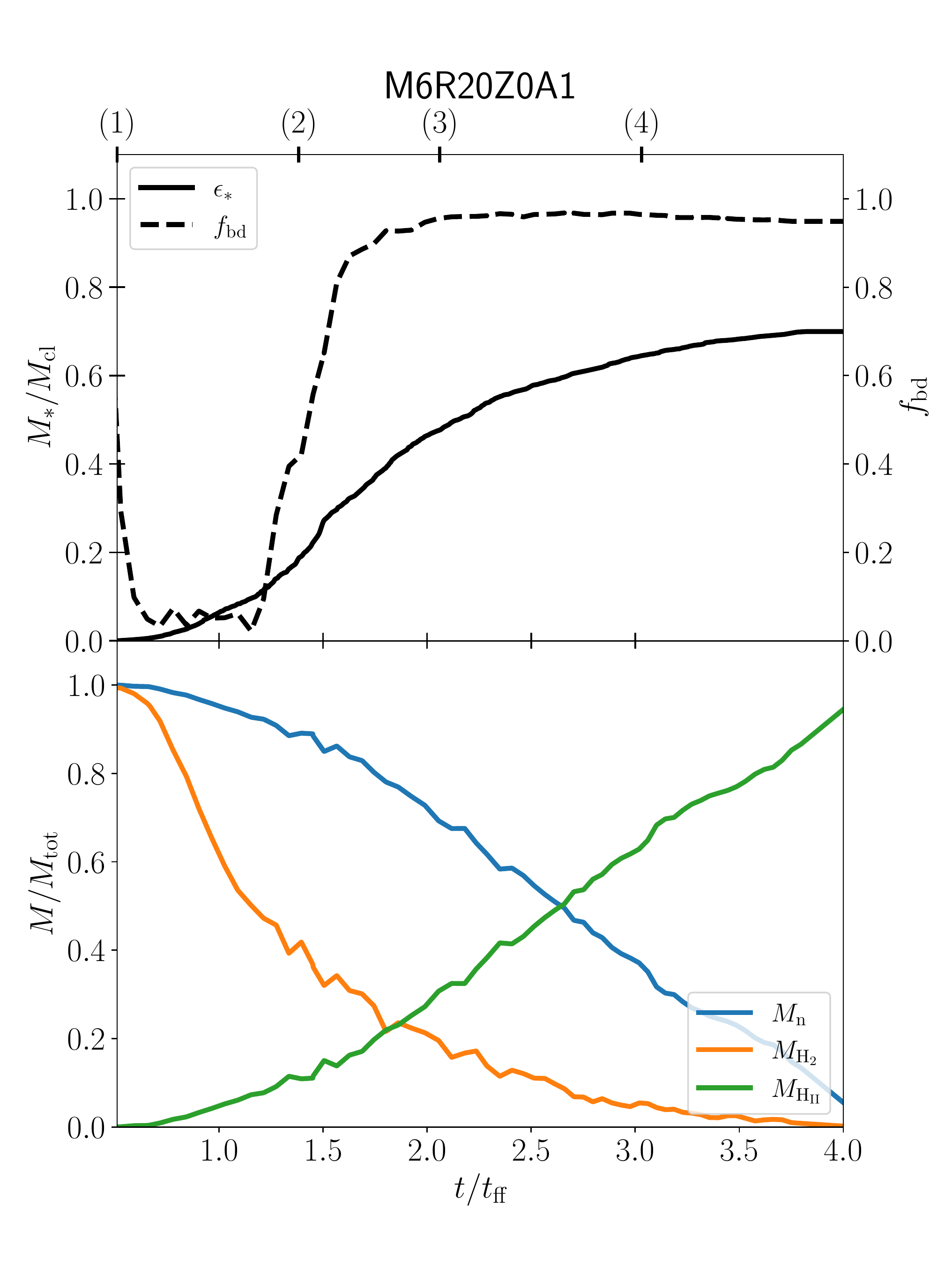}
    	\end{center}
    \end{minipage}
    \begin{minipage}{0.45\hsize}
    	\begin{center}
    		\includegraphics[width=80mm]{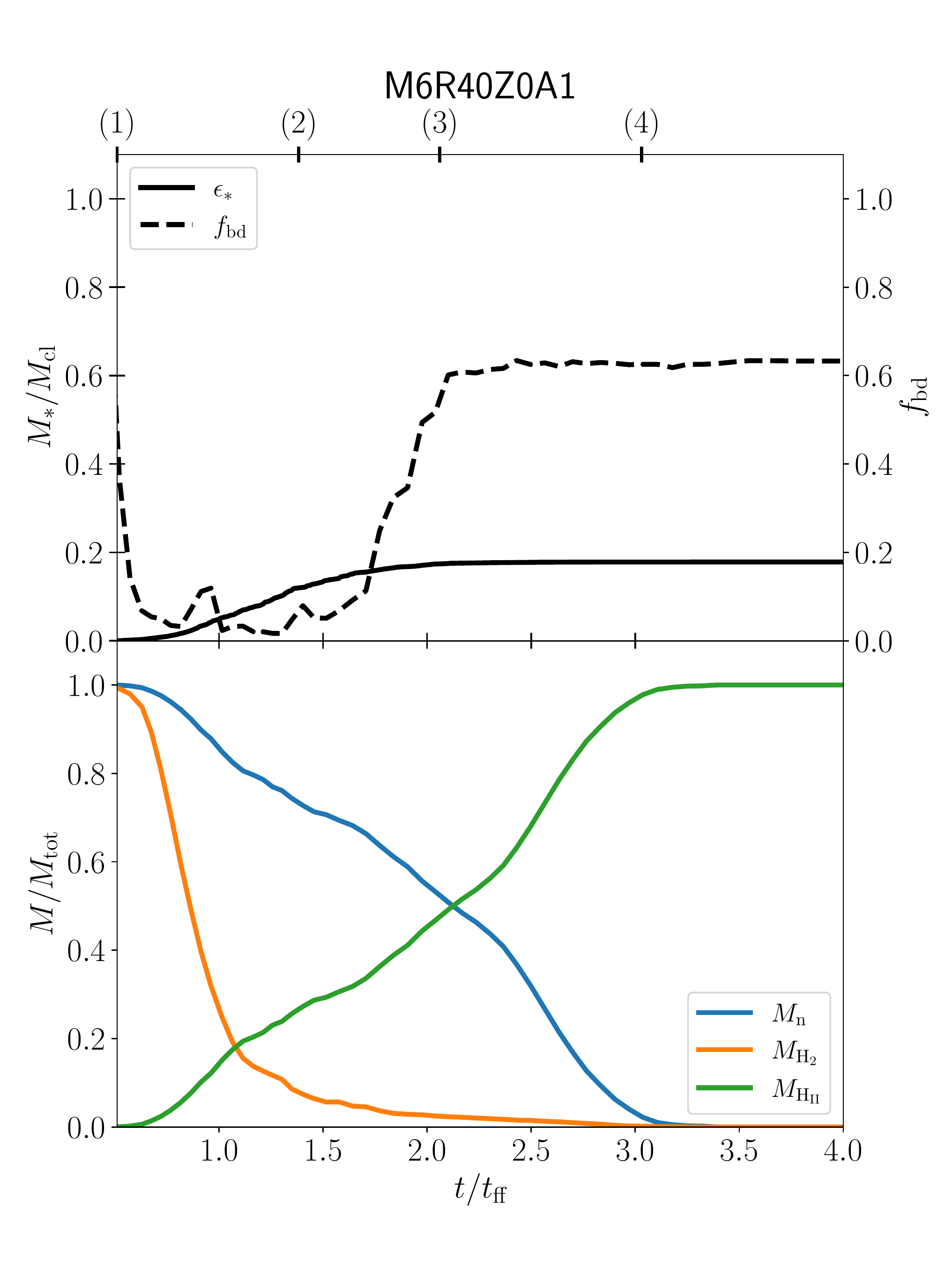}
    	\end{center}
    \end{minipage}
    \caption{
    Upper panels: The time evolution of the stellar mass normalized by the initial cloud one and the bound fraction in the models of $(M_{\rm cl}, R_{\rm cl}, Z) = (10^6~M_{\odot}, 20~{\rm pc}, Z_{\odot})$ (left) and $(10^6~M_{\odot}, 40~{\rm pc}, Z_{\odot})$ (right). Solid and dashed black lines represent the total stellar masses and the bound fractions.
    Lower panels: The masses of neutral atomic hydrogen ($M_{\rm n}$, blue), molecular hydrogen ($M_{\rm H_2}$, orange), and ionized gas ($M_{\rm HII}$, green) normalized by total gas mass.
    The labels of (1)-(4) represent the four epochs as shown in Figure \ref{fig_sigma_m6r20z1} and \ref{fig_sigma_m6r40z1}. 
   }
    \label{fig_mass_m6r20z1al1}
\end{figure*}

We first describe the model of $(M_{\rm cl}, R_{\rm cl}, Z_{\rm cl}) = (10^6M_{\odot}, 20{\rm pc}, Z_{\odot})$ as the fiducial one. 
In Figure \ref{fig_sigma_m6r20z1}, we show the time evolution of this cloud.
In the early phase, the turbulent motion controls the gas dynamics and induces the formation of the filamentary structures in which stars form.
Yet, most new born stars distributed at the center of the cloud and are tightly bound at $t\sim 1.3 \, t_{\rm ff}$.
The star cluster at the center keeps the compactness until the end of the simulation.
In this model, some remains even at $t=3~t_{\rm ff}$.

The second column of Figure \ref{fig_sigma_m6r20z1} shows the electron column density, representing the spatial distributions of H{\sc ii} regions. 
The emissivity of ionizing photons increases with the total stellar mass, 
and the H{\sc ii} regions gradually expand. 
At $t=1.3 \, t_{\rm ff}$, the H{\sc ii} regions are localized around the stars.
Then, the entire volume is highly ionized at $t>2 ~t_{\rm ff}$.

The spatial distributions of $\rm H_2$ and $\rm CO$ molecules also change with the star formation (see the third and fourth columns of Figure \ref{fig_sigma_m6r20z1}).
After the onset of the star formation, FUV photons dissociate these molecules.
Especially, the molecules are efficiently photodissociated in the low-density regions.
On the other hand, the molecules in the high-density filaments can survive against the radiative feedback for a longer time because of the dust-shielding effect \citep{2020MNRAS.497.3830F}.
However, most of the molecules disappear finally because they are consumed by the star formation.

Figure \ref{fig_mass_m6r20z1al1} presents the stellar mass and the bound fraction of the star cluster as a function of time.
Here, we adopt a similar procedure in \citet{2017A&A...605A.119S} to estimate the bound fraction.
We calculate the gravitational binding energy and the kinetic energy of sink particles, and then remove the unbound particles.
We repeat the above processes until all remained sink particles are gravitationally bounded.
The star formation begins when half of the free-fall time elapses. 
The total stellar mass reaches 0.1 of the cloud mass at $t\sim 1.3 \, t_{\rm ff}$.
At the time, the bound fraction rapidly increase. 
The increase of the bound fraction at ${\rm SFE} \gtrsim 0.1$
is consistent with the previous study \citep{2017A&A...605A.119S}.
At $t \sim 1.5 \, t_{\rm ff}$, the bound fraction exceeds 0.9, and it is almost constant until the end of the simulation.
The star formation rate (SFR) also starts to increase at $t\sim 1.3 \, t_{\rm ff}$, and it continues until $t \sim 2\, t_{\rm ff}$.
After that, radiative feedback slows down the star formation, but gas around the star cluster cannot disperse due to the deep gravitational potential.
Thus, the star formation continues for a long time until $t\sim 3.5 \, t_{\rm ff}$, finally resulting in the SFE of 0.7.

The bottom panels of Figure \ref{fig_mass_m6r20z1al1} show the time evolution of non-ionized (atomic and molecular hydrogen), ionized, and $\rm H_{2}$ molecule gas.
The hydrogen molecule abundance rapidly decreases as the star formation proceeds because of the photodissociation in the low-density regions.
It becomes $\lesssim 0.2$ when half of the cloud is converted into stars.
At $t \gtrsim 3.5\, t_{\rm ff}$, all hydrogen molecules are dissociated and more than eighty percent of the gas is ionized. In this phase, the gravitational collapse is prevented due to the pressure of the ionized gas, resulting in the quenching of the star formation.

\subsubsection{Diffuse cloud model}

\begin{figure*}
    \begin{center}
    \includegraphics[width=170mm]{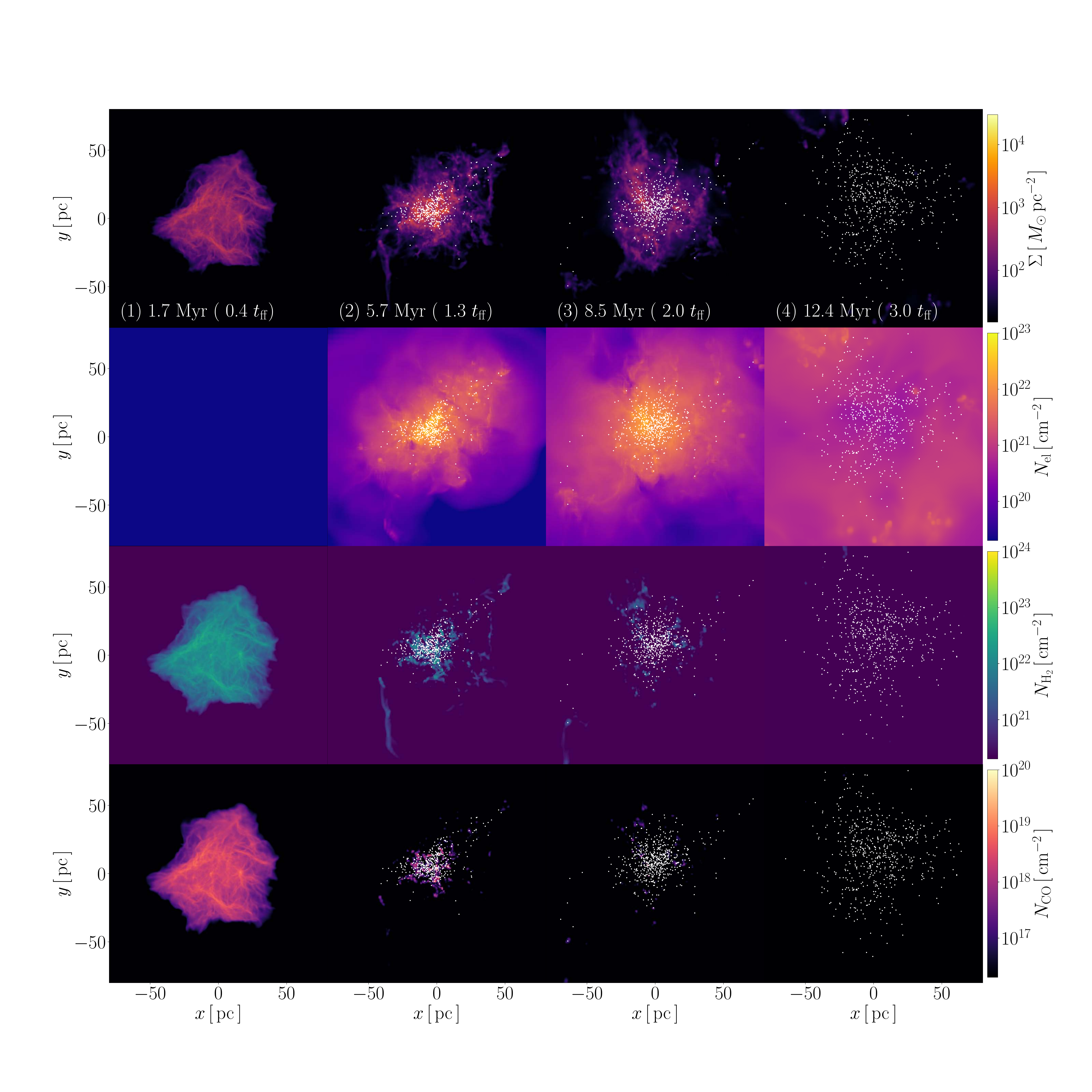}
    \end{center}
    \caption{
    Same as Figure \ref{fig_sigma_m6r20z1}, but for the case with $(M_{\rm cl}, R_{\rm cl}, Z) = (10^{6}~M_{\odot}, 40~{\rm pc}, Z_{\odot})$.
    The snapshots are at $t=1.7 (0.4~t_{\rm ff})$, $5.7 (1.3~t_{\rm ff})$, $8.5 (2 ~t_{\rm ff})$, and $12.4~{\rm Myr} (3 ~ t_{\rm ff})$.
   }
    \label{fig_sigma_m6r40z1}
\end{figure*}

To investigate the impacts of the cloud compactness on the star formation, we simulate a diffuse cloud with $(M_{\rm cl}, R_{\rm cl}, Z) = (10^6~M_{\odot}, 40~{\rm pc}, Z_{\odot})$ and compare with the fiducial model.
Figure \ref{fig_sigma_m6r40z1} shows the evolution of this cloud as in Figure \ref{fig_sigma_m6r20z1}.
Same as the fiducial model, the turbulent motions drive the filamentary structures in the early phase.
The star formation begins at $t\sim 0.5~t_{\rm ff}$.
In this case, the stellar distribution is more extended, not concentrated at the center, unlike the fiducial compact case.
(see Fig \ref{fig_sigma_m6r40z1}-3 and 4).
As the stellar mass increases, the gas is evacuated due to the radiative feedback. The shallower gravitational potential well can not hold the hot ionized gas. Therefore, once the cloud is ionized, it is dispersed ($t \sim 3~t_{\rm ff}$).

In this diffuse case, the dust column densities in the filaments are not enough to shield FUV photons. Therefore, the photodissociation fronts rapidly expand.
At $t \sim 1.3~t_{\rm ff}$, more than ninety per cent of hydrogen molecules are dissociated as shown in Figure \ref{fig_sigma_m6r40z1}-(2).
After $t \sim 2.0 ~ t_{\rm ff}$, all hydrogen molecules disappear, resulting in the quenching of the star formation.

The stellar mass and the bound fraction are presented in the right panels of Figure~\ref{fig_mass_m6r20z1al1}.
The star formation begins at $t \sim 0.5 ~t_{\rm ff}$ and continues until $t \sim 2 ~ t_{\rm ff}$.
The final SFE is 0.18 which is lower than the fiducial case by a factor of $\sim 4$.
Unlike the fiducial case, the SFR does not change with the time for $t \sim 0.5 - 2.0~t_{\rm ff}$. 
The bound fraction is much smaller than 0.1 at $t<1.5~t_{\rm ff}$, but it rapidly increases once the SFE exceeds $\sim 0.15$.
At the end of star formation, the bound fraction becomes $\sim 0.6$ and remains constant until the end of the simulation.
This trend is consistent with the previous case.

\subsubsection{low-metallicity cases}

\begin{figure*}
    \begin{center}
    \includegraphics[width=170mm]{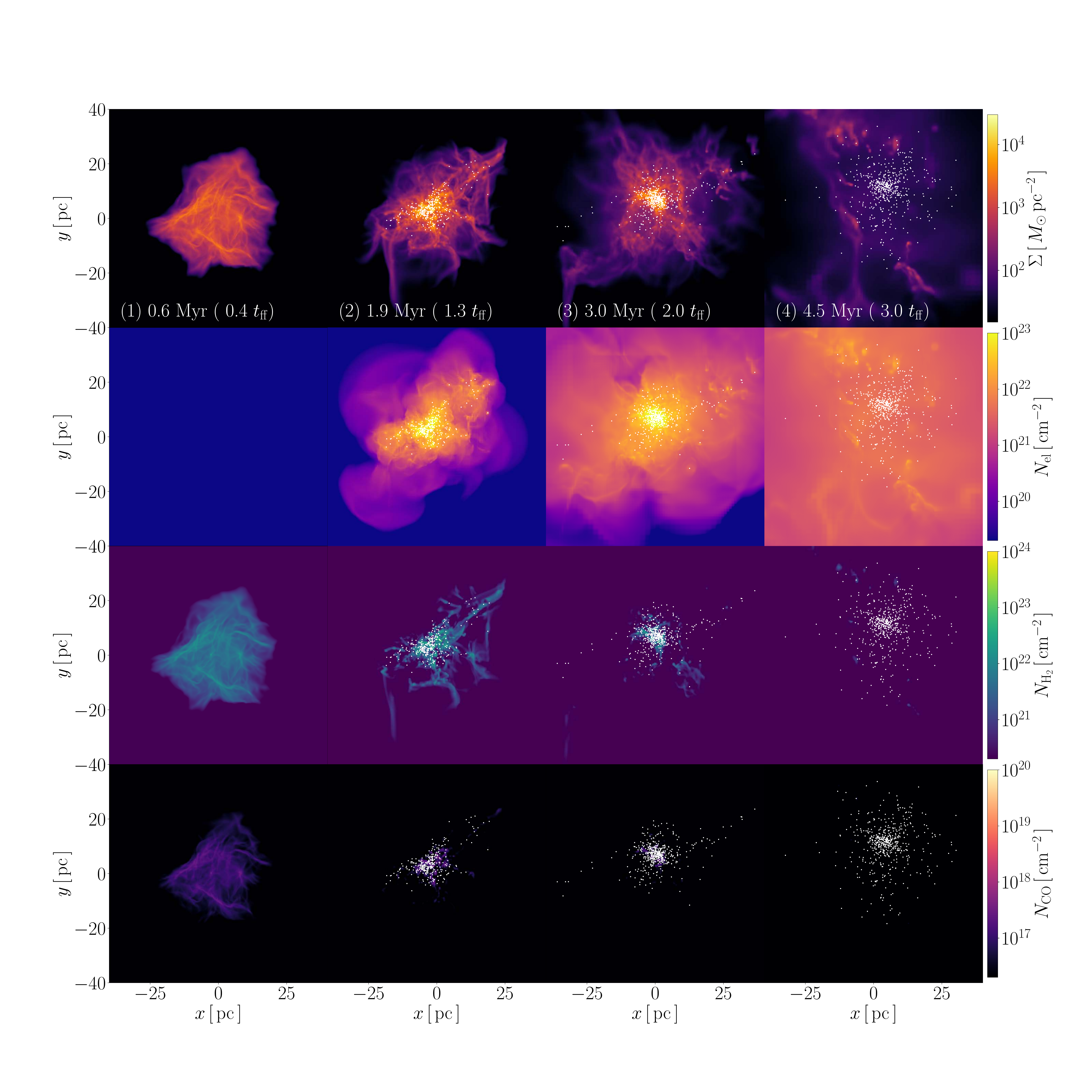}
    \end{center}
    \caption{
    Same as Figure \ref{fig_sigma_m6r20z1} but for the case with $(M_{\rm cl}, R_{\rm cl}, Z) = (10^{6}~M_{\odot}, 20~{\rm pc}, 10^{-2}Z_{\odot})$.
   The snapshots are at $t = 0.6 (0.4~t_{\rm ff})$, $1.9 (1.3~t_{\rm ff})$, $3 (2 ~t_{\rm })$, and $4.5~{\rm Myr} (3.1~t_{\rm ff})$.
    }
    \label{fig_sigma_m6r20z-2}
\end{figure*}

As shown in \citetalias{2020MNRAS.497.3830F}, the SFE sensitively depends on the metallicity of clouds. If the metallicity is low, the temperature of ionized gas becomes higher and the dust-shielding of UV radiation is ineffective, resulting in the lower SFE. However, the previous study did not take massive compact clouds into account. Therefore, here 
we show the results of the low-metallicity cloud with $(M_{\rm cl}, R_{\rm cl}, Z) = (10^{6}~M_{\odot}, 20~{\rm pc}, 10^{-2}Z_{\odot})$.

As in the fiducial model, the low-metallicity cloud begins to form stars at $t \sim 1.3~t_{\rm ff}$. Then, the dense massive cluster form at the center, and most of the stars are bound with $f_{\rm bd} = 0.89$.
However, the SFE ($\epsilon_{*} = 0.28$) is smaller than that of the fiducial model ($\epsilon_{*} = 0.7$) due to the stronger feedback.

Figure \ref{fig_sigma_m6r20z-2} also shows the column densities of ${\rm H_2}$ and ${\rm CO}$ molecules.
In this case, we assume that the gas is atomic initially.
Also, molecular formation on dust grains is insufficient to make gas fully molecular.
Thus, the molecules only exist in the high-density regions, and the maximum column density of $\rm H_2$ molecules are 10 times smaller than that of the case with $Z=Z_{\odot}$.
After the onset of the star formation, FUV photons start to dissociate molecules in the low-density regions.
In \citetalias{2020MNRAS.497.3830F}, we showed that FUV photons completely photodissociate molecules in the lower-mass clouds.
The star formation occurs from atomic gas at the low-metallicity \citep{2012ApJ...759....9K}.  
On the other hand, the molecules remain in the high density filaments in the cloud with $M_{\rm cl} = 10^{6} M_{\odot}$, as shown in Figure \ref{fig_sigma_m6r20z-2}.
In \citetalias{2020MNRAS.497.3830F}, we estimated the optical depth of the filaments for FUV photons, considering the Jeans unstable filaments as 
\begin{align}
	\tau_{\rm sf} = 4.6 \left(\frac{\Sigma}{800~M_{\odot}{\rm pc^{-2}}} \right) \left( \frac{\alpha_{0}}{1} \right) \left( \frac{Z}{10^{-2}Z_{\odot}} \right). \label{tau_filament}
\end{align}
The star-forming filaments are optically-thick for FUV photons, and dust shielding prevents photodissociation even at $Z=10^{-2}Z_{\odot}$.
Therefore, the molecular gas remains until the high-density filaments are completely destroyed.

\subsection{Dependence of SFEs and SFRs on cloud properties}\label{star_formation_each_cloud}
\subsubsection{Star formation efficiencies and lifetimes of clouds} \label{star_form_efficiency}

\begin{figure}
    \begin{center}
    	\includegraphics[width=\columnwidth]{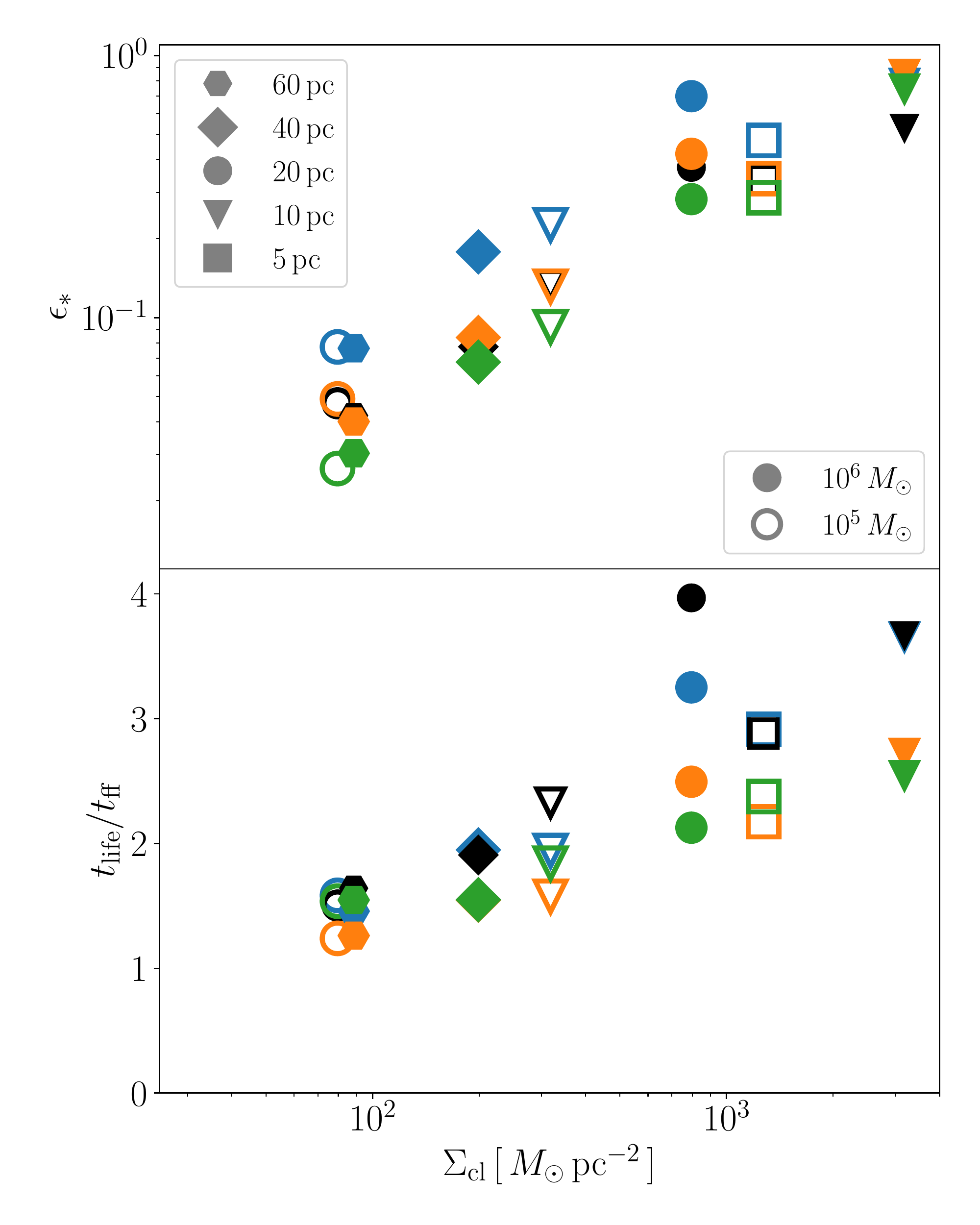}
    \end{center}
    \caption{
  	Upper panel: The SFEs of clouds as a function of surface densities. 
  	Each symbol represents the different cloud masses: $M_{\rm cl} = 10^5~M_{\odot}$ (open) and $10^{6} ~M_{\odot}$ (filled), and with the different radii $R_{\rm cl} = 5 ~{\rm pc}$ (square), $10~{\rm pc}$ (triangle), $20~{\rm pc}$ (circle), $40~{\rm pc}$ (diamond), and $60~{\rm pc}$ (hexagon).
  	Each color shows the different metallicities: $Z=10^{-2}Z_{\odot}$ (green), $10^{-1}Z_{\odot}$ (orange), and $Z_{\odot}$ (blue).
	The black symbols represent the cases with $\alpha_{0} =2$.
  	Lower panel: The ratio of the cloud lifetimes to the free-fall times.
       }
    \label{fig_SFEs}
\end{figure}

By using the total stellar mass $M_{*}$ at the end of the simulations, we estimate the SFEs as $\epsilon_* = M_{*}/M_{\rm cl}$.
Figure \ref{fig_SFEs} shows the SFEs as the function of the cloud surface densities. 
As in \citetalias{2020MNRAS.497.3830F}, the SFE gradually increases with the surface density \citep[see also,][]{2010ApJ...710L.142F, 2016ApJ...829..130R,2017MNRAS.471.4844G,2018MNRAS.475.3511G, 2018ApJ...859...68K, 2019MNRAS.489.1880H}. 
In the case with $Z=Z_{\odot}$, the SFE is $0.08$ at $\Sigma_{\rm cl}=80~M_{\rm \odot} {\rm pc^{-2}}$, which is similar with the results in the previous works of \citet{2018ApJ...859...68K} and \citet{2019MNRAS.489.1880H}.
Comparing with the results in \citetalias{2020MNRAS.497.3830F}, 
we simulate conditions with higher turbulent velocity.
The higher turbulent motions make the cloud less bound, and the star formation rate decreases.
In such a case, more gas is evaporated before it is converted into stars. 
Thus, the SFE is about half as small as that of our previous work. 
We will discuss the impacts of the turbulent motion in detail at Sec. \ref{virial_parameter}.

For the higher surface densities, the SFEs can exceed 0.1. For the clouds with $M_{\rm cl} = 10^5 ~M_{\odot}$, the SFE achieves 0.47 at $\Sigma_{\rm cl} = 1300 ~M_{\odot} \, {\rm  pc^{-2}}$ (M5R5Z0A1).
In the cases with the massive clouds of $M_{\rm cl} = 10^6\, M_{\odot}$, the SFEs  become higher, and 0.7 for $800~M_{\odot}\, {\rm pc^{-2}}$ (M6R20Z0A1) and 0.78 for $3200~M_{\odot}\, {\rm pc^{-2}}$ (M6R10Z0A1). 
In these two clouds, the cloud escape velocities exceed $20~{\rm km/s}$ that is higher than the sound speed of the H{\sc ii} regions.
In such cases, the deep gravitational potential can hold the ionized gas and allows further star formation \citep{2012MNRAS.427.2852D, 2013MNRAS.430..234D}.

In the massive compact clouds, the radiation pressure plays a main role in suppressing the star formation instead of the photoionization feedback \citep[e.g.,][]{2009ApJ...703.1352K, 2010ApJ...710L.142F}.
The radiation pressure is due to the direct light from the radiation sources and dust thermal emission (IR emission).
We can roughly estimate the conditions that radiation pressure suppresses star formation, assuming the spherical shell accelerated by the radiation pressure.
The radiation pressure of the direct light on the shell is estimated as
\begin{align}
	F_{\rm rad} = \frac{L}{c} = \frac{\epsilon_* M_{\rm cl} l_{*}}{c}, \label{radiation_pressure}
\end{align}
where $\epsilon_*$ is the SFE, and $l_*$ is the luminosity per unit mass defined as $L = l_{*}M_{*}$.
We estimate the gravitational force on the shell at the outer boundary of the cloud as \citep{2016ApJ...819..137K}
\begin{align}
	F_{\rm grav} = \frac{GM_{\rm sh} (M_* + M_{\rm sh}/2)}{R_{\rm cl}^2} = \frac{GM_{\rm cl}^2 }{2 R_{\rm cl}^2} \left( 1 - \epsilon_*^2 \right), \label{gravity}
\end{align}
where the self-gravity of the shell is included. 
The shell mass is $M_{\rm sh} = (1 - \epsilon_*) M_{\rm cl}$.
By taking the balance between the radiation pressure (Equation~\ref{radiation_pressure}) and the gravitational force (Equation~\ref{gravity}), the SFE is estimated as \citep{2016ApJ...819..137K, 2019ARA&A..57..227K}, 
\begin{align}
	\frac{\epsilon_*}{1 - \epsilon_*^2} = \frac{\Sigma _{\rm cl}}{\Sigma_{\rm PR, UV}}, \label{SFE_RP}
\end{align}
where $\Sigma_{\rm cl}$ is the cloud's surface density, and 
\begin{align}
	\Sigma_{\rm RP, UV} = \frac{2 l_*}{\pi G c } \simeq 3.9 \times 10^3 ~M_{\odot}{\rm pc^{-2}}.\label{Sigma_RPUV}
\end{align}
The radiation pressure can evacuate the gas if the SFE exceeds the above.
Note that, the SFE can be larger than that given in Equation \eqref{Sigma_RPUV} because there is a time lag of a free-fall time from the launching to the gas evacuation.
Equation \eqref{SFE_RP} indicates the radiation pressure from the direct light is effective only if $\Sigma_{\rm cl} < \Sigma_{\rm RP, UV}$. If $\Sigma_{\rm cl} > \Sigma_{\rm RP, UV}$, the SFE becomes almost unity.
All clouds in this study satisfy the condition of $\Sigma_{\rm cl} < \Sigma_{\rm RP, UV}$.
However, the surface density of the clouds with $(M_{\rm cl}, R_{\rm cl}) = (10^6 ~M_{\odot}, 10~{\rm pc})$ is close to this limit.
Thus, the star formation hardly stops, and the SFE is close to 80 per cent.
The radiation pressure from dust thermal emission is only important if the optical depth of the IR light is larger than unity as
\begin{align}
	\tau_{\rm IR} = \rho \kappa_{\rm IR} R_{\rm cl} = 0.5 \left( \frac{\Sigma}{3.200 ~M_{\odot} \, {\rm pc^{-2}}} \right) \left( \frac{\kappa_{\rm IR}}{1 ~{\rm cm^{2} g^{-1}}} \right) > 1. \label{optical_depth_IR}
\end{align}
In our simulations, the initial states of clouds are optically thin for IR photons even at the solar metallicity.
Thus, the radiation pressure of IR photons is unlikely to be effective.
However, as the cloud collapses, the surface density can be larger than $10^{4}~M_{\odot} {\rm pc^{-2}}$ locally as shown in Figure \ref{fig_sigma_m6r20z1}.
In these high-density regions, the IR radiation pressure works in suppressing star formation.
If an individual star is resolved in simulations, the IR radiation pressure also plays a role in regulating the accretion rate onto a proto-star \citep[e.g.,][]{1987ApJ...319..850W, 2009Sci...323..754K, 2010ApJ...722.1556K, 2018MNRAS.473.4754F, 2020MNRAS.497..829F}. Our current simulations do not take this process into account. 
Note that \citet{2015ApJ...809..187S} argued that the IR radiation pressure works only for cases with the opacity $\kappa_{\rm IR} \gtrsim 15 ~{\rm cm^2} \, {\rm g^{-1}}$ which corresponds to the metallicity higher than the solar abundance.
Hence, the radiation pressure from stellar light is a dominant feedback in regulating star formation.

The SFE decreases at lower-metallicity. 
In the cases of the clouds with $\Sigma_{\rm cl} < 10^{3}~M_{\odot}{\rm pc^{-2}}$, the SFEs for $Z=10^{-2}Z_{\odot}$ are lower than that for $Z=Z_{\odot}$ by a factor of $\sim 3$.
In this case, the dust shielding of ionizing photons is ineffective \citep{2020MNRAS.497.3830F}.
Therefore, the propagation of the ionization front is significantly faster, resulting in the early quenching of the star formation.
On the other hand, if $\Sigma > 10^3~M_{\odot}{\rm pc^{-2}}$, the SFE is not sensitive to the metallicity.
In these compact clouds, the main feedback is the radiation pressure due to the photon absorption by neutral hydrogen or interstellar dust.
Since the optical depth of the dust for UV photons is larger than unity even at $Z=10^{-2}Z_{\odot}$, the strength of the radiation pressure does not change significantly irrespective of the metallicity.

\begin{table*}
    \caption{Simulation results}
    \label{tab_results}
    \centering

\begin{tabular}{|l|l|l|l|l|l|l|l|l|l|} \hline \hline

model & $\epsilon_*$  & $\epsilon_{\rm ff, 0}$ & $\epsilon_{\rm ff}$ & $t_{\rm life} \, [\, {\rm Myr} \, ]$ & $t_{\rm dr} \, [ \, {\rm Myr} \, ]$ & $M_{\rm bd} \, [ \, {M_{\odot}} \, ]$ & $f_{\rm bd}$ & $r_{\rm h} \, [ \, {\rm pc} \, ]$ & $t_{\rm rh} \, [ \, {\rm Myr} \, ]$ \\ \hline
M5R5Z0A1& $0.47$ & $0.15$ & $0.30$  $^{\rm a}$ & $1.71 (2.91t_{\rm ff})$ & $1.24 (2.12t_{\rm ff})$ & $4.38 \times 10^4$ & $0.92$ & $0.88$ & $1.09 \times 10^2$ \\
M5R5Z0A2& $0.33$ & $0.11$ & $0.19$  $^{\rm a}$ & $1.69 (2.88t_{\rm ff})$ & $1.24 (2.12t_{\rm ff})$ & $2.91 \times 10^4$ & $0.87$ & $1.06$ & $1.22 \times 10^2$ \\
M5R5Z-1A1& $0.34$ & $0.15$ & $0.22$ & $1.27 (2.17t_{\rm ff})$ & $0.83 (1.41t_{\rm ff})$ & $2.77 \times 10^4$ & $0.82$ & $0.87$ & $8.87 \times 10$ \\
M5R5Z-2A1& $0.29$ & $0.11$ & $0.21$  $^{\rm a}$ & $1.39 (2.38t_{\rm ff})$ & $0.92 (1.57t_{\rm ff})$ & $2.38 \times 10^4$ & $0.83$ & $0.99$ & $1.01 \times 10^2$ \\
M5R8Z0A1& $0.32$ & $0.13$ & $0.24$  $^{\rm a}$ & $2.81 (2.37t_{\rm ff})$ & $1.88 (1.59t_{\rm ff})$ & $2.72 \times 10^4$ & $0.86$ & $1.75$ & $2.51 \times 10^2$ \\
M5R10Z0A1& $0.23$ & $0.11$ & $0.17$ & $3.22 (1.94t_{\rm ff})$ & $1.94 (1.17t_{\rm ff})$ & $1.54 \times 10^4$ & $0.68$ & $2.50$ & $3.43 \times 10^2$ \\
M5R10Z0A2& $0.13$ & $0.05$ & $0.10$  $^{\rm a}$ & $3.86 (2.33t_{\rm ff})$ & $2.65 (1.6t_{\rm ff})$ & $1.81 \times 10^2$ & $0.01$ & $0.69$ & $1.07 \times 10$ \\
M5R10Z-1A1& $0.13$ & $0.08$ & $0.14$ & $2.61 (1.58t_{\rm ff})$ & $1.37 (0.83t_{\rm ff})$ & $6.15 \times 10^2$ & $0.05$ & $1.63$ & $5.64 \times 10$ \\
M5R10Z-2A1& $0.09$ & $0.05$ & $0.08$ & $3.05 (1.84t_{\rm ff})$ & $1.77 (1.07t_{\rm ff})$ & $1.43 \times 10^2$ $^{\rm b}$ & $0.02$  $^{\rm c}$ & - $^{\rm d} $ & -  $^{\rm e}$  \\
M5R12Z0A1& $0.16$ & $0.09$ & $0.16$ & $3.70 (1.70t_{\rm ff})$ & $2.06 (0.94t_{\rm ff})$ & $5.95 \times 10^3$ & $0.37$ & $2.39$ & $2.23 \times 10^2$ \\
M5R20Z0A1& $0.08$ & $0.05$ & $0.08$ & $7.43 (1.58t_{\rm ff})$ & $3.89 (0.83t_{\rm ff})$ & $5.39 \times 10^2$ & $0.07$ & $3.51$ & $1.70 \times 10^2$ \\
M5R20Z0A2& $0.05$ & $0.03$ & $0.05$ & $7.04 (1.50t_{\rm ff})$ & $3.70 (0.79t_{\rm ff})$ & $1.09 \times 10^2$ & $0.02$ & $0.19$ & $1.38$ \\
M5R20Z-1A1& $0.05$ & $0.04$ & $0.09$ & $5.81 (1.24t_{\rm ff})$ & $2.19 (0.47t_{\rm ff})$ & $4.84 \times 10^2$ & $0.10$ & $2.29$ & $8.71 \times 10$ \\
M5R20Z-2A1& $0.03$ & $0.02$ & $0.03$ & $7.21 (1.54t_{\rm ff})$ & $3.39 (0.72t_{\rm ff})$ & $1.71 \times 10^2$ $^{\rm b}$ & $0.06$  $^{\rm c}$ & - $^{\rm d} $ & -  $^{\rm e}$  \\
M6R10Z0A1& $0.78$ & $0.20$ & $0.38$  $^{\rm a}$ & $1.91 (3.65t_{\rm ff})$ & $1.46 (2.79t_{\rm ff})$ & $7.51 \times 10^5$ & $0.96$ & $1.28$ & $6.14 \times 10^2$ \\
M6R10Z0A2& $0.53$ & $0.14$ & $0.18$  $^{\rm a}$ & $1.92 (3.66t_{\rm ff})$ & $1.49 (2.84t_{\rm ff})$ & $4.87 \times 10^5$ & $0.92$ & $1.69$ & $7.75 \times 10^2$ \\
M6R10Z-1A1& $0.84$ & $0.29$ & $0.46$  $^{\rm a}$ & $1.42 (2.72t_{\rm ff})$ & $0.95 (1.82t_{\rm ff})$ & $8.19 \times 10^5$ & $0.97$ & $0.69$ & $2.52 \times 10^2$ \\
M6R10Z-2A1& $0.74$ & $0.28$ & $0.40$  $^{\rm a}$ & $1.33 (2.54t_{\rm ff})$ & $0.86 (1.63t_{\rm ff})$ & $7.32 \times 10^5$ & $0.99$ & $0.66$ & $2.27 \times 10^2$ \\
M6R175Z-2A1& $0.41$ & $0.18$ & $0.29$ & $2.56 (2.11t_{\rm ff})$ & $1.54 (1.27t_{\rm ff})$ & $3.82 \times 10^5$ & $0.94$ & $1.90$ & $8.36 \times 10^2$ \\
M6R20Z0A1& $0.70$ & $0.20$ & $0.39$  $^{\rm a}$ & $4.82 (3.25t_{\rm ff})$ & $3.50 (2.36t_{\rm ff})$ & $6.70 \times 10^5$ & $0.96$ & $1.73$ & $9.17 \times 10^2$ \\
M6R20Z0A2& $0.37$ & $0.09$ & $0.12$  $^{\rm a}$ & $5.88 (3.97t_{\rm ff})$ & $4.63 (3.12t_{\rm ff})$ & $3.53 \times 10^5$ & $0.94$ & $2.33$ & $1.10 \times 10^3$ \\
M6R20Z-1A1& $0.42$ & $0.16$ & $0.28$  $^{\rm a}$ & $3.70 (2.49t_{\rm ff})$ & $2.48 (1.67t_{\rm ff})$ & $4.05 \times 10^5$ & $0.96$ & $1.98$ & $9.09 \times 10^2$ \\
M6R20Z-2A1& $0.28$ & $0.13$ & $0.19$ & $3.15 (2.13t_{\rm ff})$ & $1.94 (1.31t_{\rm ff})$ & $2.50 \times 10^5$ & $0.88$ & $3.53$ & $1.78 \times 10^3$ \\
M6R25Z0A1& $0.62$ & $0.21$ & $0.40$  $^{\rm a}$ & $5.73 (2.77t_{\rm ff})$ & $3.90 (1.88t_{\rm ff})$ & $6.01 \times 10^5$ & $0.97$ & $2.03$ & $1.12 \times 10^3$ \\
M6R25Z-2A1& $0.16$ & $0.08$ & $0.13$ & $3.85 (1.86t_{\rm ff})$ & $2.27 (1.10t_{\rm ff})$ & $5.54 \times 10^4$ & $0.34$ & $5.74$ & $1.99 \times 10^3$ \\
M6R30Z0A1& $0.53$ & $0.20$ & $0.35$  $^{\rm a}$ & $6.77 (2.49t_{\rm ff})$ & $4.42 (1.62t_{\rm ff})$ & $5.15 \times 10^5$ & $0.97$ & $2.11$ & $1.11 \times 10^3$ \\
M6R30Z-2A1& $0.12$ & $0.06$ & $0.09$ & $5.21 (1.92t_{\rm ff})$ & $3.16 (1.16t_{\rm ff})$ & $2.40 \times 10^3$ & $0.02$ & $0.87$ & $3.51 \times 10$ \\
M6R32.5Z0A1& $0.41$ & $0.16$ & $0.24$  $^{\rm a}$ & $7.23 (2.35t_{\rm ff})$ & $4.68 (1.53t_{\rm ff})$ & $3.90 \times 10^5$ & $0.96$ & $2.38$ & $1.18 \times 10^3$ \\
M6R35Z0A1& $0.27$ & $0.12$ & $0.18$ & $7.23 (2.11t_{\rm ff})$ & $4.52 (1.32t_{\rm ff})$ & $2.39 \times 10^5$ & $0.88$ & $5.23$ & $3.14 \times 10^3$ \\
M6R35Z-2A1& $0.09$ & $0.05$ & $0.10$ & $5.25 (1.53t_{\rm ff})$ & $2.69 (0.79t_{\rm ff})$ & $1.65 \times 10^3$ $^{\rm b}$ & $0.02$  $^{\rm c}$ & - $^{\rm d} $ & -  $^{\rm e}$  \\
M6R40Z0A1& $0.18$ & $0.09$ & $0.13$ & $8.17 (1.95t_{\rm ff})$ & $5.08 (1.21t_{\rm ff})$ & $7.77 \times 10^4$ & $0.44$ & $10.0$ & $5.26 \times 10^3$ \\
M6R40Z0A2& $0.08$ & $0.04$ & $0.06$ & $8.00 (1.91t_{\rm ff})$ & $5.30 (1.26t_{\rm ff})$ & $7.68 \times 10^2$ & $0.01$ & $4.27$ & $2.57 \times 10^2$ \\
M6R40Z-1A1& $0.08$ & $0.05$ & $0.09$ & $6.48 (1.54t_{\rm ff})$ & $3.66 (0.87t_{\rm ff})$ & $1.64 \times 10^3$ & $0.02$ & $3.97$ & $2.99 \times 10^2$ \\
M6R40Z-2A1& $0.07$ & $0.04$ & $0.08$ & $6.49 (1.55t_{\rm ff})$ & $3.39 (0.81t_{\rm ff})$ & $1.75 \times 10^3$ $^{\rm b}$ & $0.03$  $^{\rm c}$ & - $^{\rm d} $ & -  $^{\rm e}$  \\
M6R60Z0A1& $0.08$ & $0.05$ & $0.09$ & $11.2 (1.46t_{\rm ff})$ & $5.87 (0.76t_{\rm ff})$ & $1.41 \times 10^3$ $^{\rm b}$ & $0.02$  $^{\rm c}$ & - $^{\rm d} $ & -  $^{\rm e}$  \\
M6R60Z0A2& $0.04$ & $0.02$ & $0.04$ & $12.6 (1.64t_{\rm ff})$ & $7.80 (1.01t_{\rm ff})$ & $9.21 \times 10^2$ $^{\rm b}$ & $0.02$  $^{\rm c}$ & - $^{\rm d} $ & -  $^{\rm e}$  \\
M6R60Z-1A1& $0.04$ & $0.03$ & $0.06$ & $9.72 (1.26t_{\rm ff})$ & $4.67 (0.61t_{\rm ff})$ & $1.56 \times 10^3$ $^{\rm b}$ & $0.04$  $^{\rm c}$ & - $^{\rm d} $ & -  $^{\rm e}$  \\
M6R60Z-2A1& $0.03$ & $0.02$ & $0.03$ & $11.9 (1.55t_{\rm ff})$ & $6.38 (0.83t_{\rm ff})$ & $1.61 \times 10^3$ $^{\rm b}$ & $0.05$  $^{\rm c}$ & - $^{\rm d} $ & -  $^{\rm e}$  \\

\hline 
\end{tabular}
    \begin{minipage}{1 \hsize}
     Notes. Column 1: model names, Column 2: star formation efficiency, Column 3: star formation rate per free fall time, Column 4: specific star formation rate per free fall time in the duration time of star formation $t_{\rm dr}$, Column5: lifetime of clouds, Column 6: duration time of star formation, Column 7: gravitationally bounded mass, Column 8: gravitational bound fraction of star cluster, Column 9: half mass radius, Column 10: half mass relaxation time.
     
     $^{\rm a}$ We estimate the parameters $\epsilon_{\rm ff}$ from the star formation rates in the period between the start of the star formation and $1~t_{\rm ff}$ later. 
     
     $^{\rm b, c}$ All star particles are not gravitationally bounded, and we obtain the gravitational bounded mass and the bound fractions from the maximum mass of sink particles.
     
     $^{\rm d, e}$ The half mass radius and the relaxation time are not evaluated if all stellar particles are gravitationally unbounded. 
   \end{minipage}
\end{table*}


The bottom panel of Figure \ref{fig_SFEs} shows the cloud lifetimes.
The cold gas and molecules inside it are almost dispersed by radiative feedback at the end of the star formation.
Therefore, the end of the cloud lifetimes is almost the same as the end of the star formation.
Here, we define the cloud lifetimes as the period from the starting time of the simulations to the time when the total stellar mass reaches 95 percent of the final stellar mass.
The lifetimes increase slightly in the higher surface density clouds.
Also, we estimate the duration time of the star formation defined as the time required for the total stellar mass to increase from 5 percent to 95 percent of the final stellar mass (Table \ref{tab_results}).
The star formation mainly occurs after one free-fall time in all cases and continues for one free-fall time in the clouds with $\Sigma_{\rm cl} < 10^3~M_{\odot} {\rm pc^{-2}}$.
In the cases with $\Sigma_{\rm cl} = 3200~M_{\odot} {\rm pc^{-2}}$, the clouds are not dispersed, and the star formation continues until $t\sim 3.3 t_{\rm ff}$. 
In the low-metallicity environments, the lifetimes are shorter than that of the solar metallicity.

\begin{figure}
    \begin{center}
    	\includegraphics[width=\columnwidth]{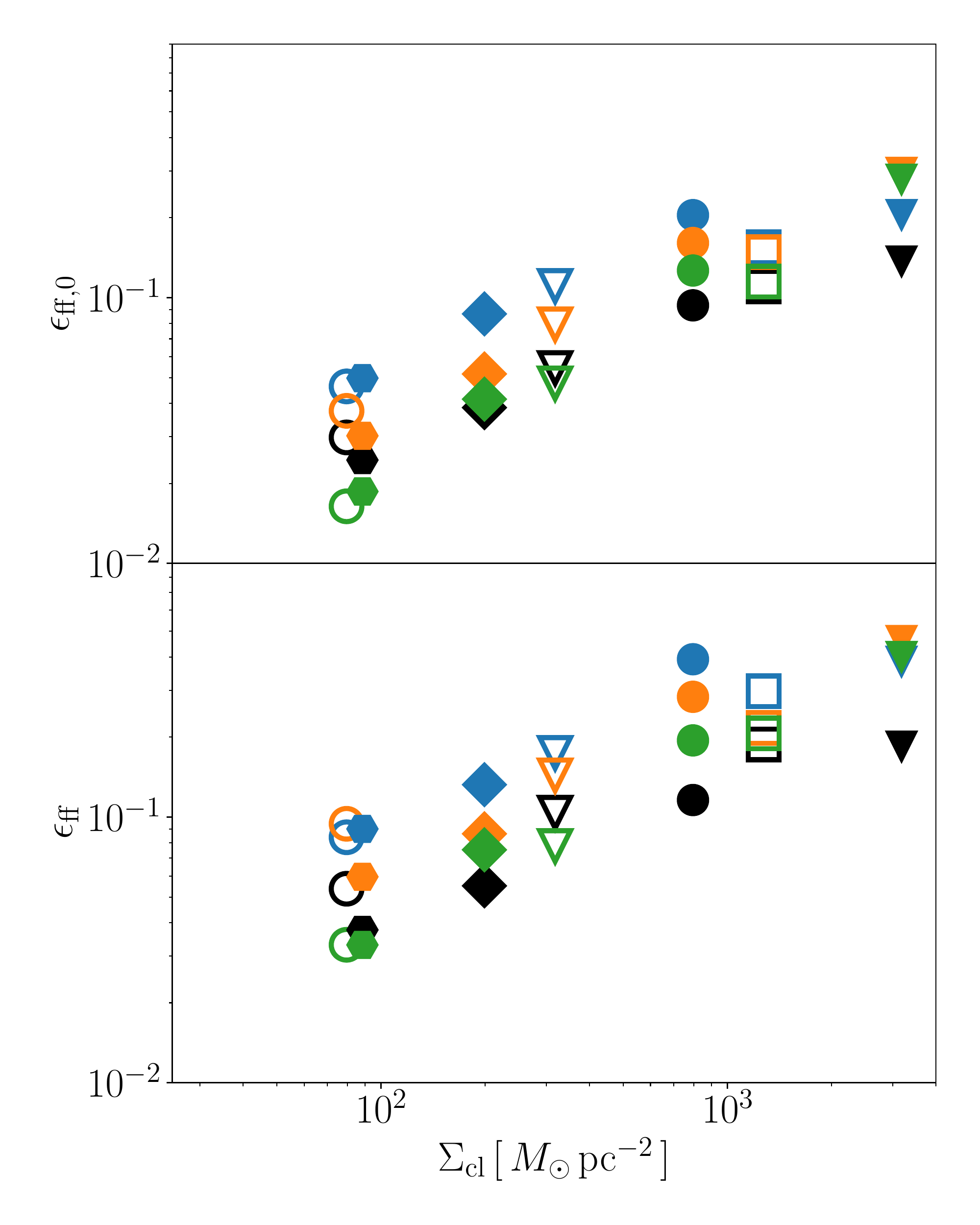}
    \end{center}
    \caption{
    The star formation parameters $\epsilon_{\rm ff, 0}$ and $\epsilon_{\rm ff}$ as the function of the surface densities.
    The upper panel shows the fraction of the cloud mass transformed into stars per free-fall time averaged in the lifetimes of clouds $(t_{\rm life})$ (Eq. \ref{def_mdot}).
    The lower panel represents the same fraction but averaged in the duration time of the star formation $(t_{\rm dr})$ (Eq. \ref{def_mdot2}).
    The styles of symbols are the same as in Figure \ref{fig_SFEs}.
}
    \label{fig_epsff}
\end{figure}

The star formation rate (SFR) is also related to the initial conditions of clouds.
In observations, the SFR is speculated by measuring properties of star clusters and cloud life times for various clouds. By introducing a conversion efficiency from the gas into stars $\epsilon_{\rm ff, 0}$, the SFR is expressed as
\begin{align}
\dot M_{*, 0} &=   \frac{M_*}{t_{\rm life}} = \epsilon_{\rm ff, 0} \frac{ M_{\rm cl}}{t_{\rm ff}} , 
\label{def_mdot} 
\end{align}
where $t_{\rm life}$ is the cloud lifetime and $t_{\rm ff} = \sqrt{3 \pi /(32 G \rho)}$ is the free-fall time.
This estimate corresponds to take into account the various evolution stages of clouds over the lifetime. However,
as shown in Figure \ref{fig_mass_m6r20z1al1}, most stars form within a shorter time scale than the cloud lifetime.  
Therefore, we here evaluate the SFR by using the duration time of the star formation ($t_{\rm dr}$):
\begin{align} 
\dot M_* = \frac{M_*}{t_{\rm dr}} = \epsilon_{\rm ff} \frac{M_{\rm cl}}{t_{\rm ff}}. \label{def_mdot2} 
\end{align}
The above $\epsilon_{\rm ff}$ is related to the star formation efficiency as
\begin{align}
  \epsilon_{\rm ff} = \epsilon_* \frac{t_{\rm ff}}{t_{\rm dr}}. \label{eps_ff}
\end{align}
As shown in Figure \ref{fig_mass_m6r20z1al1}, the duration time of the star formation is much longer when the stellar core formation occurs.
As discussed in Appendix \ref{appendix_resolution_study}, our current simulations may somewhat overestimate the impact of the radiation pressure. However, even in such a case, the radiation pressure cannot evacuate the gas and allows the gas accretion onto the stellar core until its mass becomes massive enough. Consequently, the overestimated radiation pressure does not alter the SFE, while it makes the duration time of star formation longer.   
Here, we underestimate the SFR of the main phase of the star formation.
Thus, we recalculate $\epsilon_{\rm ff}$ in the periods from the starting time of the star formation to $1.0~t_{\rm ff}$ later if the duration time is longer than $1.5~t_{\rm ff}$. 
Figure \ref{fig_epsff} presents $\epsilon_{\rm ff,0}$ and $\epsilon_{\rm ff}$ in each cloud. 
Because of $t_{\rm dr}$ shorter than $t_{\rm life}$, $\epsilon_{\rm ff}$ is higher than $\epsilon_{\rm ff, 0}$ by a factor of $\sim 2$.
Both panels show that the SFRs increase with the higher surface densities.
Besides, the SFRs are smaller as the metallicity decreases. 
In Section \ref{rapid_increase_of_SFE}, we further discuss the increase of the SFRs in the compact clouds.

\subsection{Rapid increase of SFE}\label{rapid_increase_of_SFE}

As in \citetalias{2020MNRAS.497.3830F}, if the clouds are disrupted due to the photoionization feedback, the SFE is expressed as the power-law function of the surface density $(\propto \Sigma^{-1/2} )$.
However, in the case of massive compact clouds, the SFE is unlikely to obey the simple power-law function because the photoionization does not suppress the star formation.
We perform the additional simulations of the $10^6~M_{\odot}$ clouds at $Z=Z_{\odot}$ with $R_{\rm cl} = 25$, $30$, $32.5$, and $35~{\rm pc}$ to investigate the relation between the SFE and the surface density.
We further perform the simulations of the $10^5~M_{\odot}$ clouds at $Z=Z_{\odot}$ with $R_{\rm cl} = 8$, $12~{\rm pc}$, and the $10^{6}~M_{\odot}$ clouds at $Z=10^{-2}~Z_{\odot}$ with $R_{\rm cl} = 17.5$, $25$, $30$, $35 ~{\rm pc}$ to consider the dependency on the cloud masses and metallicity.  

\begin{figure}
    \begin{center}
    	\includegraphics[width=\columnwidth]{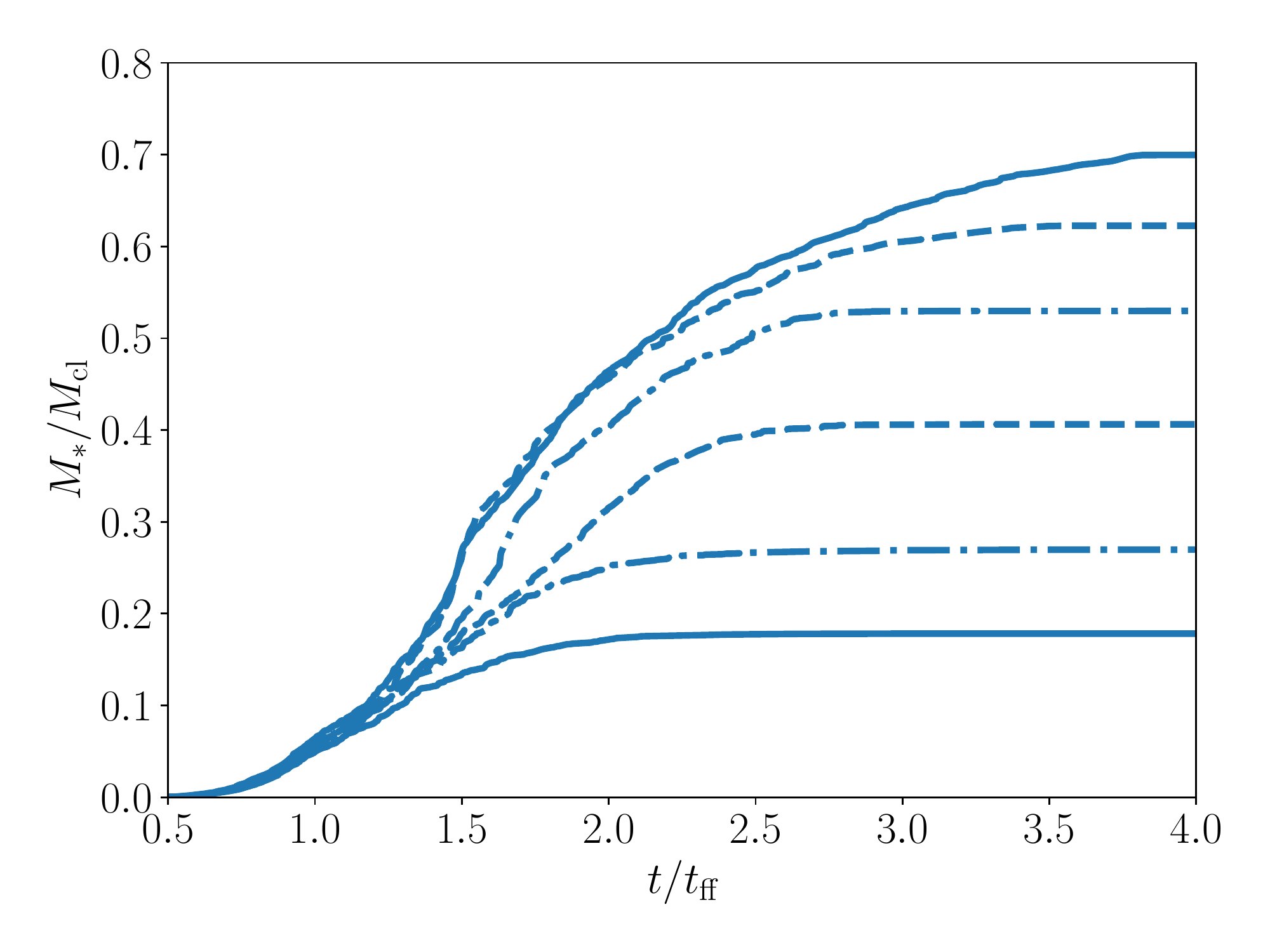}
    \end{center}
    \caption{
    The time evolution of the stellar mass in the case with $(M_{\rm cl}, \alpha_{0}) = (10^6~M_{\odot},1)$.
    Each line represents the different initial cloud radius: $R_{\rm cl} = 20$, $25$, $30$, $32.5$, $35$, and $40~{\rm pc}$ from top to bottom.
}
    \label{fig_m6series}
\end{figure}

\begin{figure}
    \begin{center}
    	\includegraphics[width=\columnwidth]{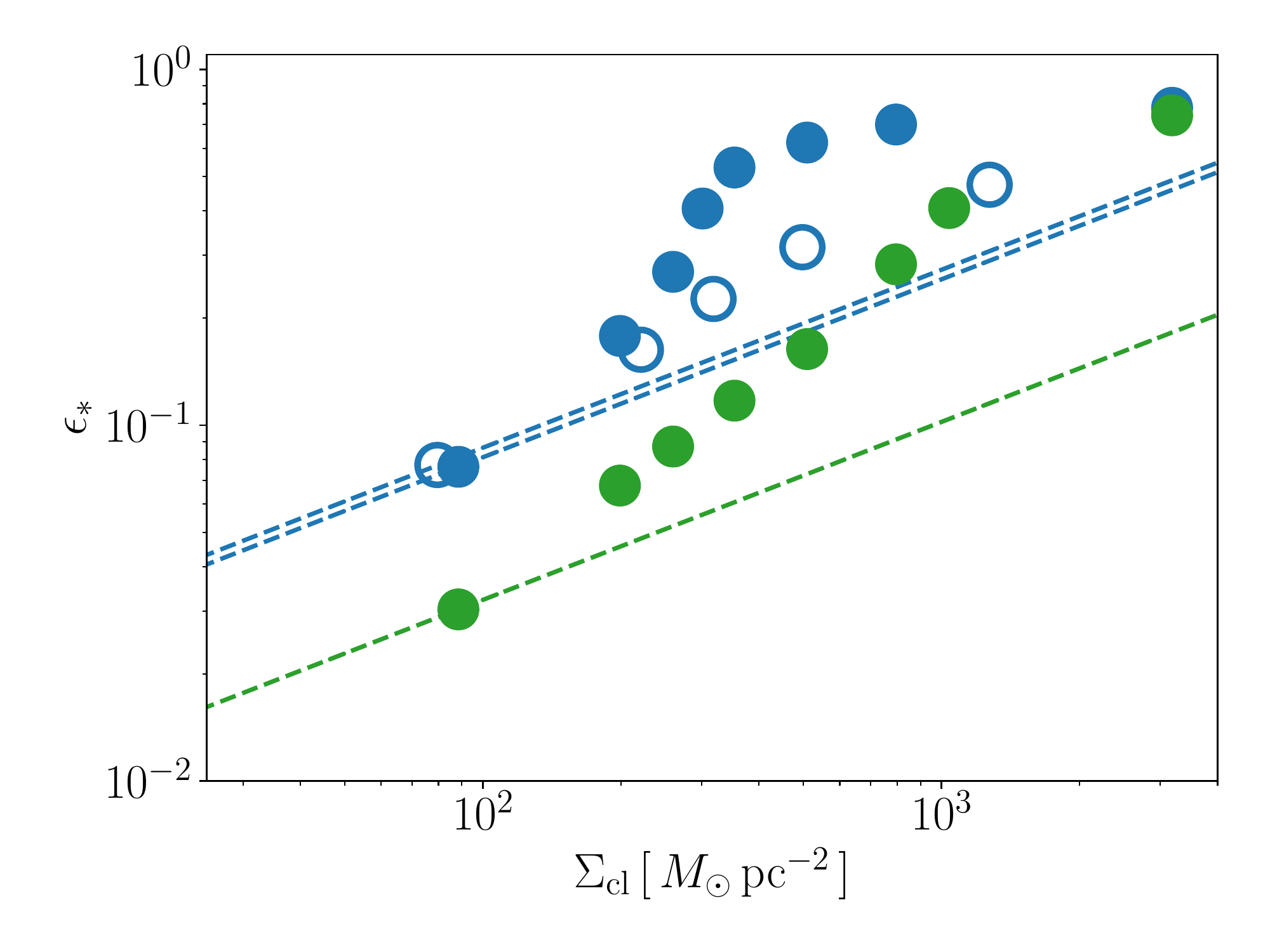}
    \end{center}
    \caption{
    The SFEs in the cases with $(M_{\rm cl}, Z) = (10^6~M_{\odot},Z_{\odot})$ (blue filled circles), $(10^5~M_{\odot},Z_{\odot})$ (blue open circles), and  $(10^6~M_{\odot},10^{-2}~Z_{\odot})$ (green filled circles) as the function of the surface densities.
    The dashed lines show the power law functions as $\epsilon_* \propto \Sigma_{\rm cl}^{1/2}$ with different normalization factors related to physical properties of clouds.
}
    \label{fig_SFEs_m6}
\end{figure}

Figure \ref{fig_m6series} shows the time evolutions of the total stellar masses in the clouds of $10^{6}~M_{\odot}$ with the different initial radii.
All models show the similar stellar masses until $t\sim 1.3 ~t_{\rm ff}$.
Then, the growth curves of the stellar masses look quite different.
In the cases with $R_{\rm cl} \gtrsim 35~{\rm pc}$, the stellar masses increase slowly with almost constant SFRs until $t\sim 2~t_{\rm ff}$. Then the star formation is quenched and the stellar masses become constant.
On the other hand, the SFRs significantly increase at $t \sim 1.3~t_{\rm ff}$ in the more compact clouds.
Besides, the duration times of the star formation are longer, resulting in the SFEs higher than 0.4. 
Note that, although the surface density of M6R30Z0A1 is only 1.3 times larger than that of M6R35Z0A1, the SFE is double.

In Figure \ref{fig_SFEs_m6}, we present the dependence of the SFEs on the surface densities.
We find that the SFEs can be fit with the power law function of $\epsilon_{*} \propto \Sigma_{\rm cl}^{1/2}$ 
at $\Sigma_{\rm cl} \lesssim 200~M_{\odot}{\rm pc^{-2}}$.
The SFE jumps at $\Sigma \sim 300 ~M_{\odot} {\rm pc^{-2}}$ in the cases with $(M_{\rm cl}, Z) = (10^{6}~M_{\odot}, 1)$, and achieves $\sim 0.7$ at $\sim 800~M_{\odot} {\rm pc^{-2}}$.
In the other cases, the SFEs are gradually apart from the power-law lines.

\begin{figure*}
    \begin{center}
    	\includegraphics[width=190mm]{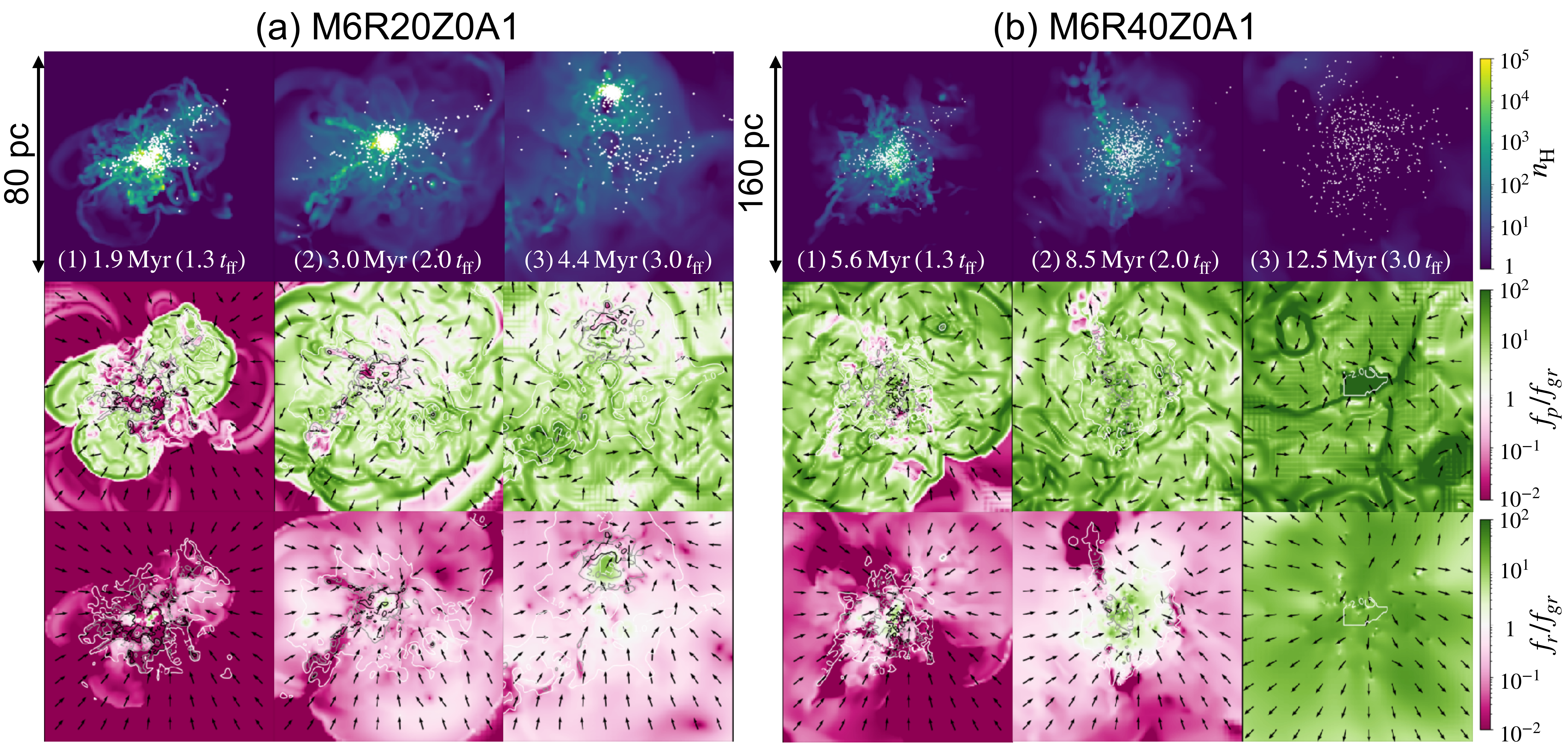}
    \end{center}
    \caption{
    The ratio of gas pressure $(f_{\rm p})$ and radiation one ($f_{\rm r}$) to the gravitational force $(f_{\rm gr})$ on the plane crossing the center of mass of the star clusters.
    Each panel shows the number density of hydrogen $(n_{\rm H})$, $f_{\rm p}/f_{\rm gr}$, and $f_{\rm r}/f_{\rm gr}$ from top to bottom.
    Each figure shows the cases with models of M6R20Z0A1 (left) and M6R40Z0A1 (right).
    The snapshots are taken from $t=1.3~t_{\rm ff}$, $2~t_{\rm ff}$, and $3~t_{\rm ff}$.
    The white contours in middle and bottom panels indicate the distributions of number densities of gas.
    White dots in top panels represent the potions of sink particles.
}
    \label{fig_force_compare}
\end{figure*}

At the time when the SFE starts to jump, the stars and the gas concentrate at the centers of the clouds, resulting in deeper gravitational potential.
This can make the bound fraction of the star cluster larger as shown in Figure \ref{fig_mass_m6r20z1al1}, resulting in the formation of stellar cores.
In this phase, surrounding gas accretes to the central regions and form stars efficiently.   
The rapid increase of the SFR starts when the SFE achieves $\sim 0.1$
regardless of the cloud model, and it is consistent with the previous study \citep[e.g.,][]{2017A&A...605A.119S}.  
In the cases of the diffuse clouds, the radiative feedback can evacuate the gas from the central region and suppress the formation of the stellar cores.

The conditions of the stellar core formation are determined by the competition between the gravitational force and the radiative feedback.
In Figure \ref{fig_force_compare}, we show the spatial distributions of the ratio of thermal pressure $(f_{\rm p})$ and radiation pressure $(f_{\rm r})$ to the gravitational force $(f_{\rm gr})$ in the models of M6R20Z0A1 and M6R40Z0A1.
The gravitational force includes the contributions from gas and stars.
In the case of M6R20Z0A1, the core forms at $t \sim 1.3 ~t_{\rm ff}$.
At this time, the H{\sc ii} regions start to expand, and thermal pressure is dominant in the low-density regions.
However, the gravitational force overcomes thermal pressure in the high-density regions around the core.
Also, the radiation pressure is weaker than the gravitational force until $t \sim 2 ~t_{\rm ff}$.
Thus, the high-density gas remains in the central region, and the star formation proceeds rapidly as shown in Figure \ref{fig_m6series}.
Then, the radiation pressure increases with the stellar mass and evacuates the gas at $t \sim 3 ~t_{\rm ff}$, making the hole structure as shown in the figure.
In the case of M6R40Z0A1, the stars are concentrated at $t=1.3~t_{\rm ff}$ as in M6R20Z0A1.
However, the force from the thermal pressure is larger than the gravitational force even in the higher-density region.
Therefore, the gas around the star cluster is dispersed rapidly.
As a result, no stellar core is formed, and stars are widely distributed.
To summarize the above discussion, the SFE jump condition is whether the gravitational force overcomes the thermal pressure in the high-density region.
We further discuss the condition
with the semi-analytical model in Section \ref{analytical_star_formation_efficiencies}.

\begin{figure}
    \begin{center}
    	\includegraphics[width=\columnwidth]{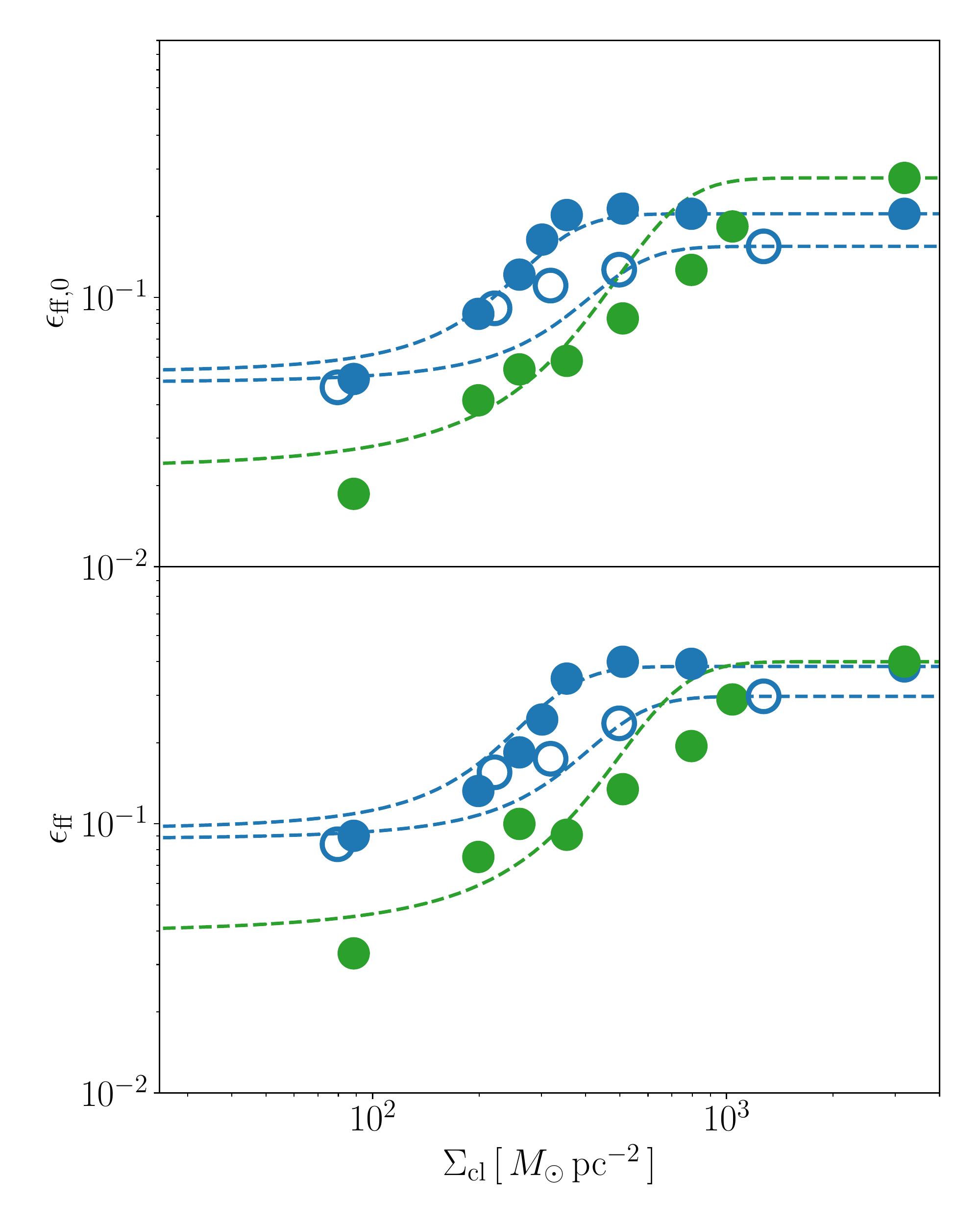}
    \end{center}
    \caption{
    Same as Figure \ref{fig_SFEs} but only for the cases with $M_{\rm cl} = 10^6 ~M_{\odot}$.
    The styles of the symbols are the same as in Figure \ref{fig_SFEs_m6}.
    The dashed lines show the fitting function of Eq.\eqref{fit_function_eps}.
    The fitting parameters are tabulated in Table \ref{parameter_analytical_estimate}.
}
    \label{fig_epsff_m6}
\end{figure}

As discussed in Section \ref{star_form_efficiency}, the SFRs is characterized by the parameter $\epsilon_{\rm ff, 0}$ or $\epsilon_{\rm ff}$.
Figure \ref{fig_epsff_m6} shows the dependencies of  $\epsilon_{\rm ff, 0}$ and $\epsilon_{\rm ff}$ on the surface densities. 
Both $\epsilon_{\rm ff, 0}$ and $\epsilon_{\rm ff}$ are like step functions transiting at the specific surface densities.
We make a fitting function as
\begin{align}
    \epsilon_{\rm ff} = \epsilon_{\rm ff, 1} + \frac{\epsilon_{\rm ff, 2} - \epsilon_{\rm ff,1 }}{ 1 + e^{- f(\Sigma_{\rm cl})}}, \label{fit_function_eps}
\end{align}
where $\epsilon_{\rm ff, 1}$ and $\epsilon_{\rm ff, 2}$ are the fitting parameters for the low and high surface densities.
We use the values at the most compact and most diffuse cases for $\epsilon_{\rm ff, 1}$ and $\epsilon_{\rm ff, 2}$.
In Equation \eqref{fit_function_eps}, $f(\Sigma_{\rm cl})$ is defined as
\begin{align}
    f(\Sigma_{\rm cl}) = a \frac{\Sigma_{\rm cl} - \Sigma_{\rm th}}{\Sigma_{\rm th}}, \label{fit_functin_eps2}
\end{align}
where we use $a=5$ as the gain of the sigmoid function.
In Equation \eqref{fit_functin_eps2}, $\Sigma_{\rm th}$ is the threshold surface density of the stellar core formation.
We further consider the dependence of this threshold surface density on the cloud mass and metallicities in Section \ref{analytical_star_formation_efficiencies}, and the details of the fitting parameters are summarized in Table \ref{parameter_analytical_estimate}.

\subsection{Dependence on virial parameter}\label{virial_parameter}
\begin{figure}
    \begin{center}
    	\includegraphics[width=\columnwidth]{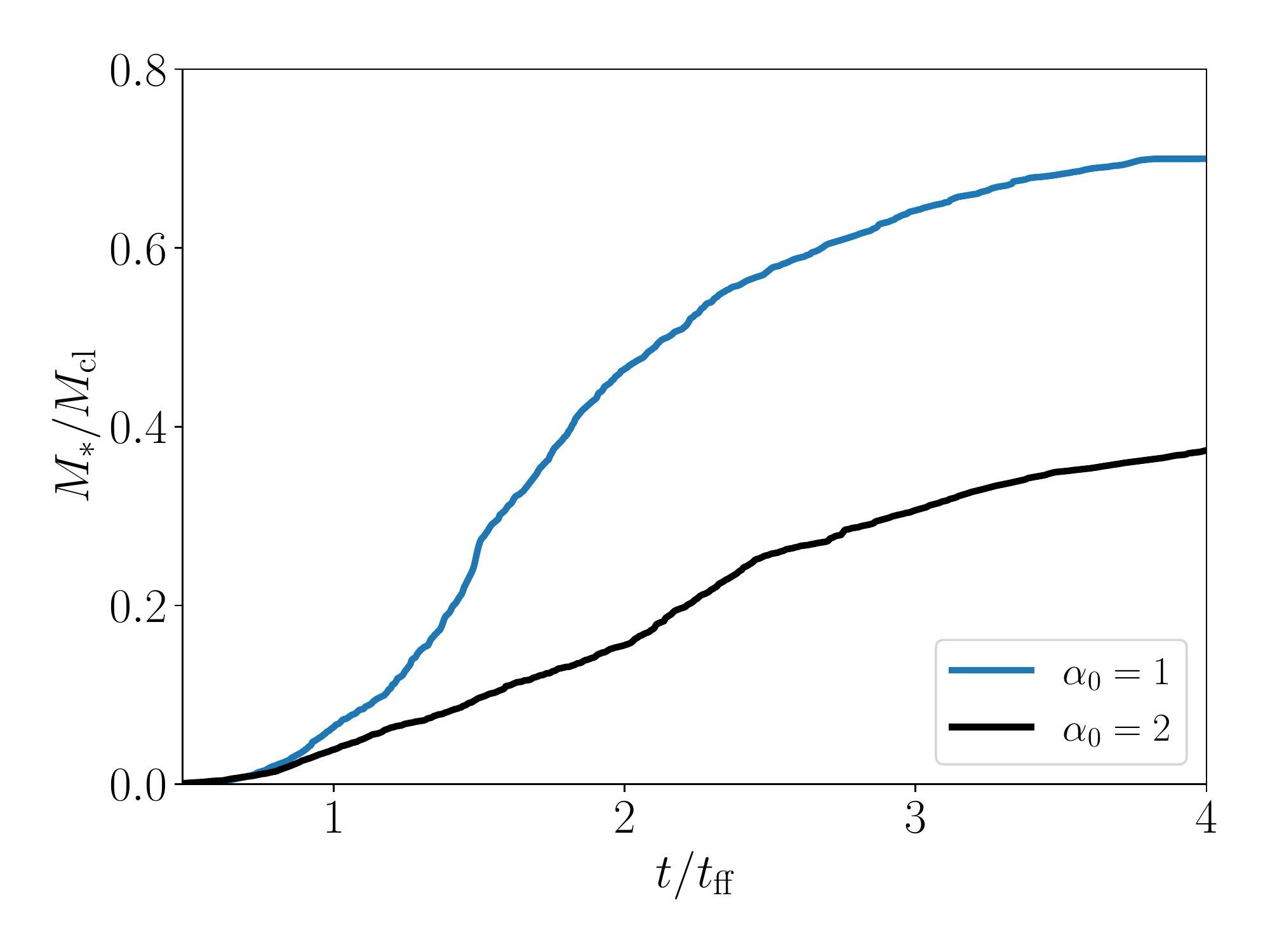}
    \end{center}
    \caption{
    The time evolution of the stellar mass in the case with $(M_{\rm cl}, R_{\rm cl}, Z) = (10^{6}\, M_{\odot}, 20\, {\rm pc}, Z_{\odot})$.
    Blue and black lines represent the cases with $\alpha_0 = 1$ and $\alpha_0 = 2$, respectively.
}
    \label{fig_alpha1}
\end{figure}

Turbulent motions prevent the gravitational collapse of clouds.
Therefore, the SFRs are likely to decrease for a larger virial parameter $\alpha_{0}$.
In Figure \ref{fig_alpha1}, we show the evolution of the total stellar mass with the different virial parameters $\alpha_0 = 1$ and $2$ in the clouds with $(M_{\rm cl}, R_{\rm cl}, Z) = (10^6~M_{\odot}, 20~{\rm pc}, Z_{\odot})$.
In the model of $\alpha_0 = 1$, the star formation rate rapidly increases at $t \simeq t_{\rm ff}$ via the core formation.
In the case of $\alpha_0 = 2$, the kinetic energy is comparable to the gravitational one.
Accordingly, the turbulent motions can prevent the cloud collapse, resulting in the lower SFR.

\begin{figure}
    \begin{center}
    	\includegraphics[width=\columnwidth]{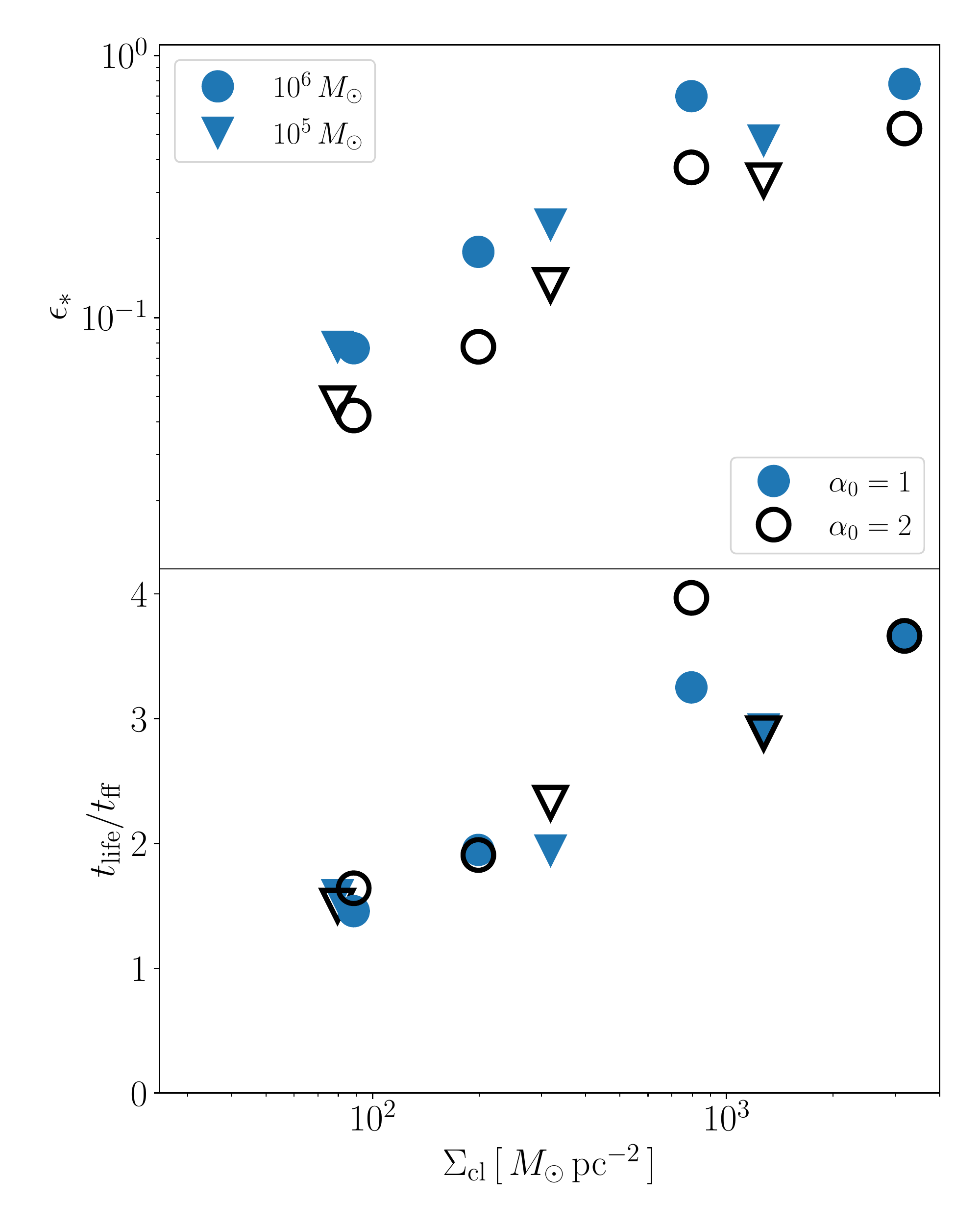}
    \end{center}
    \caption{
    Same as Figure \ref{fig_SFEs} but for the cases of $(M_{\rm cl}, R_{\rm cl}, Z) = (10^6 \, M_{\odot}, 20 \, {\rm pc}, Z_{\odot})$ with $\alpha_0 = 1$ (blue) and $2$ (black). 
}
    \label{fig_SFEs2}
\end{figure}

The slow star formation gives time for the expansion of H{\sc ii} regions, which disrupts the star-forming clumps.
Besides, the shear motions of turbulence decrease local gas density.
In the lower density regions, the H{\sc ii} regions expand rapidly.
Therefore, the SFEs are smaller for the higher virial parameter \citep{2021ApJ...911..128K}.
Figure \ref{fig_SFEs2} shows the SFEs as the function of surface densities with $\alpha_0 = 1$ and $2$.
The SFEs of $\alpha_0 = 2$ are lower than that of $\alpha_0 = 1$ by a factor of $\sim 1.5- 2.0$.

We also investigate the cloud lifetimes for the different virial parameters as shown in Figure ~\ref{fig_SFEs2}.
At $\Sigma_{\rm cl} \lesssim 200 ~M_{\odot} {\rm pc^{-2}}$, the cloud lifetime does not change with the virial parameter.
At $200~M_{\odot}{\rm pc^{-2}} < \Sigma_{\rm cl} \lesssim 10^3~M_{\odot}{\rm pc^{-2}}$, the lifetime of $\alpha=2$ becomes longer than that of $\alpha=1$.  
In such a case, the turbulent motions delay the star formation, but the radiative feedback cannot easily blow away gas from the massive clouds.
Therefore, the star formation continues until most of the gas is exhausted, resulting in the longer lifetime. 
At the higher surface density, the cloud lifetimes are slightly different.
In these cases, 
radiative feedback prevents gas from converting to stars around the stellar cores.
This process mainly determines the duration time of the star formation.

The turbulent motion also has impacts on the properties of star clusters like the bound fraction.
For example, the bound fraction of M6R40Z0A2 is much lower than the same mass and radius cloud with a lower virial parameter $\alpha=1$.
In this model, the SFE is lower than $\sim 0.1$ that is the critical value to induce the formation of a high-density stellar core. 
Once the surface density of the clouds exceeds $800\, M_{\odot}{\rm pc^{-2}}$, 
the SFEs are much higher than 0.1 even with $\alpha=2$.
Therefore, the star clusters with the high bound fractions can form irrespective of the virial parameter.

\subsection{Dependency on escape velocity}\label{dpdce_on_esc_velocity}

Previous studies suggested that the escape velocity from a cloud was a key parameter to understand the impacts of the photoionization feedback on cloud disruption and star formation. 
\citep{2012MNRAS.427.2852D, 2012ApJ...758L..28B}. 
According to their simulation results, the photoionization feedback is ineffective if the escape velocity is higher than the sound speed of ionized gas, $\sim 10~\rm km~s^{-1}$.
On the other hand, our simulations show that the SFEs mainly depends on the initial surface densities of the clouds, as shown in Figure \ref{fig_SFEs} and \ref{fig_SFEs2}.
Although the escape velocity of the clouds with $(M_{\rm cl}, R_{\rm cl}) = (10^6~M_{\odot}, 40~{\rm pc})$ is $15~{\rm km/s}$, the clouds are dispersed rapidly.
The SFEs are lower than $0.1$ except for M6R40Z0A1 of which the value is $0.18$.
Thus, the escape velocity is not only the parameter to understand the suppression of star formation.
In section \ref{analytical_arguments}, we will further discuss this by developing a semi-analytical model.

\subsection{Properties of star clusters }\label{prop_star_cluster}

\begin{figure}
    \begin{center}
    	\includegraphics[width=\columnwidth]{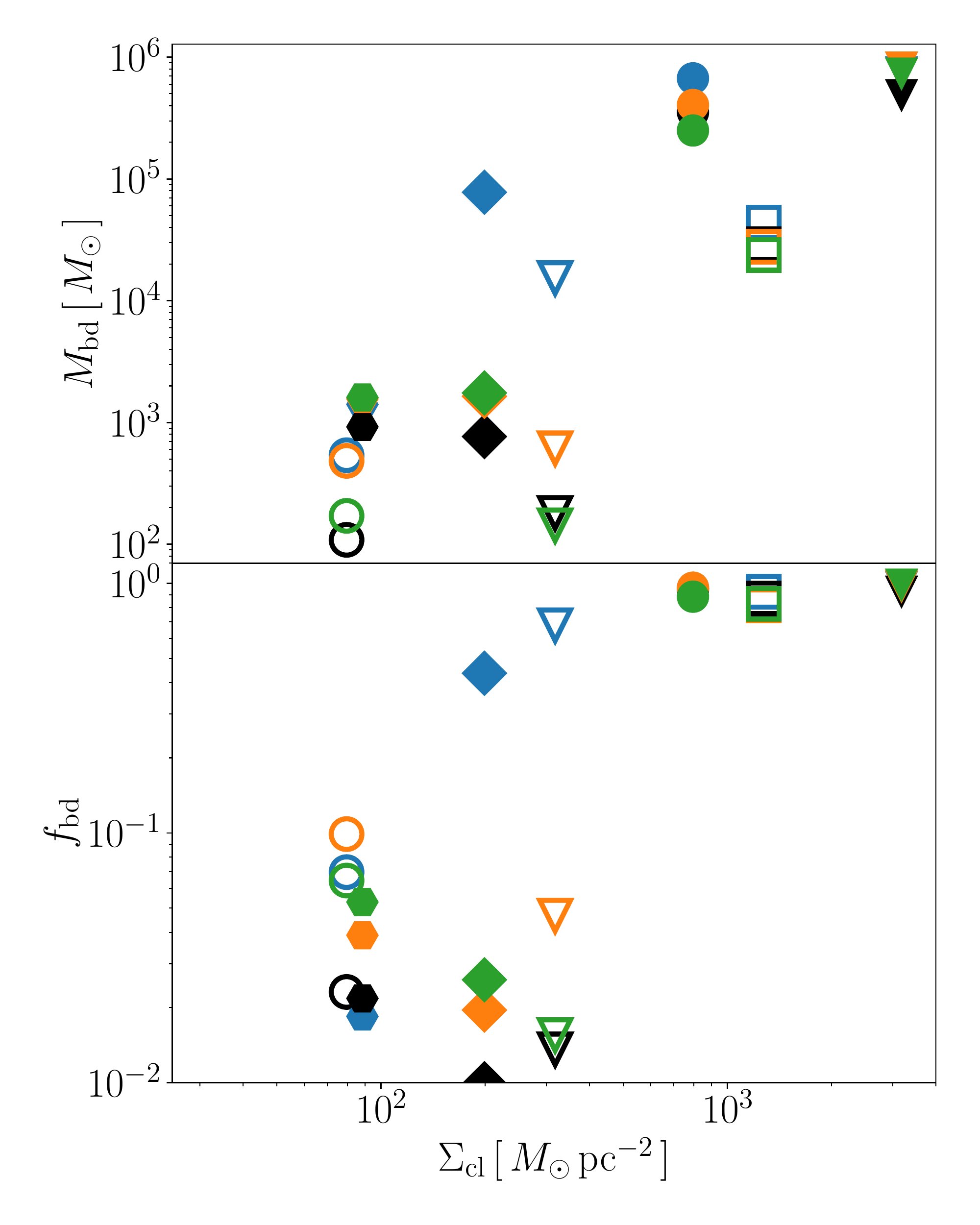}
    \end{center}
    \caption{
    The gravitationally bounded masses and bound fractions of star clusters at the cloud lifetimes $t=t_{\rm life}$ as the function of the surface densities.
    The styles of symbols are the same as Figure \ref{fig_SFEs}.
}
    \label{fig_bds}
\end{figure}

A significant fraction of stars can escape from their natal cluster.
Therefore the final mass of the cluster becomes lower than the initial state. 
Figure \ref{fig_bds} shows the masses of bounded stars at the end of the cloud lifetimes.
At $\Sigma_{\rm cl} \gtrsim 800\, M_{\odot} {\rm pc^{-2}}$, the young massive clusters (YMCs, $M_{\rm bd} > 10^4 \, M_{\odot}$) form in all cases.
On the other hand, the cluster mass reaches $10^4\, M_{\odot}$ only in the two models of $Z=Z_{\odot}$ (M5R10Z0A1 and M6R40Z0A1) at $\Sigma < 800 \, M_{\odot} {\rm pc^{-2}}$.
YMCs do not form from the low-metallicity and low surface density clouds. 
In particular, the total stellar mass exceeds $10^4\, M_{\odot}$ in the models of M6R40Z-1A1 and M6R40Z-2A1, but most of the stars are dispersed.

The bound fractions $(f_{\rm bd})$ are related to the compactness of clouds.
Figure \ref{fig_bds} shows the bound fractions at the end of the cloud lifetime.
At the solar metallicity, the bound fraction increases from 0.05 to $> 0.9$ with the surface density in the range of $\Sigma_{\rm cl} = 80-800~M_{\odot}{\rm pc^{-2}}$.
In the cases of the low-metallicty clouds, the bound fractions are lower than that of the solar metallicity at $\Sigma_{\rm cl} \leqq 320~M_{\odot}{\rm pc^{-2}}$, but become almost similar ($\gtrsim 0.9$) if $\Sigma \geqq 800~M_{\odot}{\rm pc^{-2}}$.

\begin{figure}
    \begin{center}
    	\includegraphics[width=\columnwidth]{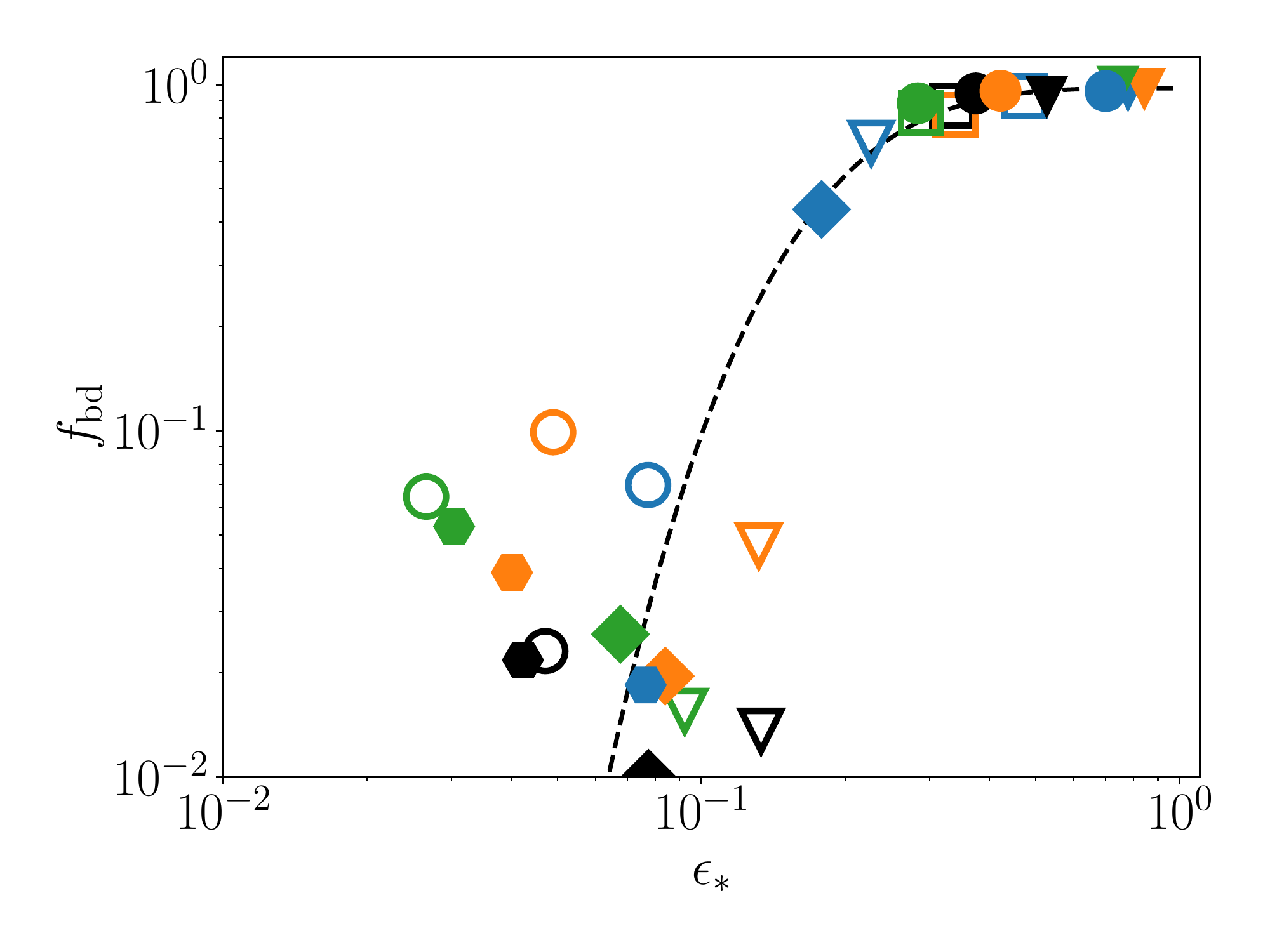}
    \end{center}
    \caption{
    The bound fractions of star clusters at the cloud lifetime $t=t_{\rm life}$ as the function of the star formation efficiency.
    The styles of symbols are the same as Figure \ref{fig_SFEs}.
    The dashed line is the fitting function given as Equation \eqref{fitting_bd}.
}
    \label{fig_sfe_bds}
\end{figure}

Figure \ref{fig_sfe_bds} presents the bound fraction as the function of the SFEs.
In the cases with the SFEs lower than 0.1, the star clusters can not keep holding their own stars, resulting in the disperse of them and the low bound fractions.
Once the SFE exceeds 0.1, the bound fraction abruptly increases and achieves $\gtrsim 0.9$ at ${\rm SFE} \gtrsim 0.3$.
These results are consistent with previous numerical simulations
\citep[e.g.,][]{2019MNRAS.487..364L,2020arXiv200804453G}.
We make a fitting function as
\begin{align}
  f_{\rm bd} = a_{1} 10^{-|\log_{10}(\epsilon_* )|^{a_2}}. \label{fitting_bd}
\end{align}
The best fit parameters are $(a_1,a_2)=(3.9, 0.98)$.
This fitting formula is steeper with decreasing SFEs than the analytical model of \cite{2019MNRAS.487..364L} that assumed the Maxwellian velocity distribution of stars.
The bound fraction at $\epsilon_{*} \lesssim 0.1$ scatters significantly.  
This trend was also reported in \citet{2020arXiv200804453G}. 
Note that, our estimate is likely to be inaccurate if $\epsilon_* \ll 0.1$ because a sink particle is modelling bounded several stars. In this case, the actual bound fraction can be smaller than our results.

\citet{2007MNRAS.380.1589B} showed that the relation between the SFE and the bound fraction depends on the timescale of gas removal.
In their results, the SFE is needed to be higher than 0.33 to form the bounded cluster if the gas is removed instantaneously.
The critical SFE decreases to 0.15 if the removal timescale is similar to the crossing time of the star cluster.
In our simulations, the removal timescale is comparable to the free-fall time of the clouds, and the critical SFEs roughly match their results.

\begin{figure}
    \begin{center}
    	\includegraphics[width=\columnwidth]{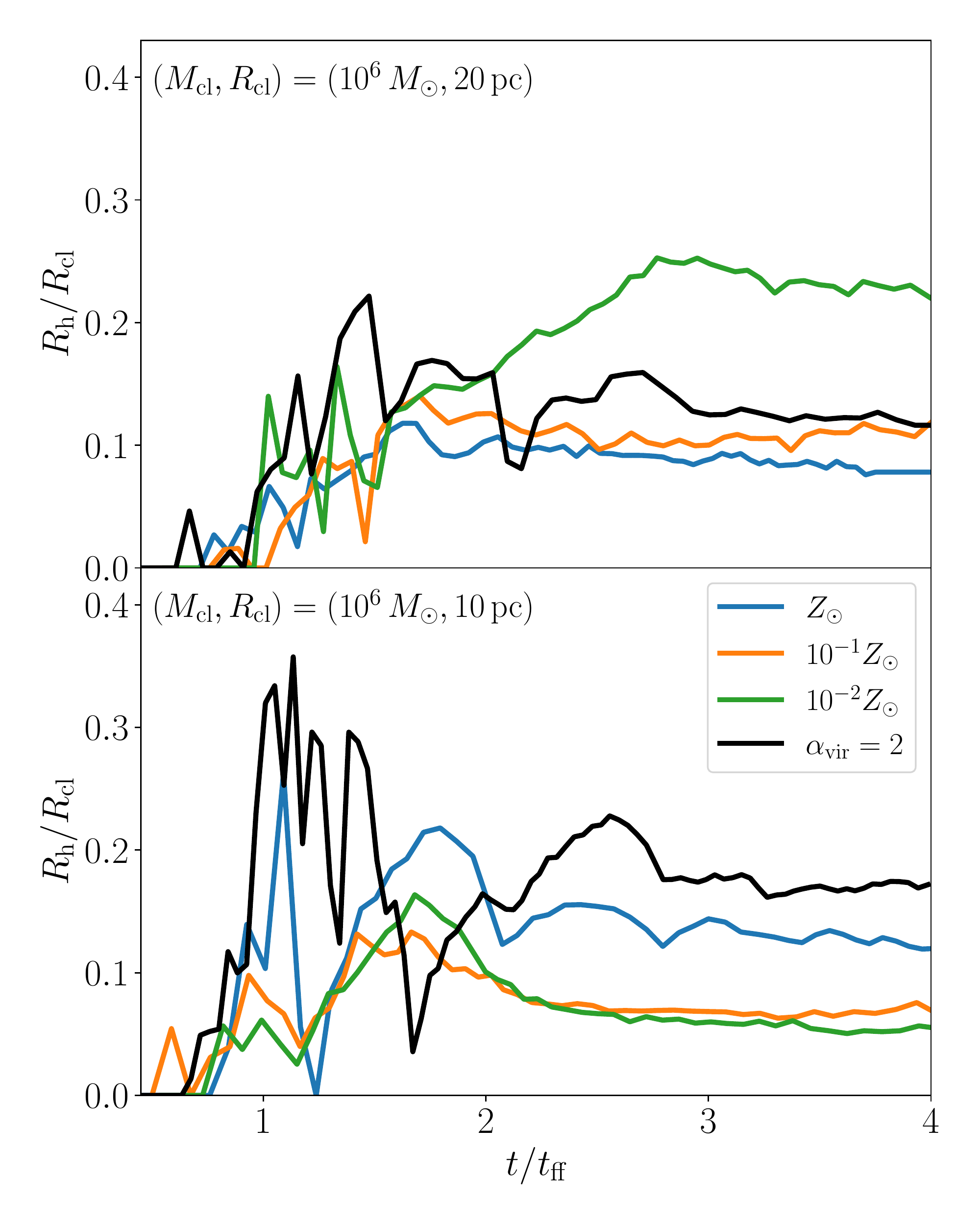}
    \end{center}
    \caption{
    The evolution of half mass radius in the models of $(M_{\rm cl}, R_{\rm cl}) = (10^{6}\, M_{\odot}, 20\, {\rm pc})$ (top) and $(M_{\rm cl}, R_{\rm cl}) = (10^{6}\, M_{\odot}, 10\, {\rm pc})$ (botom).
    Each line shows the different metallicity: $Z=Z_{\odot}$ (blue), $10^{-1}Z_{\odot}$ (orange), and $10^{-2}Z_{\odot}$ (green). 
    The black lines represent the cases for $\alpha_{\rm vir} = 2$ and $Z=Z_{\odot}$. 
}
    \label{fig_Rh_evolv}
\end{figure}

Figure \ref{fig_Rh_evolv} shows the evolution of half mass radii in the cases with $(M_{\rm cl}, R_{\rm cl}) = (10^6 \, M_{\odot}, 20 \, {\rm pc})$ and $(10^6 \, M_{\odot}, 10 \, {\rm pc})$.
As shown in \citetalias{2020MNRAS.497.3830F}, the radius of a star cluster rapidly increases in the cases of the low-surface density clouds. Whereas the bound fraction of the star clusters in the massive compact clouds is high ($> 0.9$) and the radii of them do not increase as shown in the figure.
At $t \lesssim 1.0 t_{\rm ff}$, the spatial distributions of stars are stochastically determined due to the turbulent motions, which induces the fluctuation of the radii. 
Once the bound fraction exceeds $\sim 0.1$, the radii become almost constant and result in the values of $\sim 0.05 - 0.2$ times the initial cloud ones.

\begin{figure}
    \begin{center}
    	\includegraphics[width=\columnwidth]{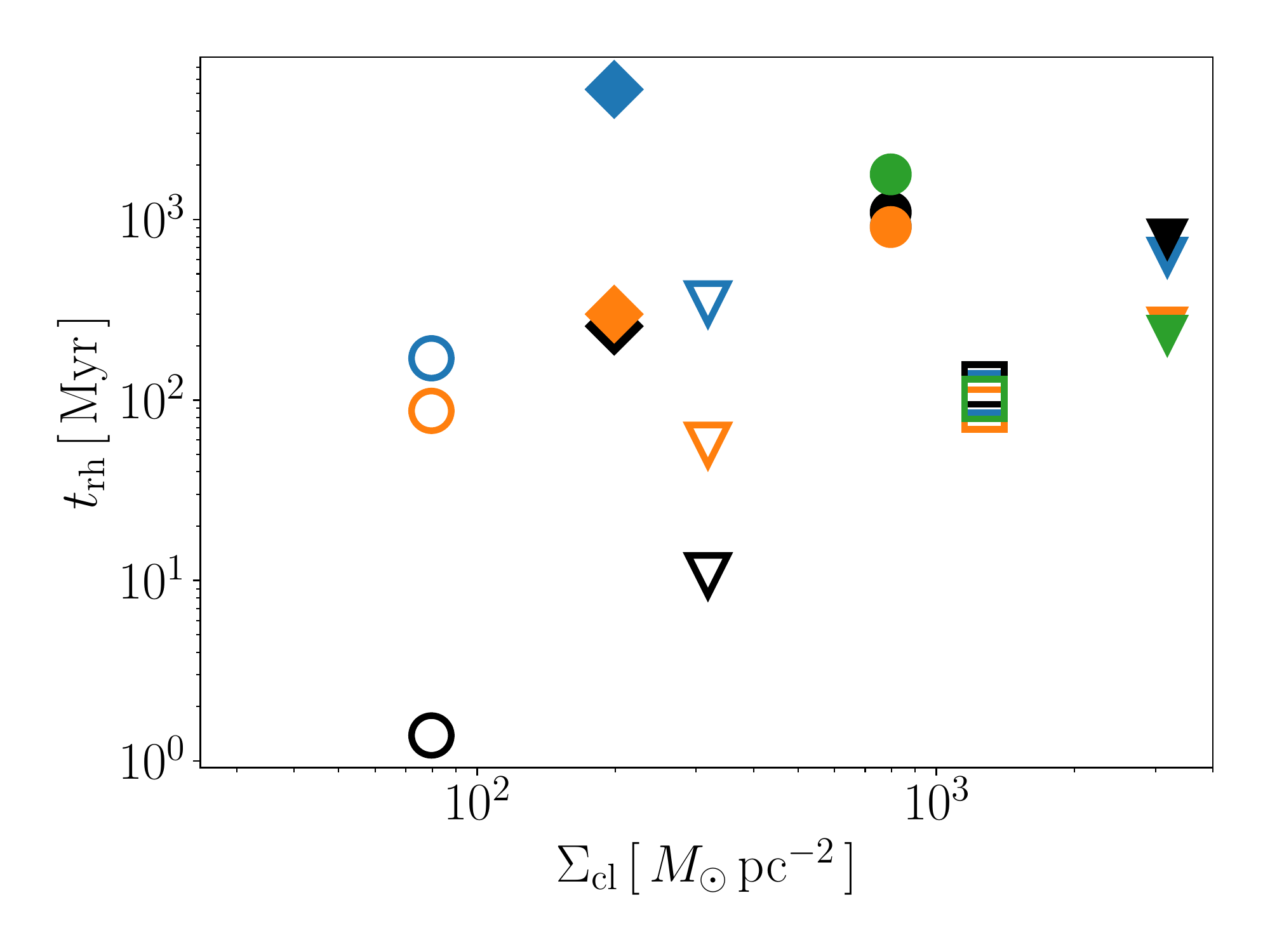}
    \end{center}
    \caption{
    The relaxation time $t_{\rm rh}$ of star cluster at $t=t_{\rm life}$ as the function of the surface densities.
    The  relaxation time $t_{\rm rh}$ is given as Eq. \eqref{eq_trh}.
    The styles of symbols are the same as Figure \ref{fig_SFEs}.
}
    \label{fig_trh}
\end{figure}

The structure of a bound star cluster would change via 2-body interaction between stars.
The time-scale is characterized by the relaxation time
given as \citep{1987degc.book.....S, 2019MNRAS.486.3942K}
\begin{align}
    t_{\rm rh} = 0.0477 \frac{N}{(G \rho_{\rm hm})^{1/2} \log (0.4 N)}, \label{eq_trh}
\end{align}
where $\rho_{\rm hm}$ is the stellar mass density inside the half-mass radius $r_{\rm h}$.
The number of stars $N$ is obtained as $N = M_{\rm bd}/ \langle m \rangle$ where $\langle m \rangle$ is the mean stellar mass.
In this work, we assume the Chabrier IMF $(0.1 ~M_{\odot} \leqq M_* \leqq 150~M_{\odot})$ to sink particles.
Therefore, the mean stellar mass is estimated as $\langle m \rangle = 0.75~M_{\odot}$. 
Figure \ref{fig_trh} shows the relaxation time of each models. 
In the cases of $(M_{\rm cl}, Z, \alpha_{0})=(10^5~M_{\odot}, Z_{\odot}, 2)$ and $(M_{\rm cl}, Z) = (10^5~M_{\odot}, 10^{-1}Z_{\odot})$, the relaxation times are shorter than $10^2~{\rm Myr}$.
As shown in \citet{2014MNRAS.439.1003F}, the star cluster can induce the core-collapse within $\sim 0.2 \times ~\rm t_{\rm rh}$. 
Due to the core collapse, the stellar density at the center increases, which induces the formation of stellar-binary systems. In particular, if the core collapse occurs within a few Myr, massive stars can migrate to the center and form the binary before the end of their lives. Such a star cluster can be a potential formation site of the binaries that induce the gravitational-wave events \citep{2016PhRvL.116f1102A}.
In our simulations, the star clusters formed in the clouds with  $M=10^{5}~M_{\odot}$ and $\alpha_0 = 2$ or $Z=10^{-1}Z_{\odot}$ can be  the formation sites of the gravitational wave sources.  

Note that, since the sink particle masses are much higher than the average stellar mass of the IMF, the relaxation time of these sink particles might be underestimated. 
For further discussions on this problem, higher resolution simulations are awaited.

\section{Analytical arguments}\label{analytical_arguments}

\subsection{Star formation efficiencies}\label{analytical_star_formation_efficiencies}

As shown in Section \ref{results}, the SFEs mainly depend on the surface densities ($\Sigma$).
In \citetalias{2020MNRAS.497.3830F}, we constructed the semi-analytical model to describe the dependence of the SFEs on the initial surface density and metallicity. 
We assumed that the duration time of the star formation is equal to the propagation time of the expanding shell outside the cloud. 
Then, the total stellar mass $M_* = \epsilon_* M_{\rm cl}$ is calculated by multiplying $t_{\rm HII}$ with the star formation rate $\dot M_*$ in the clouds.
However, this model is not reasonable for massive-compact clouds where the photoionization feedback alone cannot quench the star formation.
Therefore, we develop a new semi-analytical model taking the cases of massive-compact clouds into account.

Here, we use a expanding shell model as in
\citet{2009ApJ...703.1352K,2016ApJ...819..137K}. 
We assume that a uniform density sphere whose mass and radius are $M_{\rm cl}$ and $R_{\rm cl}$.
The mass of the expanding shell is $M_{\rm sh} = M_{\rm cl} (1-\epsilon_*) (r_{\rm sh}/R_{\rm cl})^3$, where $r_{\rm sh}$ and $\epsilon_*$ are the radius and the SFE.
Here, we assume that the SFE is constant in the entire volume.
The radiation pressure and gravitational force on the shell are estimated from equation \eqref{radiation_pressure} and \eqref{gravity}.
The equation of motion of the expanding shell is represented as \citep{2009ApJ...703.1352K, 2016ApJ...819..137K}  
\begin{align}
  \frac{d}{dt} \left( M_{\rm sh} \dot r_{\rm sh} \right) = 4 \pi r_{\rm sh}^2 \rho_{\rm i} c_{\rm i}^2 + \frac{ L_{*}}{ c} - \frac{G M_{\rm sh} (M_* + M_{\rm sh}/2)}{r_{\rm sh}^2}, \label{eom_of_shell}
\end{align} 
where $\rho_{\rm i}$ and $c_{\rm i}$ are the density and the sound speed in H{\sc ii} regions. 
The second and third terms are radiation force and gravity.
We assume that the shell is optically thick for direct light and thin for thermal radiation from dust.
The first term on the right-hand side is the thermal pressure on the shell from the H{\sc ii} regions.
The density of the H{\sc ii} regions is determined by the balance between the ionization and recombination as 
\begin{align}
  n_{\rm i} = \left( \frac{\rho_{\rm i} c_{\rm i}^2}{k_{\rm B}T_{\rm i}} \right) = \left( \frac{3 f_{\rm ion} S_{\rm ion}}{4 \pi r_{\rm sh}^3 \alpha_{\rm B} } \right)^{1/2}, \label{number_density_HIIregion}
\end{align}
where $T_{\rm i}$, $S_{\rm ion}$ and $\alpha_{\rm B}$ are the temperature of ionized gas, the emissivity of ionizing photons and the recombination rate coefficient $\alpha_{\rm B} = 2.6 \times 10^{-13}(T_{\rm i} /10^4\,{\rm K})^{-0.8} \, {\rm cm^3 s^{-1}}$ \citep{1989agna.book.....O}.
In the H{\sc ii} regions, dust grains absorb a part of ionizing photons.
We adopt the absorption rate by hydrogen atoms as $f_{\rm ion} = 1-0.27 \, (Z/Z_{\odot})$ \citep{1997ApJ...476..144M, 2009ApJ...703.1352K}.

In this model, we assume that the gas converts into stars after the shell passes with the conversion rate $\epsilon_{*, 0}$. 
Here, we can regard $\epsilon_{*, 0}$ as the SFE $\epsilon_{*}$. Therefore, we simply use the notation $\epsilon_{*}$ for the conversion rate hereafter.
The mass of stars formed in the shell is given as $M_* = \epsilon_*/(1-\epsilon_*) M_{\rm sh}$. 
The luminosity and the photon emissivity are estimated as
\begin{align}
    L_* = \frac{\epsilon_*}{1-\epsilon_*} l_* M_{\rm sh}, \label{luminosity_rate}
\end{align}
\begin{align}
    S_{\rm ion} = \frac{\epsilon_*}{1-\epsilon_*} s_* M_{\rm sh}, \label{emissivity_ion} 
\end{align}
where the luminosity and the emissivity per unit mass are given as $l_* = 1.3 \times 10^3 \, L_{\odot} \, {M_{\odot}}^{-1}$ and $s_* = 7.5 \times 10^{46} \, {\rm s^{-1}} \, {M_{\odot}^{-1}}$.

We introduce a characteristic radius $r_{\rm ch, r}$ at where the radiation force is equivalent to the thermal pressure of H{\sc ii} regions , 
using equations \eqref{eom_of_shell}, \eqref{number_density_HIIregion}, \eqref{luminosity_rate} and \eqref{emissivity_ion}:
\begin{align}
  r_{\rm ch, r} &= \frac{ c k_{\rm B} T_{\rm i}}{ l_*} \left( \frac{ 12 \pi f_{\rm ion} s_* R_{\rm cl}^3} {\alpha_{\rm B} M_{\rm cl} \epsilon_*}  \right)^{1/2} \nonumber \\
  & \simeq 26 \,{\rm pc} ~ \left( \frac{\epsilon_*}{0.1} \right)^{-1/2}  \left( \frac{\Sigma_{\rm cl}}{80 \, M_{\odot} \, {\rm pc}^{-2} }\right)^{-3/4}  \left( \frac{M_{\rm cl}}{10^5 \, M_{\odot}}\right)^{1/4}  \left( \frac{T_{\rm i}}{8000\,{\rm K}} \right)^{7/5}  , \label{rl_estimate}
\end{align}
where we use $f_{\rm ion} = 0.73$ in the second equation.
We also define another characteristic radius $r_{\rm ch,g}$ for the equilibrium between the thermal pressure and the gravity force as
\begin{align}
  r_{\rm ch,g} &= \left[ \frac{2 l_* R_{\rm cl}^3 }{G c M_{\rm cl}} \left( \frac{\epsilon_*}{1 - \epsilon_*^2} \right) r_{\rm ch, r} \right]^{1/2} \nonumber \\
  & \simeq 51 \, {\rm pc} \left( \frac{\epsilon_*}{0.1} \right)^{1/4}  \left( \frac{\Sigma_{\rm cl}}{80 \, M_{\odot}  {\rm pc}^{-2} }\right)^{-9/8} \left( \frac{M_{\rm cl}}{10^5 M_{\odot}}\right)^{3/8}   \left( \frac{T_{\rm i}}{8000\,{\rm K}} \right)^{7/10}. \label{rg_estimate}
\end{align} 
Next, we rewrite the equation \eqref{eom_of_shell} by introducing dimensionless parameters:
\begin{align}
  x_{\rm sh} = r_{\rm sh} / r_{\rm ch, r}, \label{dimensionless_para}
\end{align}
and 
\begin{align}
  \tau = t/t_{\rm ch}, \label{dm_tau}
\end{align}
where
\begin{align}
  t_{\rm ch} &= \left[ \frac{c }{ l_*} \left( \frac{1 - \epsilon_*}{\epsilon_*} \right) r_{\rm ch, r} \right]^{1/2} \nonumber \\
  & \simeq 2.9 \, {\rm Myr} \,  \left( \frac{\epsilon_*}{0.1} \right)^{-3/4}  \nonumber \\
  & \hspace{1cm} \left( \frac{\Sigma_{\rm cl}}{80 \, M_{\odot}  {\rm pc}^{-2} }\right)^{-3/4} \left( \frac{M_{\rm cl}}{10^5  M_{\odot}}\right)^{1/2}  \left( \frac{T_{\rm i}}{8000{\rm K}} \right)^{7/10}. \label{tl}
\end{align}
Substituting equation \eqref{dimensionless_para} and \eqref{dm_tau} into equation \eqref{eom_of_shell}, we rewrite the equation of motion as
\begin{align}
  \frac{d}{d \tau} \left( x_{\rm sh}^3 \dot x_{\rm sh} \right) = x_{\rm sh}^2 (1 + x_{\rm sh} - R_{\rm g}^2 x_{\rm sh}^2), \label{ndm_eom}
\end{align}
where $R_{\rm g} = r_{\rm ch, r} / r_{\rm ch, g}$.
On the right-hand side, each term represents the contributions from the thermal pressure, the radiation force, and the gravity force.
At $x_{\rm sh} \ll 1$, the thermal pressure plays a dominant role in the acceleration of the shell. 
The radiation force becomes important only if the cloud radius is larger than $r_{\rm ch, r}$.
Additionally, the thermal pressure cannot overcome the gravity in the regions outside $r_{\rm g}$.
Here, we assume that the duration time of the star formation is equal to the crossing time of the shell over the cloud $(t_{\rm exp})$ as 
\begin{align}
  M_* = \epsilon_* M_{\rm cl} = \dot M_{*} t_{\rm exp}. \label{stellar_mass}
\end{align}
where we use the SFR defined in Equations \eqref{def_mdot2}.
The expanding time $t_{\rm exp}$ also depends on the SFE $\epsilon_*$.
We need to solve Equation \eqref{ndm_eom} and \eqref{stellar_mass} consistently.
Furthermore, the parameter of SFR $\epsilon_{\rm ff}$ is given by Equation \eqref{fit_function_eps}. 
Note that, however the dependence of the threshold surface density $\Sigma_{\rm th}$ on the cloud mass and metallicity is still uncertain.

If the cloud radius $(R_{\rm cl})$ is smaller than $r_{\rm ch, r}$ and $r_{\rm ch, g}$, the thermal pressure alone contributes the shell dynamics.
Therefore, as a first step, we consider the case only with the thermal pressure.
In this case, we can estimate the SFEs analytically.
Considering only the first term in the right-hand side of Equation \eqref{ndm_eom},  the self-similar solution is derived as  $x_{\rm sh} = \tau / \sqrt{3}$ and $\dot x_{\rm sh} = 1/\sqrt{3}$.
The shell arrives at the surface of the cloud with the time-scale:
\begin{align}
t_{\rm exp, th} &=  \frac{\sqrt{3} t_{\rm ch} R_{\rm cl}}{r_{\rm ch, r}} \nonumber \\
&= 3.9 \, {\rm Myr} \, \left( \frac{\epsilon_*}{0.1} \right)^{-1/4} \left( \frac{\Sigma_{\rm cl}}{80 \, M_{\odot}  {\rm pc}^{-2} }\right)^{1/4}  \left( \frac{T_{\rm i}}{8000{\rm K}} \right)^{-7/10}  . \label{exp,sh}
\end{align}
Substituting $t_{\rm exp, th}$ into \eqref{stellar_mass}, we obtain the SFE as
\begin{align}
  \epsilon_{*, {\rm th}} \simeq 0.09 \left( \frac{\epsilon_{\rm ff}}{0.1} \right)^{4/5} \left( \frac{\Sigma_{\rm cl}}{80 \, M_{\odot} \, {\rm pc}^{-2} }\right)^{1/2} \left(\frac{M_{\rm cl}}{10^5 \, M_{\odot}}\right)^{1/10} \left( \frac{T_{\rm i}}{8000 \, {\rm K}} \right)^{-14/25}. \label{SFE_HII}
\end{align}
This is same as the model derived in \citetalias{2020MNRAS.497.3830F} where we assumed the central point sources.
The SFEs mainly depends on the surface densities, and it is slightly changed with the cloud mass.
The shell velocity does not depend on the position of the shell in this solution.
We obtain the shell velocity as
\begin{align}
  v_{\rm sh, th} & = \frac{r_{\rm ch, r}}{\sqrt{3} t_{\rm ch}} \nonumber \\
  &= 4.9~{\rm km/s} \left(\frac{\epsilon_{\rm ff}}{0.1}\right)^{1/5} \nonumber \\ 
 & \hspace{0.5cm}   \left(\frac{\Sigma_{\rm cl}}{80 \, M_{\odot} \, {\rm pc}^{-2} }\right)^{-1/4} \left( \frac{M_{\rm cl}}{10^5 \, M_{\odot}} \right)^{3/20} \left( \frac{T_{\rm i}}{8000 \, {\rm K}} \right)^{14/25}. \label{vsh_th}
\end{align}

As discussed in Section \ref{rapid_increase_of_SFE}, the parameter $\epsilon_{\rm ff}$ can be fit with the step function transiting at the threshold density $\Sigma_{\rm th}$.
If the velocity of the expanding shell is smaller than the escape velocity from the core, the shell is likely to fall back and enhance the star formation at the core. 
As shown in Figure \ref{fig_epsff_m6}, the thermal pressure mainly contributes to accelerating the shell around the core.
Assuming that the escaping shell from the core has the same velocity as the self-similar solution given as Equation \eqref{vsh_th}, the condition of the SFR enhancement is given as 
\begin{align}
    v_{\rm th} < v_{\rm esc, core} = \sqrt{\frac{2 G M_{\rm core}}{R_{\rm core}}}, \label{eq_condition_SFRup}
\end{align}
where $M_{\rm core}$ and $R_{\rm core}$ are the core mass and radius.
As discussed in Section \ref{rapid_increase_of_SFE}, the core formation starts when the total stellar mass exceeds 0.1 times the cloud mass.
As shown in Figures \ref{fig_mass_m6r20z1al1} and \ref{fig_Rh_evolv}, the bound fraction $(f_{\rm bd})$ and the half-mass radius $(R_{\rm h})$ are typically $f_{\rm bd} \sim 0.1$ and $R_{\rm h} \sim 0.1$ at this epoch, regardless of the cloud model.
We adopt this values in Equation \eqref{eq_condition_SFRup}.
Substituting $M_{\rm core} = 10^{-2}~M_{\rm cl}$ and $R_{\rm core} = 0.1~R_{\rm cl}$, we obtain the condition of the surface density for the SFR enhancement as
\begin{align}
    \Sigma_{\rm cl} > \Sigma_{\rm th}, \label{eq_condition_SFR2}
\end{align}
where
\begin{align}
    \Sigma_{\rm th} \simeq 280~M_{\odot}{\rm pc^{-2}} \, \left( \frac{\epsilon_{\rm ff}}{0.1} \right)^{2/5} \left(\frac{M_{\rm cl}}{10^6M_{\odot}} \right)^{-1/5} \left( \frac{T_{\rm i}}{8000~{\rm K}} \right)^{28/25}. \label{eq_threshold_sigma} 
\end{align}
In the cases with $(M_{\rm cl}, Z, \alpha_0) = (10^6~M_{\odot}, Z_{\odot}, 1)$, the threshold density is estimated as $\Sigma_{\rm th}=270~M_{\odot}{\rm pc^{-2}}$. 
The SFR enhancement occurs around $\sim 300~M_{\odot} {\rm pc^{-2}}$ in the numerical simulations.
Thus, the estimate in Equation \eqref{eq_threshold_sigma} reproduces the simulation results well.

\begin{table}
    \caption{the parameters used in the analytical estimate}
    \label{parameter_analytical_estimate}
    \centering
    \begin{tabular}{|c|c|c|c|c|c|}
      \hline \hline
      $M_{\rm cl} \, [M_{\odot}]$ & $Z \, [Z_{\odot} ]$ & $\epsilon_{\rm ff, 1}$ & $\epsilon_{\rm ff, 2}$ & $T_{\rm i} \, [{\rm K}]$ & $\Sigma_{\rm th} \, [ M_{\odot} {\rm pc^{-2}} ]$  \\ \hline
      $10^5$ & $1$  & $0.08$ & $0.30$ & $8.0 \times 10^3$ & $4.1 \times 10^2$ \\
      $10^5$ & $10^{-1}$  & $0.09$ & $0.22$ & $1.6 \times 10^4$ & $9.4 \times 10^2$ \\
      $10^5$ & $10^{-2}$  & $0.03$ & $0.21$ & $2.2 \times 10^4$ & $8.8 \times 10^2$ \\
      $10^6$ & $1$  & $0.09$ & $0.38$ & $8.0 \times 10^3$ & $2.7 \times 10^2$ \\
      $10^6$ & $10^{-1}$  & $0.06$ & $0.46$ & $1.6 \times 10^4$ & $4.9 \times 10^2$ \\
      $10^6$ & $10^{-2}$  & $0.03$ & $0.40$ & $2.2 \times 10^4$ & $5.6 \times 10^2$ \\
      \hline
    \end{tabular}
    \begin{minipage}{1\hsize}
    \end{minipage}
  \end{table}

We estimate the shell crossing time and the velocity by integrating equation \eqref{ndm_eom} from $x_{\rm sh}=0$ to the cloud radius $x_{\rm cl} = R_{\rm cl}/r_{\rm ch, r}$.
Equation \eqref{ndm_eom} has the asymptotic solution at $x_{\rm sh} \ll 0$ as $x_{\rm sh} = 1/\sqrt{3} \tau$ and $\dot x_{\rm sh} = 1/\sqrt{3}$.
We use these solutions as the inside boundary conditions.
We iterate the calculations of Equation \eqref{ndm_eom} until the SFE $(\epsilon_*)$ satisfies Equation \eqref{stellar_mass}.
The SFR is given by Equations \eqref{def_mdot2} and \eqref{eps_ff}.
We adopt Equation \eqref{fit_function_eps} to obtain the parameter $\epsilon_{\rm ff}$.
The SFR sensitively depends on various processes such as  thermal pressure, radiative feedback and turbulent motions. Therefore, 
it is difficult to predict the SFR analytically. 
In this work, we use the SFRs obtained from the numerical simulations.
The parameters $\epsilon_{\rm ff, 1}$ and $\epsilon_{\rm ff, 2}$ are evaluated from the results of the most diffuse and most compact clouds for each cloud mass and metallicity.
The threshold surface density of the SFR enhancement is given by Equation \eqref{eq_threshold_sigma}.
The parameters are summarized in Table \ref{parameter_analytical_estimate}.

\begin{figure}
    \begin{center}
    	\includegraphics[width=\columnwidth]{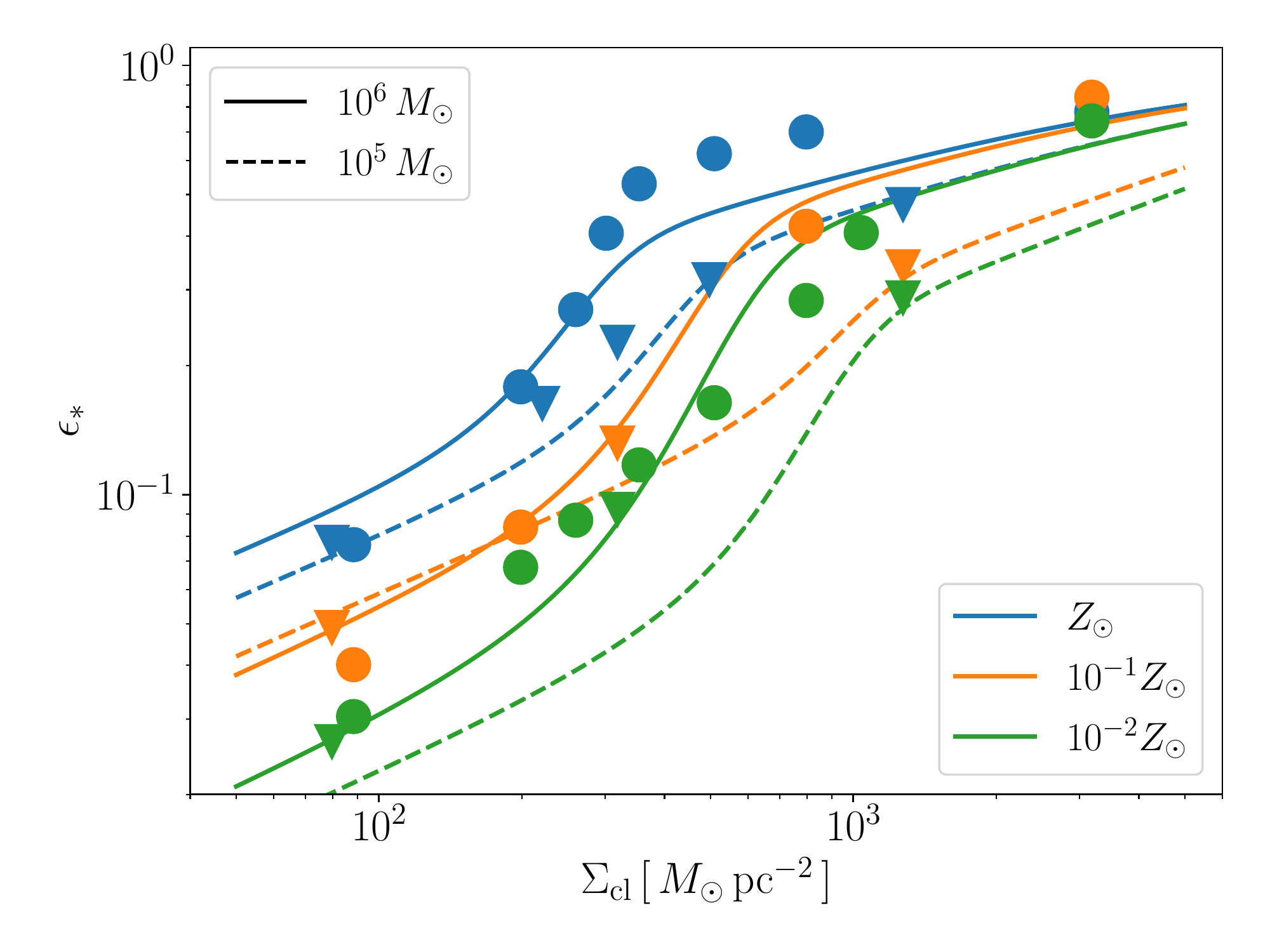}
    \end{center}
    \caption{
    The analytical estimate of star formation efficiencies for the clouds with $M_{\rm cl} = 10^6 ~{M_{\odot}}$ (solid) and $10^5 ~{M_{\odot}}$ (dashed).
    Each symbol represents the simulation results with $M_{\rm cl} = 10^6~M_{\odot}$ (circle) and $10^5~M_{\odot}$ (triangle).
    We only plot the cases with $\alpha_{0} =1$.
    The colors of the symbols are the same as the lines.
}
    \label{fig_eps}
\end{figure}

Figure \ref{fig_eps} shows the SFEs obtained by the semi-analytical model.
The semi-analytical model reproduces the simulation results well. 
In the high- and low-surface limits, the SFE is the simple power-law function of the surface density.
The SFE jump occurs at the threshold surface density $(\Sigma_{\rm th})$.
The threshold surface densities increases in the lower mass cloud, e.g., from $\Sigma_{\rm th} \simeq 270~M_{\odot}{\rm pc^{-2}}$ at $M_{\rm cl}=10^6~M_{\odot}$ to $\Sigma_{\rm th} \simeq 410~M_{\odot}{\rm pc^{-2}}$ at $M_{\rm cl}=10^5~M_{\odot}$ in the cases with $Z=Z_{\odot}$.
Also, it is sensitive to the temperature of H{\sc ii} regions, i.e., metallicity.
The thresholds of the clouds with $Z=10^{-2}~\rm Z_{\odot}$ are higher than that for $Z=~\rm Z_{\odot}$ by a factor of $\sim 2$.

\subsection{Mass of bounded objects}

\begin{figure}
    \begin{center}
    	\includegraphics[width=\columnwidth]{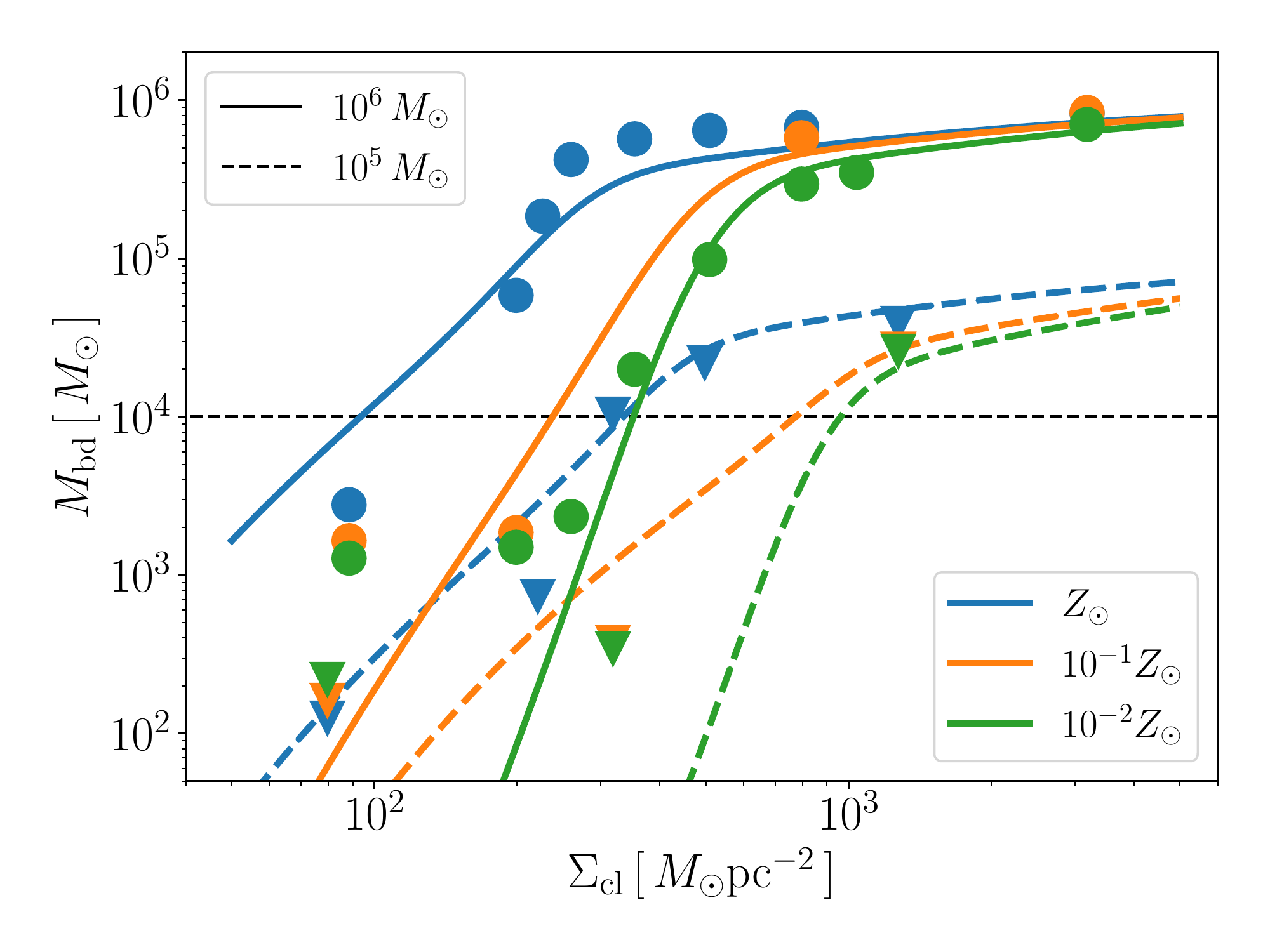}
    \end{center}
    \caption{
    The semi-analytical estimates of the gravitationally bounded stellar masses in the clouds with $M_{\rm cl} = 10^{5} \, M_{\odot}$ (dashed) and $10^6 \, M_{\odot}$ (solid) as the function of the surface densities.
    The symbols represent the simulation results, and the styles of symbols are same as Figure \ref{fig_eps}.
    The black dashed line represents $M_{\rm cl} = 10^4~M_{\odot}$.
}
    \label{fig_bds2}
\end{figure}

The gravitationally bound fraction is tightly related to the SFEs, as shown in Figure \ref{fig_bds2}.
Combining the semi-analytical model of the SFEs and the relation between the SFEs and the bound fractions given as Equation \eqref{fitting_bd}, we obtain the masses of the gravitationally bounded object as shown in Figure \ref{fig_bds2}.
We find that young massive clusters (YMC, $M_{\rm cl} > 10^{4} \, M_{\odot}$) with $Z \sim Z_{\odot}$ are likely to form in the clouds of $10^6~M_{\odot}$ with the surface densities higher than $\sim 100 \, M_{\odot} {\rm pc}^{-2}$. 
The mass and surface density are similar to the typical GMCs in the Milky Way \citep{2010ApJ...723..492R}.
On the other hand, at $Z<Z_{\odot}$, YMCs forms from clouds with $10^6~M_{\odot}$ only if the surface density is larger than $\Sigma_{\rm cl} \sim 200~M_{\odot}{\rm pc^{-2}}$ ($\sim 350~M_{\odot}{\rm pc^{-2}}$) at $Z=10^{-1}~Z_{\odot}$ ($Z=10^{-2}~Z_{\odot}$).
Besides, the condition of YMC formation for $10^5~M_{\odot}$ is strict.
Thus, we suggest that YMCs do not form frequently in low-metallicity environments.

\section{Summary and Discussion}\label{summary}

We have studied the formation processes of young massive star clusters (YMCs, $>10^4~M_{\odot}$) by performing three-dimensional radiation hydrodynamics simulations.
Our simulations include photoionization, photodissociation of molecules, radiation pressure.
We have investigated giant-molecular clouds (GMCs) with  the wide parameter ranges:
$\Sigma \simeq 80-3200~M_{\odot}{\rm pc^{-2}}$, $Z=10^{-2}Z_{\odot} - Z_{\odot}$ and the virial parameter $\alpha_0 = 1$ or $2$.
Also, we have developed a semi-analytical model that reproduces the simulation results nicely.
Our findings are summarized as follows:

\begin{description}
\item[(i)]
In the early phase of the cloud collapse, turbulent motions induce the formation of high-density filaments in which stars form.
Then, after one free-fall time, the star formation rate rapidly increases at the center of a cloud.

\item[(ii)]  
The star clusters are gravitationally bounded once the total stellar mass exceeds 0.1 times the initial cloud mass.
Simultaneously, the stellar core forms if the cloud is compact enough.
In such a case, the gravitational force from the star cluster can pull even ionized hot gas and induces long-lasting star formation.
We obtain the threshold surface density for the core formation as $\Sigma_{\rm th} \sim 300~M_{\odot} {\rm pc^{-2}}$ in the cases with the cloud mass $M_{\rm cl} = 10^{6} ~M_{\odot}$ at $Z=Z_{\odot}$.

\item[(iii)] The star formation efficiencies (SFEs) sensitively depend on the surface densities of clouds.
If the surface density is lower than the threshold value of the core formation, the photoionization feedback easily disrupts clouds, resulting in the SFEs lower than 0.2. 
In compact clouds, the star formation can continue against the photoionization feedback.
In these cases, the SFEs are $\epsilon_* \geq 0.3$, regardless of metallicities and virial parameters.

\item[(iv)] 
The SFE sensitively depends on the metallicity in the cases of diffuse clouds. The temperature of H{\sc ii} regions of low-metallicity gas is higher than that of metal-rich gas because of the lower cooling rate. Therefore, the H{\sc ii} regions rapidly expand and suppress the star formation. On the other hand, in the cases of massive compact clouds, the SFE does not change significantly with the metallicity, because the photoionization feedback is not dominant.

\item[(v)] 
Turbulent motions delay the gravitational contraction of the clouds. 
Therefore, in the cases with higher turbulence energy, the star formation proceeds slowly and the SFEs become lower. Also, we find that the star clusters can be unbounded due to the strong turbulent motion.

\item[(vi)]
We have constructed the semi-analytical model reproducing the simulation results.
We have used the expanding shell model, considering the thermal pressure of H{\sc ii} regions, radiation pressure, and gravity from the formed star cluster and the shell.
The SFEs are estimated based on the assumption in which the star formation continues only for the shell expanding. 
We also consider the enhancement of the star formation rates due to the stellar core formation.
The threshold surface density of the core formation  $(\Sigma_{\rm th})$ is analytically estimated by considering the condition that the velocity of the expanding shell is comparable to the escape velocity of the stellar core. 
This threshold density increases with the lower-metallicity or the smaller cloud, e.g., $\Sigma_{\rm th} \sim  560~M_{\odot}{\rm pc^{-2}}$ ($410~M_{\odot}{\rm pc^{-2}}$) with the cloud mass $M_{\rm cl} = 10^{6}~M_{\odot}$ and the metallicity $Z=10^{-2}Z_{\odot}$ ($10^5~M_{\odot}$ and $Z_{\odot}$).

\item[(vii)]
Combining the analytical estimation of SFEs and the bound fraction obtained from the numerical simulations, we have derived the condition for the YMC formation.
We have shown that massive and higher-metallicity clouds are favored for the formation of YMCs.
In the clouds with $M_{\rm cl} = 10^6~M_{\odot}$, the YMCs are formed if the surface density is larger than $100~M_{\odot}{\rm pc^{-2}}$.

\end{description}

In the lower-metallicity environments, higher surface densities are needed for producing YMCs. 
In particular, we predicted that YMCs are formed only in clouds with $\Sigma_{\rm cl} \gtrsim 350 ~M_{\odot} {\rm pc^{-2}}$ at $Z=10^{-2}~Z_{\odot}$ even if the cloud mass is $10^6~M_{\odot}$.
These conditions are related to the minimum metallicity of observed globular clusters \citep[GCs:][]{2018RSPSA.47470616F}.
\citet{2018RSPSA.47470616F} showed the metallicity floor of GCs in the Milky Way with $Z \sim 10^{-2.5} Z_{\odot}$ \citep[but see ,][]{2020Natur.583..768W}.
Recently \citet{2020arXiv201007395L} observed the lower metallicity GC of $Z\sim 10^{-2.9}Z_{\odot}$ in Andromeda galaxy.
\citet{2019MNRAS.486L..20K} pointed out that this floor is related to the minimum mass for GCs formation of the host galaxies and their maximum redshift.
We have suggested additional condition that the higher surface densities $(\Sigma_{\rm cl} \gtrsim 350 ~M_{\odot} {\rm pc}^{-2})$ is required for the formation of low-metallicity GCs.
Previous studies indicated that the low-metallicity massive clouds are formed in the early galaxies \citep[e.g.,][]{2016ApJ...831..204R, 2018MNRAS.475.4252A}.
We will investigate the star cluster formation in galaxies modeled by cosmological simulations in future work.

\citet{2009A&A...498L..37P} suggested that there are two types of young star clusters with $\sim 10^4~M_{\odot}$ , YMCs, and "leaky clusters."
Leaky clusters show the low-stellar density ($\sim 1-10~M_{\odot}{\rm pc^{-3}}$) compared with YMCs \citep[$\gtrsim10^3~M_{\odot}{\rm pc^{-3}}$,][]{2010ARA&A..48..431P}.
\citet{2016ApJ...817....4F} showed that these two populations had different surface densities at birth \citep[also see, ][]{2011A&A...536A..90P}.
In our simulation with $10^{6}~M_{\odot}$ clouds at $Z=Z_{\odot}$, the SFEs jump occurs in the narrow surface density range.
This SFE jump is likely to separate star clusters into the two populations.
Note that, in the low-metallicity cases, it is difficult for the lower surface density clouds to form star clusters with $10^4~{M_{\odot}}$ due to the low SFE.
Therefore, we indicate that these two populations do not emerge in the low-metallicity environments.

In this work, we have focused on radiative feedback. There are other feedback processes such as stellar wind, outflow, and supernovae. These feedback processes are also likely to regulate star formation.
For example, the stellar wind pushes out the ambient gas. This makes low-density bubbles where H{\sc ii} regions can expand rapidly.
Recently, \citet{2020MNRAS.tmp.3491G} showed the combination effects of the stellar wind and the photoionization processes \citep[see also,][]{2014MNRAS.442..694D, 2020MNRAS.tmp.3491G, 2020MNRAS.497.4718D}.
The collimated outflow from stars also has impacts on the star formation via the injection of turbulence energy
\citep{2006ApJ...640L.187L, 2007ApJ...662..395N}.
Recently, \citet{2021MNRAS.502.3646G} found that the outflow has a crucial role in regulating the IMF.
They also found that the low-surface density cloud $(\sim 60~M_{\odot} {\rm pc^{-2}})$ with $2 \times 10^4~M_{\odot}$ wad disrupted only by the outflow. 
The properties of the outflow sensitively depend on the physical states of the star-forming regions. 
The direction of the outflow from the single star depends on both angular momentum and magnetic fields \citep[e.g.,][]{2020ApJ...898..118H}, and the outflow is failed in massive and weakly magnetized massive core \citep{2020MNRAS.499.4490M}.
Some cases in our simulations have cloud lifetimes longer than the lifetimes of OB stars. In this case, SNe should turn on and evacuate the gas \citep[e.g.,][]{2016MNRAS.463.3129G}.
The magnetic field can affect the gas structures and suppress the star formation rate by a factor of $\sim 2$ \citep[e.g.,][]{2012ApJ...761..156F}.
It also suppresses shell expansion caused by radiative feedback or stellar wind \citep[e.g.,][]{2018NatAs...2..896O, 2021ApJ...911..128K}.
In future works, we will incorporate these feedback effects in our simulations and investigate the impacts of multiple feedback. 

In this work, we have assumed the constant mass-to-light ratio for sink particles.
This assumption is reasonable only for massive star clusters with a mass larger than $10^4~M_{\odot}$ \citep{2016ApJ...819..137K}.
In practice, in the cases of clouds with $\lesssim 10^{5}~M_{\odot}$, the masses of star clusters are below $10^4~M_{\odot}$, which indicates that the expected number of massive stars can be lower than unity. 
In such a case, formation sites and timings of radiation sources depend on the models. 
\citet{2019MNRAS.488.2970G} showed that the SFEs could change with models of star formation by a factor of $\sim 3$ for the clouds with $M_{\rm cl}=10^5~M_{\odot}$ and $R_{\rm cl} = 20~{\rm pc}$.
Also, total emissivities of star clusters depend on the resolution in the cases with diffuse clouds of $M_{\rm cl}=10^5~M_{\odot}$ (see Appendix \ref{appendix_resolution_study}).
Thus, we suggest that more sophisticated modeling of star formation should be developed in studying star clusters in clouds with $\lesssim 10^{5}~M_{\odot}$ in the future.

We have followed the formation and dissociation of CO molecules.
We have found that the spatial distribution and total abundance of CO molecules depended on the metallicity. Also, the CO distribution changes with the cloud evolution. Therefore, the surface brightness maps of CO lines can be a tool to understand the physical properties of star-forming clouds.
The CO observations in the low-metallicity galaxies (such as LMC/SMC) are progressing \citep[e.g.,][]{2015ApJ...807L...4F, 2017ApJ...844...98M, 2019ApJ...886...15T}.
We plan to model the CO map to compare with the observations in a future study.

\section*{Acknowledgements}

The authors wish to express their cordial thanks to Profs. Masayuki Umemura and Ken Ohsuga for their continual interest, advice, and encouragement.  
We would like to thank Michiko Fujii, Tomoaki Matsumoto, Takashi Hosokawa, Kazuyuki Sugimura, Kazutaka Kimura, Kazuyuki Omukai and Shu-ichiro Inutsuka for useful discussions and comments.
The numerical simulations were performed on the Cray XC50 (Aterui II) at the Center for Computational Astrophysics of National Astronomical Observatory of Japan and Yukawa-21 at Yukawa Institute for Theoretical Physics in Kyoto University.
This work is supported in part by the MEXT/JSPS KAKENHI Grant Number 17H04827, 20H04724, National Astronomical Observatory of Japan (NAOJ) ALMA Scientific Research Grant Number 2019-11A, and JST Fusion Oriented REsearch for disruptive Science and Technology (FOREST) (HY). 

\section*{Data Availability}

The data underlying this article will be shared on reasonable request to the corresponding author.



\bibliographystyle{mnras}
\bibliography{article} 




\appendix

\section{Chemical network}\label{thermochemical_processes}

In table \ref{chem_process}, we summarize the chemical reaction related with ${\rm H}$, ${\rm H_2}$, ${\rm H}^+$, ${\rm H}^-$, ${\rm H_2}^+$, and ${\rm e}$.
We also include $\rm CO$, $\rm C^{+}$, $\rm O$, ${\rm O^+}$, and ${\rm O}^{2+}$ as the metal species.
We use the chemical network of \citet{1997ApJ...482..796N} for CO formation as in \citet{2020MNRAS.497.3830F}.

\begin{table*}
    \caption{Chemical Reactions }
    \label{chem_process}
    \centering
    \begin{tabular}{lllc}
      \hline \hline
      Number & Reaction & Rate Coefficient & Reference \\
      \hline
      ${\rm H0}$ & ${\rm H} + {\rm e} \rightarrow {\rm H}^{+} + 2 e$ & $k_{\rm H0} = \exp [ - 32.71396786 + 13.536556 \times  \left( \ln T \left({\rm ev} \right) \right) -5.73932875 \left( \ln T \left({\rm ev} \right) \right)^{2}$ & 1 \\
      & & $+1.563154998 \left( \ln T \left({\rm ev} \right) \right)^{3} -0.2877056   \left( \ln T \left({\rm ev} \right) \right)^{4} + 3.48255977 \times 10^{-2}  \left(\ln T \left({\rm ev} \right) \right)^{5} $ & \\
      & & $ -2.63197617 \times 10^{-3}  \left( \ln T \left({\rm ev} \right) \right)^{6} + 1.11954395 \times 10^{-4}  \left( \ln T \left({\rm ev} \right) \right)^{7} - 2.03914985 \times 100
      ^{-6}  \left( \ln T \left({\rm ev} \right) \right)^{8}  $ & \\
       ${\rm H1}$ & ${\rm H^{+}} + e \rightarrow {\rm H} + \gamma$ & $k_{\rm H1, A} = 1.269 \times 10^{-13} (315614/T)^{1.503} [1+(604625/T)^{0.47}]^{-1.923}  $ \hspace{2mm} (case A) & 2, 3 \\
      & & $k_{\rm H1, B} = 2.753 \times 10^{-14}(315614/T)^{1.5} [1+(115188/T)^{0.407}]^{-2.242}$ \hspace{3.3mm} (case B)& \\

 	  ${\rm H2}$ & ${\rm H^{-}} + {\rm H} \rightarrow {\rm H_{2}} + {\rm e} $ & $k_{\rm H2} = 1.35 \times 10^{-9} \left[ T^{0.098493} + 0.32852 T^{0.5561} + 2.771 \times 10^{-7} T^{2.1826} \right]$ & 4 \\
      & & $/ \left[ 1 + 6.191 \times 10^{-3} T^{1.0461} + 8.9712 \times 10^{-11} T^{3.0424} + 3.2576 \times 10^{-14} T^{3.7741}  \right]$ & \\

      ${\rm H3}$ & ${\rm H_2} + {\rm H}^{+} \rightarrow {\rm H_2}^+ + {\rm H}$ & $k_{\rm H3} = 3 \times 10^{-10} \exp(-21050/T) $ \hspace{10mm} $ T < 10^{4}~{\rm K}$ & 5 \\
      & & \hspace{7mm}  $1.5 \times 10^{-10} \exp(-14000/T)$ \hspace{7.5mm} $T>10^{4}~{\rm K}$ & \\
      
      ${\rm H4}$ & ${\rm H_2} + {\rm e} \rightarrow 2 {\rm H} + {\rm e}$ & $k_{\rm H4} = 4.4 \times 10^{-10} T^{0.35} \exp(-102000/T)$ & 5 \\

      ${\rm H5}$ & ${\rm H2} + {\rm H} \rightarrow 3 {\rm H}$ & see the reference & 6 \\

      ${\rm H6}$ & $3 {\rm H} \rightarrow {\rm H_2} + {\rm H}$ & $k_{\rm H6} = 6 \times 10^{-32} T^{-0.25} + 2 \times 10^{-31} T ^{-0.5}$ & 7 \\

      ${\rm H7}$ & $2 {\rm H} + {\rm H_2} \rightarrow 2 {\rm H_2}$ & $k_{\rm H7} = k_{\rm H6}/8$ & 8 \\

 	  ${\rm H8}$ & $2 {\rm H_{2}} \rightarrow 2 {\rm H} + {\rm H_{2}}$ & $k_{\rm H8} = k_{\rm high}^{1 - a} k_{\rm low}^{a}$ & 8 \\
      & & \hspace{2mm} $k_{\rm low} = 1.18 \times 10^{-10} \exp(-6.95 \times 10^{4} / T)$ &  \\
      & & \hspace{2mm} $k_{\rm high} = 8.125 \times 10^{-8} T^{-1/2} \exp(-5.2 \times 10^{4} / T) \left[ 1 - \exp(-6000/T) \right]$ & \\
      & & \hspace{2mm} $a = \left( 1 + n / n_{\rm cr} \right)^{-1}$, $\log_{10} (n_{\rm cr}) = 4.845 - 1.3 \log_{10} \left(T / 10^{4} \right) + 1.62 \left[ \log_{10} \left( T / 10^{4}  \right) \right]^2$ & \\

      ${\rm H9}$ & ${\rm H} + {\rm e} \rightarrow {\rm H}^{-} + \gamma$ & $k_{\rm H9} = 1.4 \times 10^{-18} T^{0.928} \exp(-T/16200)$ & 5 \\

      ${\rm H10}$ & $2 {\rm H} \rightarrow {\rm H}^{+} + {\rm e} + {\rm H}$ & $k_{\rm H10} = 1.7 \times 10^{-4} k_{\rm H0}$ & 5 \\

      ${\rm H11}$ & ${\rm H}^{-} + {\rm e} \rightarrow {\rm H} + 2 {\rm e}$ & $k_{\rm H11} = \exp [ - 18.01849334 + 2.3608522 \times  \left( \ln T \left({\rm ev} \right) \right)-0.28274430 \left( \ln T \left({\rm ev} \right) \right)^{2} +1.62331664 \times 10^{-2}  \left( \ln T \left({\rm ev} \right) \right)^{3}  $& 1  \\
		&& \hspace{0.5cm} $-3.36501203 \times 10^{-2}   \left( \ln T \left({\rm ev} \right) \right)^{4} +1.17832978 \times 10^{-2}  \left( \ln T \left({\rm ev} \right) \right)^{5}  -1.65619470 \times 10^{-3}  \left( \ln T \left({\rm ev} \right) \right)^{6} $ & \\
        & & \hspace{0.5cm} $+1.06827520 \times 10^{-4}  \left( \ln T \left({\rm ev} \right) \right)^{7} - 2.63128581 \times 10^{-6}  \left( \ln T \left({\rm ev} \right) \right)^{8} $ & \\
        
      ${\rm H12}$ & ${\rm H}^{-} + {\rm H}^{+} \rightarrow {\rm H_2}^{+} + {\rm e}$ & $k_{\rm H12} = 6.9 \times 10^{-9} / T^{0.35}$ \hspace{5mm} $T< 8000~{\rm K}$ & 5 \\
      & & \hspace{8mm} $9.6 \times 10^{-7} / T^{0.9} $ \hspace{6mm} $T>8000~{\rm K}$ & \\

      ${\rm H13}$ & ${\rm H}^{-} + {\rm H}^{+} \rightarrow 2 {\rm H}$ & $k_{\rm H13}= 6.3 \times 10^{-8} + 5.7 \times 10^{-6} / \sqrt{T} - 9.2 \times 10^{-11} \sqrt{T} + 4.4 \times 10^{-13} T$ & 5 \\

      ${\rm H14}$ & ${\rm H} + {\rm H}^{+} \rightarrow {\rm H_2}^{+} + \gamma$ & $k_{\rm H14} = 10^{-19.38-1.523 \log T+1.118(\log T)^2 - 0.1269(\log T)^3}$ & 5 \\

      ${\rm H15}$ & ${\rm H_2}^{+} + {\rm H} \rightarrow {\rm H_2} + {\rm H}^+$ & $k_{\rm H15} = 6.4 \times 10^{-10}$ & 5 \\

      ${\rm H16}$ & ${\rm H_2}^{+} + {\rm e} \rightarrow 2 {\rm H}$ & $k_{\rm H16} = 2 \times 10^{-7} / \sqrt{T}$ & 5 \\

      ${\rm H17}$ & ${\rm H_2}^{+} + {\rm H}^{-} \rightarrow {\rm H_2} + {\rm H}$ & $k_{\rm H17} = 2.3 \times 10^{-7} / \sqrt{T/300}$ & 9 \\

	  ${\rm H18}$ & ${\rm 2 H } + {\rm grain} \rightarrow {\rm H_{2}} $ &  $k_{\rm H18} = 6.0 \times 10^{-17} \sqrt{T/300} f_{a} \left(Z/Z_{\odot} \right) [1.0 + 4.0 \times 10^{-2} \sqrt{T+T_{\rm gr}} + 2.0 \times 10^{-3} T + 8.0 \times 10^{-6} T^{2}]^{-1}$ &  10\\
      & & \hspace{2mm} $f_{a} = [1.0 + \exp (7.5 \times 10^{2} (1/75 - T_{\rm gr}^{-1}))]^{-1}$ & \\
      
      ${\rm M1}$ & ${\rm C^+} + {\rm O} \rightarrow {\rm CO}$ & see the references & 11,12 \\

      ${\rm RH1}$ & ${\rm H} + \gamma \rightarrow {\rm H}^{+} + {\rm e}$ & $R_{\rm HI}$ & Eq. \eqref{A2_16}  \\

      ${\rm RH2}$ & ${\rm H_2} + \gamma \rightarrow 2 {\rm H}$ & $R_{\rm H_2}$ & Eq. \eqref{A22_1}\\

      ${\rm RH3}$ & ${\rm H}^{-} + \gamma \rightarrow {\rm H} + {\rm e}$ & $R_{\rm H^{-}}$ & 13 \\

      ${\rm RCO}$ & ${\rm CO} + \gamma \rightarrow C^{+} + {\rm O}$ & $R_{\rm CO}$ & Eq. \eqref{A22_1} \\

      \hline
    \end{tabular}
     \begin{minipage}{1\hsize}
     References. (1) \citet{1997NewA....2..181A} (2) \citet{1992ApJ...387...95F} (3) \citet{2007ApJ...666....1G} (4) \citet{2010Sci...329...69K} (5) \citet{1998A&A...335..403G} (6) \citet{1998ApJ...499..793M} (7) \citet{2013ApJ...773L..25F} (8) \citet{1983ApJ...271..632P} (9) \citet{1991A&A...242..241M} (10) \citet{1985ApJ...291..722T} (11) \citet{1997ApJ...482..796N} (12) \citet{2018ApJ...857...57N} (13) \citet{2016ApJ...824..119H}
     \end{minipage}
   \end{table*}

   \section{Thermal processes}\label{thermal_processes}

  In table \ref{thermal_process}, we summarize the heating and cooling processes included in our simulations.
  We incorporate line cooling of $\rm H_2$ rovibrational transitions and metal line cooling of C{\sc ii}, CO, O{\sc i}, O{\sc ii}, and O{\sc iii}.
  We use the fitting function derived in \citet{2015MNRAS.453.2901G} for the line cooling from the
  $\rm H_2$ rovibrational transition, and the escape probability tabulated by \citet{2018MNRAS.473.4754F}.
  The cooling rates of C{\sc ii}, O{\sc i}, O{\sc ii}, and O{\sc iii} are estimated by solving the statistical equilibrium of each energy level as \citet{2020MNRAS.497..829F}. 
  Here, we assume that the ionization rate of O{\sc i} is the same as H{\sc i} because the ionization energies of O{\sc i} and H{\sc i} are almost the same.
  We solve equilibrium state between O{\sc ii} and O{\sc iii} as \citet{2020MNRAS.497..829F}.
  The energy transport between gas and dust grain is formulated as the function of dust temperature $T_{\rm d}$ \citep{2005ApJ...626..627O}.

   \begin{table*}
    \caption{Thermal processes.}
    \label{thermal_process}
    \centering
    \begin{tabular}{lllc}
      \hline 
      Number & Process & Rate ($\rm erg \, cm^{-3} \, s^{-1}$) & Reference \\
      \hline
        \multicolumn{4}{|c|}{Heating}\\ 
        1 & ${\rm H_2}$ formation & $\Gamma_{1} = [ \, 3.73 (1+ n_{\rm cr} / n_{\rm H})^{-1} k_{\rm H2} n({\rm H}) n({\rm H^{-}}) + 4.48 (1+ n_{\rm cr} / n_{\rm H})^{-1} k_{\rm H6} n^{3}({\rm H})  $ & 1,2 \\
        & & \hspace{7mm} $+ \, \left( 0.2 + 4.2 (1+ n_{\rm cr} / n_{\rm H})^{-1} \right) k_{\rm H18} n^{2}({\rm H})] ~ \rm{eV}  $ &  \\
        & &  $n_{\rm cr} = 10^{6}T^{-1/2}/ \left\{ 1.6 n({\rm H}) / n_{\rm H} \exp [-(400/T)^{2}] + 1.4 n({\rm H_2})/n_{\rm H} \exp[-1200/(T+1200)]  \right\} ~ {\rm cm^{-3}} $ & \\
        2 & ${\rm H}$ photoionization & $\Gamma_{2}$ & Eq. \eqref{gamma2_heat} \\
        3 & $\rm H_2$ photodissocian & $\Gamma_{3}$, see the reference & 1 \\

        \multicolumn{4}{|c|}{Cooling}\\

        1 & ${\rm H_2}$ dissociation & $\Lambda_{1} = 4.48 \, [ \, k_{\rm H4} n({\rm H_2}) n({\rm e}) + k_{\rm H5} n({\rm H_2}) n({\rm H}) + k_{\rm H8} n^{2}({\rm H_2})  \, ] \, {\rm eV}$ & 1,2 \\

        2 & ${\rm H}$ ionization & $\Lambda_2 = 13.6 k_{\rm H0} n({\rm H}) n({\rm e}) {\rm eV}$ & 1,2 \\

        3 & ${\rm H}$ recombination & $\Lambda_3 = \exp[ \ln 10 \times (-26.02 + 0.9187 \log_{10} T - 3.733 (\log_{10} T))^2 + 0.1174 (\log_{10} T))^3$ & \\
        & & \hspace{7mm} $ - 0.01707 (\log_{10} T))^4 + 8.119 \times 10^{-4} (\log_{10} T))^5 ] n({\rm H^+}) n({\rm e}) $ & 3, 4 \\

        4 & ${\rm H^{-}}$ free-bound & $\Lambda_{4} = 0.755 k_{\rm H9} n({\rm H}) n({\rm e})$ & 5 \\

        5 & ${\rm H}$ excitation  & $\Lambda_{5} = 7.50 \times 10^{-19} [1 + (T/10^5)^{1/2}]^{-1}\exp[-118348/T]n({\rm e})n({\rm H^{+}})$ & 6 \\

        6 & ${\rm He}^+$ excitation & $\Lambda_{6} = 5.54 \times 10^{-17} T^{-0.397} [1 + (T/10^5)^{1/2}]^{-1}\exp[-473638/T] n({\rm e})n({\rm He^{+}}) $ & 6 \\

        7 & Free-free & $\Lambda_{7} = 1.426 \times 10^{-27} T^{1/2} g_{\rm ff} (T) n(\rm H^+) n(\rm e) $& 7 \\

        & & $g_{\rm ff} (T) = 0.79464 + 0.1243 \log_{10} T$ \hspace{5mm} $T < 3.2 \times 10^5~{\rm K}$ & \\
        & & \hspace{7.3mm} $= 2.13164 - 0.1240 \log_{10} T$ \hspace{5mm} $T > 3.2 \times 10^5~{\rm K}$ & \\ 

        8 & Compton   & $\Lambda_{8} = 1.017 \times 10^{-37} T_{\rm CMB}^4 (T - T_{\rm CMB}) n({\rm e})$ & 6 \\
        9 & Line cooling & $\Lambda_{9} = \Lambda_{\rm H_2} + \Lambda_{\rm CII} + \Lambda_{\rm CO} + \Lambda_{\rm OI} + \Lambda_{\rm OII} + \Lambda_{\rm OIII}$ & \\
        10 & Gas-grain heat transfer & $\Lambda_{10} =  5.83 \times 10^{-8} n_{\rm H} \rho ( T / 10^{3})^{1/2} [ 1 - 0.8 \exp(- 75/ T)] ( T - T_{\rm d} ) ( Z / Z_{\odot} )  $ & 1,2 \\

      \hline
    \end{tabular}
     \begin{minipage}{1\hsize}
     References. (1) \citet{1979ApJS...41..555H} (2) \citet{2000ApJ...534..809O} (3) \citet{1992ApJ...387...95F} (4) \citet{2017MNRAS.469...62S} (5) \citet{2016ApJ...824..119H} (6) \citet{1992ApJS...78..341C} (7) \citet{1987ApJ...318...32S}
     \end{minipage}
   \end{table*}

\section{Radiation Transfer}\label{RT_Moment_solver}

We use the moment-based radiative transfer (RT) with M1-closure \citep[e.g,][]{2013MNRAS.436.2188R,2015MNRAS.449.4380R,2019MNRAS.485..117K} for extreme ultraviolet (EUV), far-ultraviolet (FUV), and infrared (IR) photons.
In this method, the computational cost does not depend on the number of radiative sources.
Therefore, it is suitable for calculating massive clouds where a lot of stars are likely to form.
The moment-based technique has been introduced in recent simulation studies of the star cluster formation \citep[e.g.,][]{2015ApJ...809..187S, 2016ApJ...829..130R, 2017MNRAS.471.4844G, 2019MNRAS.489.1880H}.

\subsection{Momentum-based RT equations}

In the RT calculations, we solve the specific intensity $I_{\nu}(\bf{x}, t, \nu, \bf{n})$, which is the radiative energy ($dE_{\nu}$) crossing the point $\bf{x}$ at time $t$, per unit area $dA$, per unit time $dt$, per unit solid angle $d \Omega$ around the direction $\bf{n}$, and per unit frequency range $d\nu$ as
\begin{align}
    dE_{\nu} = I_{\nu} dA dt d\Omega d\nu . \label{def_of_intensity}
\end{align}
The equation of RT is 
\begin{align}
    \frac{1}{c} \frac{\partial I_{\nu}}{\partial t} + \bm{n} \cdot \nabla I_{\nu} =  j_{\nu} - \alpha_{\nu} I_{\nu},  \label{A1}
\end{align}
where $j_{\nu}$ is the emissivity, 
and $\alpha_{\nu}$ is the absorption coefficient.

We obtain the zeroth and first moment equations from integration of the RT equation \eqref{A1} over all solid angle as
\begin{align}
     \frac{\partial E}{\partial t} + \nabla \cdot {\bf{F}} = S - \alpha_{\rm E} \tilde{c} E, \label{zeros_moment}
\end{align}
\begin{align}
    \frac{1}{\tilde{c}} \frac{\partial {\bf{F}}}{\partial t} + \tilde{c} \nabla \cdot {\bf{P}} = - \alpha_{\rm F} {\bf{F}}, \label{first_moment}
\end{align}
where $E$, $\bf{F}$, and $\bf{P}$ are the radiation energy density, the flux and the radiation pressure tensor \citep{2013MNRAS.436.2188R}, and $\tilde{c}$ represents the reduced light speed in the numerical simulations (also see Sec \ref{RT_timestep}).
In Equation \eqref{first_moment}, $S$ is the source term, $\alpha_{\rm E}$ and $\alpha_{\rm F}$ are the energy density and flux weighted absorption coefficients.

The radiation pressure tensor is defined with the Eddington tensor $(\bf{D})$ and the radiation energy density ($E$) as 
\begin{align}
  {\bf P} = E {\bf D}. \label{eddington_tensor}
\end{align}
We need to calculate the RT equation \eqref{A1} to obtain the accurate values of the Eddington tensor \citep[the variable Eddington tensor method,][]{1984oup..book.....M,2018MNRAS.473.4754F}.
However, direct calculations of Equation \eqref{A1}
are computationally expensive.
Thus, we adopt the M1 closure scheme in which 
the Eddington tensor $\bf{D}$ is approximated as \citep{1984JQSRT..31..149L}:
\begin{align}
  {\bf D} = \frac{1-\chi}{2} {\bf I} + \frac{3 \chi -1}{2} {\bf n} \otimes {\bf n}, \label{M1closure}
\end{align}
where
\begin{align}
  {\bf n} = \frac{{\bf F}}{|{\bf F}|}, ~ \chi = \frac{3 + 4 f^2}{5 + 2 \sqrt{4-3f^2}}, ~ {\rm and} ~ f = \frac{|{\bf F}|}{\tilde{c} E}. \label{M1c2}
\end{align}
This approximation is accurate for 
the optically thick or thin limits ($f \rightarrow 0 $, ${\bf{D}} \rightarrow \frac{1}{3} {\bf I}$ or $f \rightarrow 1 $, ${\bf{D}} \rightarrow {\bf n} \otimes {\bf n}$).
However, it fails to follow the radiation transport accurately
when fluxes from 
different directions collide.
In this work, a lot of stellar particles distributes at the center of cloud and most of radiation energy propagates radially. Therefore, the colliding effect is unlikely to be serious \citep[see also,][]{2017MNRAS.470..224T}.

We estimate the photoionization and the photodissociation rates from the number density of EUV and FUV photons.
In this case, it is convenience to calculate transfer of the photon number density directly instead of Equation \eqref{zeros_moment} and \eqref{first_moment} for EUV and FUV photons as \citep{2013MNRAS.436.2188R}
\begin{align}
    \frac{\partial N_{\gamma}}{\partial t} + \nabla \cdot {\bf F_{\gamma}} = \dot N_{\gamma, *} - \bar{\alpha} \tilde{c} N_{\gamma}, \label{A2_1} 
\end{align}
\begin{align}
    \frac{\partial {\bf F}_{\gamma}}{\partial t} + \tilde{c}^2 \nabla \cdot {\bf P}_{\gamma} = - \bar{\alpha} \tilde{c} {\bf F}_{\gamma}, \label{A2_2}
\end{align}
where $N_{\gamma}$, ${\bf F}_{\gamma}$, and $\dot N_{\gamma, *}$ are the photon number density, the photon number flux, and the photon injection rate from a radiation source.
The frequency-averaged absorption coefficient $\bar{\alpha}$ is given as
\begin{align}
    \bar{\alpha} = n_{\rm i} \bar{\sigma}_{i}, \label{A2_3}
\end{align}
\begin{align}
    \bar{\sigma}_{i} = \frac{\int^{\nu_2}_{\nu_1} \frac{4 \pi J_{\nu}}{h \nu} \sigma_{i} (\nu) d \nu}{\int^{\nu_2}_{\nu_1} \frac{4 \pi J_{\nu}}{h \nu} d \nu}, \label{A2_3_2}
\end{align} 
where $\sigma_{i}(\nu)$ is the frequency dependent cross-section, and $J_{\nu}$ is the mean intensity:
\begin{align}
    J_{\nu} = \frac{1}{4 \pi} \int I_{\nu} d \Omega. \label{A2_4}
\end{align}
As in Equation \eqref{eddington_tensor}, the radiation pressure tensor ${\bf P}_{\gamma}$ is evaluated from the M1-closure relation (see eq. \ref{M1closure}).

\subsection{Computational procedure for RT}
We solve the moment equations on
Cartesian grid.
Each grid has four variables $(E, {\bf F})$ or $(N_{\gamma}, {\bf F_{\gamma}})$.
According to \citet{2013MNRAS.436.2188R}, we adopt the operator-splitting method to advance the time step $(\Delta t)$. Here, we calculate the moment equations in the following three steps.

\subsubsection{The injection step}
In this step, we only inject the photons emitted from the radiation sources as 
\begin{align}
    N_{\gamma}^{n+1} = N_{\gamma}^{n} + \dot N_{\gamma,*} \Delta t, \label{A2_5}
\end{align}
where $\dot N_{\gamma,*}$ is the local injection rate from sources.
We estimate the injection rate with the photon emissivity $S_{\rm i}$ as
\begin{align}
    \dot N_{\gamma,*} = \frac{S_{\rm i}}{\Delta V}, \label{A2_6}
\end{align}
where $\Delta V$ is the cell volume.
As discussed in Section \ref{sec_radiation_sources}, we assume that the photon emissivities of EUV, FUV, and IR components are proportional to the sink mass in this study.

\subsubsection{The transport step}
In the transport step, we calculate the photon propagation without 
the source and absorption terms in Equations \eqref{A2_1} and \eqref{A2_2} as
\begin{align}
    \frac{\partial N_{\gamma}}{\partial t} + \nabla \cdot {\bf F}_{\gamma} = 0, \label{A2_7}
\end{align}
\begin{align}
    \frac{\partial {\bf F}_{\gamma}}{\partial t} + \tilde{c}^2 \nabla \cdot {\bf P}_{\gamma} = 0. \label{A2_8}
\end{align}
Equation \eqref{A2_7} and \eqref{A2_8} are summarized as a following vector form:
\begin{align}
    \frac{\partial \mathcal{U}}{\partial t} + \nabla \mathcal{F} (\mathcal{U}) = 0. \label{A2_9}
\end{align} 
where $\mathcal{U} = (N_{\gamma}, {\bf F_{\gamma}})$ and $\mathcal{F}(\mathcal {U}) = ({\bf F_{\gamma}}, \tilde{c}^2 {\bf P}_{\gamma})$.
In each time-step, we update $\mathcal {U}^{n}$ by solving the explicit formula of Equation \eqref{A2_9} as
\begin{align}
    & \frac{\mathcal{U}^{n+1} - \mathcal{U}^{n}}{\Delta t} \nonumber \\
    & + \frac{\mathcal{F}^{n}_{i+1/2}- \mathcal{F}^{n}_{i-1/2}}{\Delta x} 
     + \frac{\mathcal{F}^{n}_{j+1/2}-\mathcal{F}^{n}_{j-1/2}}{\Delta y}  
     + \frac{\mathcal{F}^{n}_{k+1/2}-\mathcal{F}^{n}_{k-1/2}}{\Delta z} = 0, \label{A2_10}  
\end{align}
where $n$ is the step-number, and $(i, k, j)$ represent the indices of cells in x, y, z directions.
We evaluate the intercell fluxes $\mathcal{F}^{n}_{i+1/2}$ from 
the flux between the $i$-th cell and the $(i+1)$-th cell.
In our simulations, we adopt the global Lax-Friedrich (GLF) as
\begin{align}
  \mathcal{F}^{n}_{i+1/2} = \frac{\mathcal{F}^{n}_{i} + \mathcal{F}^{n}_{i+1}}{2} - \frac{\tilde{c}}{2} \left( \mathcal{U}^{n}_{i+1} - \mathcal{U}^{n}_{i} \right). \label{A2_11}
\end{align}
The numerical diffusion of GLF is somewhat larger
than Harten-Lax-van Leer (HLL) flux function \citep{HLL1983} which is often used in RHD simulations \citep{2013MNRAS.436.2188R, 2019MNRAS.485..117K}.
However, the HLL function also induces a diffusion to a beam propagating diagonally, while it is accurate for one crossing cells vertically. 
Therefore, the HLL function tends to form an asymmetric radiation field around an isotropic radiation source. 
On the other hand, the GLF function is better to follow the isotropic radiation.

\subsubsection{The thermochemical step}
Next, we calculate the thermochemical evolution and photon absorption.
As discussed in Appendix \ref{thermochemical_processes}, we solve the equations of energy and non-equilibrium chemistry.
We estimate the photoionization and photoheating rates of neutral hydrogen ($R_{\rm HI}$, $\Gamma_{2}$), photodissociation rates of $\rm H_2$ and $\rm CO$ molecules ($R_{\rm H_2}$, $R_{\rm CO}$), and the dust absorption rate from the number densities of EUV, FUV, and IR photons.

After the calculations of thermochemical evolution, we estimate the photon absorption of each component.
Without the transfer terms, the moment equations \eqref{A2_1} and \eqref{A2_2}  become 
\begin{align}
    \frac{\partial N_{\gamma}}{\partial t} = \dot{N}_{\gamma,{\rm rec}} - \bar{\alpha} \tilde{c} N_{\gamma}, \label{A2_12}
\end{align}
\begin{align}
    \frac{\partial {\bf F}_{\gamma}}{\partial t} = - \bar{\alpha} \tilde{c} {\bf F}_{\gamma}, \label{A2_13}
\end{align}
where $\dot{N}_{\gamma,{\rm rec}}$ is the recombination rate of proton and electron, and we include this term only for EUV photon transfer.
To avoid the photon number density being negative, 
we solve the equations implicitly as
\begin{align}
    N^{n+1}_{\gamma} = \frac{1}{1+\bar{\alpha} \tilde{c} \Delta t} \left[ N_{\gamma}^{n} + \dot{N}_{\gamma,{\rm rec}} \Delta t  \right], \label{A2_14}
\end{align}
\begin{align}
    {\bf F}^{n+1}_{\gamma} = \frac{1}{1+\bar{\alpha} \tilde{c} \Delta t} {\bf F}^{n}_{\gamma}. \label{A2_15}
\end{align}

\subsubsection{Time step of RT transfer}\label{RT_timestep}

In the transport step, we solve the moment equations explicitly.
In Equation \eqref{A2_1} and \eqref{A2_2}, we introduce the reduced light speed $\tilde{c}$.
If we use the physical speed of light, the time step is too short for performing the simulations.
Thus, we use the reduced light speed approximation to keep the time step reasonable \citep{2001NewA....6..437G, 2013MNRAS.436.2188R}.
In our simulations, the outflow velocity of the gas
is comparable to the escape velocity of clouds $(< 30~{\rm km/s})$, 
and the reduced light speed needs to be larger than this value.
We adopt $\tilde{c} = 1.5 \times 10^{-4} c$ in this study.

The time step $\Delta t$ of RT transfer is determined by the Courant condition as 
\begin{align}
	\Delta t = C_{\rm RT} \frac{\Delta x}{3 \tilde{c}}, \label{courant_RT}
\end{align}
where $\Delta x$ is the cell width in each level of the AMR grid, and we adopt $C_{\rm RT} = 0.8$ \citep{2013MNRAS.436.2188R}.
We split RT transfer from the hydrodynamics solver.
If the time-step in the hydrodynamic part is larger than the above, we calculate RT transfer as the subcycle steps until the integrated time is the same as that of the hydrodynamics part.

\subsection{Computational methods of each radiative components}

In the following, we describe the details of the computational methods for each radiation component.

\subsubsection{EUV}
We consider the absorption of EUV photons by neutral hydrogen and dust grains.
The total absorption coefficient is estimated by
\begin{align}
    \bar{\alpha} = \bar{\alpha}_{\rm HI} + \bar{\alpha}_{\rm d}, \label{abs_EUV} 
\end{align}
where $\bar{\alpha}_{\rm HI}$ and $\bar{\alpha}_{\rm d}$ are the absorption coefficients for neutral hydrogen and dust grains.
We calculate these cross sections with equations \eqref{A2_3} and \eqref{A2_3_2}.
The frequency dependent cross-section of neutral hydrogen $\sigma_{\rm HI} (\nu)$ is given by \citet{1989agna.book.....O}.
We use the opacity of \citet{1993ApJ...402..441L} for the dust cross-section $\sigma_{\rm d}(\nu)$.

In the thermochemical step, we estimate the photoionization and photoheating rates ($R_{\rm HI}$, $\Gamma_{2}$) with the number density of EUV photons $N_{\gamma, \rm EUV}$:
\begin{align}
    R_{\rm HI} =  \bar{\sigma}_{\rm HI} \tilde{c} N_{\gamma, \rm EUV} , \label{A2_16}
\end{align}
\begin{align}
    \Gamma_{2} =  n_{\rm HI} \bar{\sigma}_{\rm HI} \tilde{c} N_{\gamma, \rm EUV} \bar{\gamma}_{\rm HI}, \label{gamma2_heat}
\end{align}
where $\bar{\sigma}_{\rm HI}$ is the frequency-mean cross-section defined as Equation \eqref{A2_3_2}. 
The heating rate per a hydrogen atom $\bar{\gamma}_{\rm HI}$ is estimated as
\begin{align}
    \bar{\gamma}_{\rm HI} = \frac{\int^{\nu_2}_{\nu_1} \frac{4 \pi J_{\nu}}{ h \nu} \sigma_{\rm HI} (\nu) \left[ h \left( \nu - \nu_{\rm L} \right)\right] d \nu}{\int^{\nu_2}_{\nu_1} \frac{4 \pi J_{\nu}}{ h \nu} \sigma_{\rm HI} (\nu) d \nu}. \label{A2_17}
\end{align}
The mean-cross section $\bar{\sigma}_{\rm HI}$ and the heating rate $\bar{\gamma}_{\rm HI}$ are pre-calculated based on the assumption of the fixed shape of the spectral energy distribution even after the absorption processes.

We include the emissivity from the recombination process in the injection step as
\begin{align}
    \dot N_{\gamma, \rm rec} = \left( k_{\rm H1, A} - k_{\rm H1, B} \right) n_{\rm HII} n_{\rm e} \Delta t, \label{A2_18} 
\end{align}
where $k_{\rm H1, A}$ and $k_{\rm H1, B}$ are the case-A and B recombination rates.
These values are tabulated in Table \ref{chem_process}.

\subsubsection{FUV}

We calculate the photodissociation rates ($R_{\rm H_2}$ and $R_{\rm CO}$) from FUV photon number density $N_{\gamma, \rm FUV}$ as
\begin{align}
  R_{i} = f_{{\rm shield}, i} \sigma_{i} \tilde{c} N_{\gamma, \rm FUV} \, (i={\rm H_2}, {\rm CO}), \label{A22_1}
\end{align}
where we estimate the cross sections from the reaction rates of \citet{1996ApJ...468..269D} for $\rm H_2$ and \citet{1996A&A...311..690L} for CO.
The FUV photon density also reduces because of the dust absorption. 
We here take the self-shielding effect for $\rm H_2$ and $\rm CO$ into account as the self-shielding factors $f_{{\rm shield}, i}$ in Equation \eqref{A22_1}.
The self-shielding is generally evaluated from the column density along the photon trajectory \citep{1996ApJ...468..269D}. 
Unfortunately, the photon trajectories are not followed in the moment method.
Therefore, we here estimate the column density only from local variables as
\begin{align}
    N_{i} = n_{i} \min(\lambda_{\rm J}, l_{\rm Sob}) ~ (i={\rm H_2}, {\rm CO}), \label{eq_column_density}
\end{align}
where $\lambda_{\rm J}$ is the local Jeans length and $l_{\rm Sob}$ is the Sobolev length given as
\begin{align}
    l_{\rm Sob} = \frac{v_{\rm th}}{dv/ds}, \label{eq_sobolve}
\end{align}
where $v_{\rm th}$ is the thermal velocity of molecules and $dv/ds$ is the velocity gradient.
We overestimate the column density only with the Sobolev length in static media.
We adopt the column density estimated with the local Jeans length as the upper limits of them.
We use the self-shielding factor for the $\rm H_2$ molecules formulated by \citet{2019MNRAS.484.2467W}, and the $\rm CO$ molecules tabulated by \citet{1996A&A...311..690L}.

\subsubsection{IR}
IR photons are mainly emitted from dust grains as thermal emission.
In our simulations, we separately calculate the temperatures of gas and dust ($T_{\rm g}$ and $T_{\rm d}$).
We estimate the dust temperature based on the energy equilibrium state among (1) dust thermal emission, (2) absorption of EUV photons, (3) absorption of IR photons, (4) energy transfer between gas and dust as
\begin{align}
  C_{\rm g} \left( T_{\rm d} - T_{\rm g} \right) + \kappa_{\rm d}^{0} \left( T_{\rm d} \right) a c T_{\rm d}^4 = \kappa_{\rm d}^{0} \left( T_{\rm IR} \right) \tilde{c} E_{\rm IR} + \mathcal{E}_{\rm EUV}, \label{A2_19} 
\end{align}
where $\mathcal{E}_{\rm EUV}=\kappa_{\rm d, EUV}^{0} \tilde{c} E_{\rm EUV}$ is the dust absorption rate of EUV photons, and $E_{\rm IR}$ and $E_{\rm UV}$ are the energy density of IR and UV radiation.
The coefficient of the energy transfer $C_{\rm g}$ is given as \citep{1979ApJS...41..555H, 2000ApJ...534..809O, 2005ApJ...626..627O}
\begin{align}
  C_{\rm g} = 5.83 \times 10^{-8}n_{\rm H} \left( \frac{T_{\rm g}}{10^3{\rm K}} \right)^{1/2} \left[ 1 - 0.8 \exp(-75{\rm K}/T_{\rm g}) \right]. \label{A2_20}
\end{align}
We make the pre-calculated table of 
the plank opacity for the absorption of IR radiation and the dust thermal emission as the function of IR radiation temperature $T_{\rm IR}$ and dust temperature $T_{\rm d}$, using the opacity table derived in \citet{1993ApJ...402..441L}.
The energy balance on a single dust grain is not related to the total amount of dust grains.
In Equation \eqref{A2_19}, we use the opacity ($\kappa_{\rm g}^0$) and the coefficient of energy transport $C_{\rm g}$ at $Z=Z_{\odot}$.

In the case of IR radiation, we use the moment equation of radiation energy (Eq. \ref{zeros_moment} and \ref{first_moment}) instead of that of photon number density.
In the thermochemical step, we include the dust absorption and the thermal emission as 
\begin{align}
  \frac{\partial E_{\rm IR}}{\partial t} = \rho \left[ \kappa_{\rm d}(T_{\rm d}) c a T_{\rm d}^4 - \kappa_{\rm d}(T_{\rm IR}) \tilde c E_{\rm IR} \right]. \label{A2_21}
\end{align}
The IR radiation temperature $T_{\rm IR}$ is only calculated with the IR radiation energy density, but we need to incorporate irradiation from radiation sources and the energy transport with gas for estimating dust grain temperature $T_{\rm d}$ as Equation \eqref{A2_19}.
We solve Equation \eqref{A2_21} with a semi-implicit scheme instead of calculating explicitly \citep{2015MNRAS.449.4380R, 2014ApJ...797....4K}, assuming that the energy balance on dust grains.
Then we descretize Equation \eqref{A2_19} and \eqref{A2_21} as
\begin{align}
  C_{\rm g} (T_{\rm d}^{n+1} - T_{\rm g}) + \kappa_{\rm d}^{0}(T_{\rm d}^n)ac(T_{\rm d}^{n+1})^4 = \kappa^0_{\rm d}(T_{\rm IR}^{\rm n}) \tilde{c} E_{\rm IR}^{n+1} + \mathcal{E}_{\rm EUV}, \label{A2_22_2}
\end{align}
\begin{align}
  \frac{E_{\rm IR}^{n+1} - E_{\rm IR}^{n}}{\Delta t} = \rho \left[\kappa_{\rm d}(T_{\rm d}^{n}) c a (T_{\rm d}^{n+1})^4 - \kappa_{\rm d}(T_{\rm IR}) \tilde c E_{\rm IR}^{n+1} \right], \label{A2_22} 
\end{align}
where we retain the opacity at the timestep $n$.
The sensitivity of the opacity to dust temperature is lower than the $T_{\rm d}^4$ term, and thus we assume that the dust opacity is constant at the update stage of IR radiation energy density.
Here, we define $\Delta E_{\rm IR} = E_{\rm IR}^{n+1} - E_{\rm IR}^{n}$ and $\Delta T_{\rm d} = T_{\rm d}^{n+1} - T_{\rm d}^{n}$.
It is difficult to solve the nonlinear term $(T_{\rm d}^4)$ in equation \eqref{A2_22}, and thus we linearize this term as \citep{2011A&A...529A..35C,2014ApJ...797....4K}
\begin{align}
  (T_{\rm d}^{n+1})^4 &= (T_{\rm d}^{n} + \Delta T)^4 \nonumber \\
  &\simeq (T_{\rm d}^{n})^4 + 4 (T_{\rm d}^{n})^3 \Delta T. \label{A2_23}
\end{align}
Substituting equation \eqref{A2_23} into \eqref{A2_22_2}, we obtain the relation between $\Delta T_{\rm d}$ and $\Delta E_{\rm IR}$ as
\begin{align}
  \Delta T_{\rm d} = \frac{\kappa_{\rm d}^{0}(T_{\rm IR}^n) \tilde c }{ C_{\rm g} + 4 \kappa_{\rm d}^{0}(T_{\rm d}^n)ac (T_{\rm d}^{n})^3 } \Delta E_{\rm IR}, \label{A2_24}
\end{align}
where we use the relation of equation \eqref{A2_19} at the timestep $n$.
Substituting Equations \eqref{A2_23} and \eqref{A2_24} into Equation \eqref{A2_22}, the rate of change of IR energy density
is given as
\begin{align}
  \Delta E_{\rm IR} = \frac{\rho \Delta t \left[ \kappa_{\rm d} (T_{\rm d}^{n}) c a (T_{\rm d}^n)^4 - \kappa_{\rm d} (T_{\rm IR}^{n}) \tilde{c} E_{\rm IR}^{n} \right]}{\left[ 1 + \rho \Delta t \kappa_{\rm d} (T_{\rm IR}^{n}) \tilde{c} / (1 + \chi) \right]}, \label{A2_25}
\end{align}
where $\chi$ is given as
\begin{align}
  \chi = \frac{4 \kappa_{\rm d}^{0} (T_{\rm d}^{n}) ac (T_{\rm d}^{n})^{3}}{C_{\rm g}}. \label{A2_26}
\end{align}

\subsection{Radiation Force}\label{Radiation_force}
Equations of motion and energy (eq. \ref{undo_eq} and \ref{energy_eq}) contain the radiation pressure $\bf{f}$ which is evaluated from the photon number flux $\bf{F}_{\gamma}$.
We consider the radiation pressure induced by EUV, FUV and IR photons.
The radiation pressure is formulated as 
\begin{align}
  {\bf{f}} &= \frac{y_{\rm HI} \bar{\sigma}_{\rm HI} + \bar{\sigma}_{\rm d}}{c \left( 1 + 4 y_{\rm He} \right) m_{\rm p}} \left(h \bar{\nu} \bf{F}_{\gamma} \right)_{\rm EUV} \nonumber \\
  &+\frac{ \bar{\sigma}_{\rm d}}{c \left( 1 + 4 y_{\rm He} \right) m_{\rm p}} \left(h \bar{\nu} \bf{F}_{\gamma} \right)_{\rm FUV} + \frac{\kappa_{\rm R}}{c} \left( \bf{F} \right)_{\rm IR}, \label{A2_27}
\end{align}
where $m_{\rm p}$ is proton mass and $h \bar{\nu}$ is mean energy of EUV photons.

\subsection{Test calculations: H{\sc ii} region formation}\label{test_cal}

To test the RT solver developed in this study, we perform the simulations of an H{\sc ii} bubble around a single massive star.
The properties of the radiation source are shown in Table \ref{radsource}.
Here, we set the number density of gas and metallicity as $n_{\rm H} = 10^{2} ~{\rm cm^{-3}}$ and $Z=Z_{\odot}$.
Here, we ignore radiation pressure to compare with the analytical solution of an expanding H{\sc ii} region.

\begin{table}
    \caption{properties of radiation source}
    \label{radsource}
    \centering
    \begin{tabular}{cccc}
      \hline \hline
      $M_* \, [M_{\odot}]$ & $T_{\rm eff} \, [{\rm K}]$ & $\log_{10} [S_{\rm EUV} ({\rm s^{-1}})]$ & $\log_{10} [S_{\rm FUV} ({\rm s^{-1}})]$ \\ \hline
      $40$ & $4.3 \times 10^4 ~{\rm K}$ & $49.1$ & $48.8$ \\ \hline

    \end{tabular}
    \begin{minipage}{1\hsize}
    \end{minipage}
  \end{table}

\subsubsection{Static case}

\begin{figure}
    \begin{center}
    \includegraphics[width=\columnwidth]{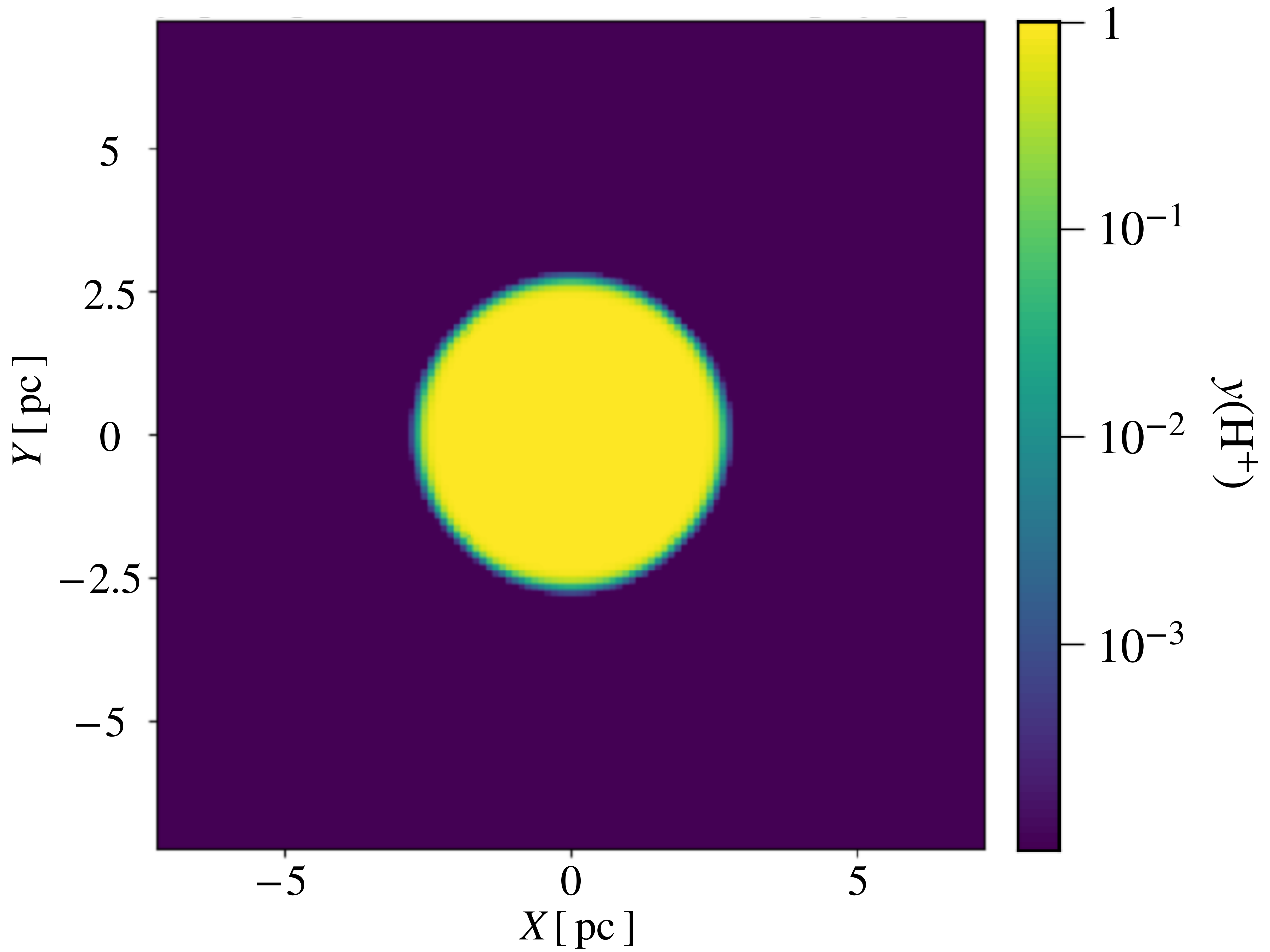}
    \end{center}
    \caption{
    Ionization structure of hydrogen around a point source in uniform density gas.
   }
    \label{zu_ypHII}
\end{figure}
\begin{figure}
    \begin{center}
    \includegraphics[width=\columnwidth]{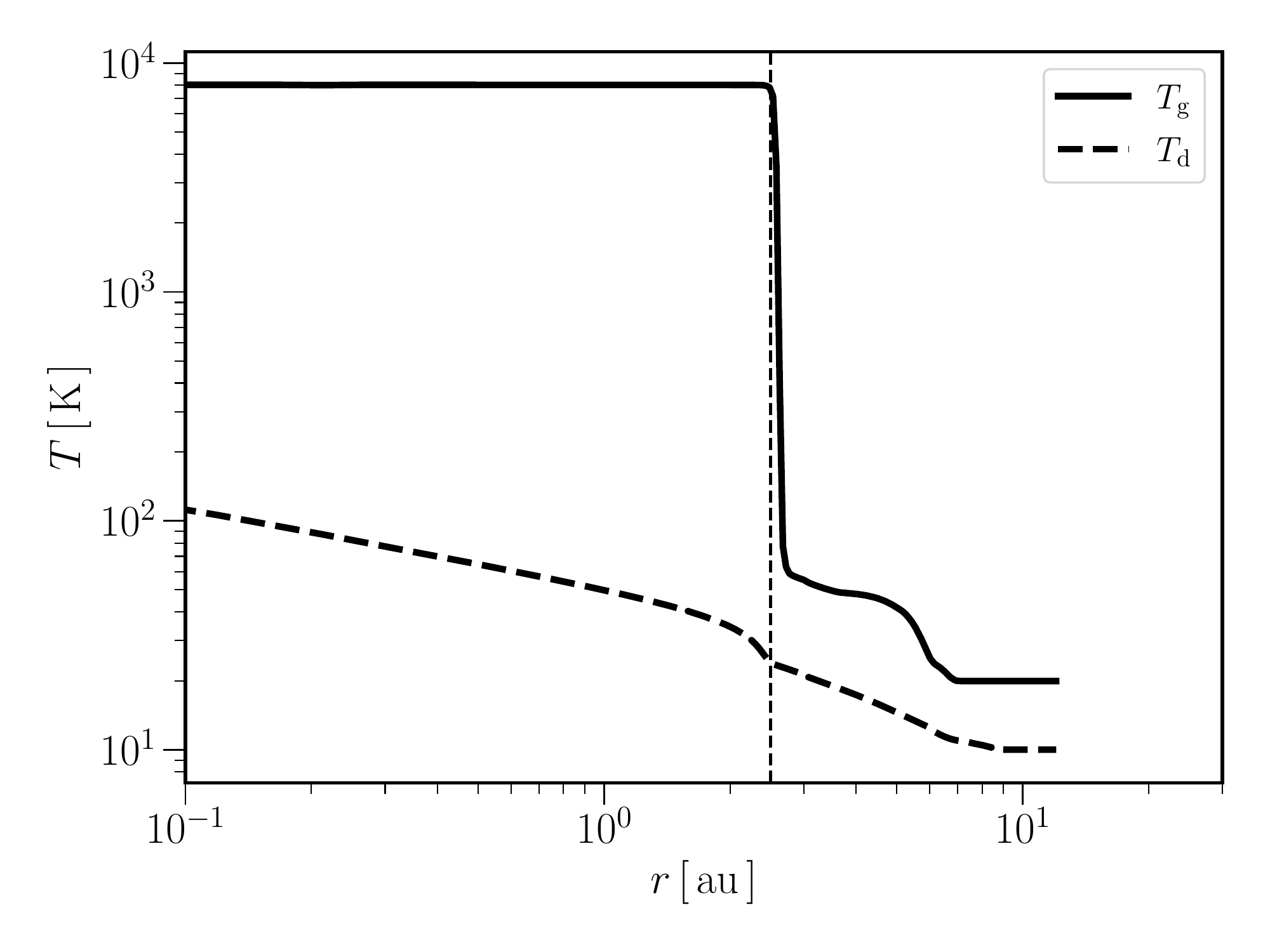}
    \end{center}
    \caption{
    Radial profiles of gas and dust temperatures  in the equilibrium state.
    The solid and dashed lines represent the gas and the dust temperatures ($T_{\rm g}$ and $T_{\rm d}$), respectively.
    The vertical dashed line shows the position of the Str\"omgren radius estimated by Eq. \eqref{Rst}.
   }
    \label{zu_THII}
\end{figure}
\begin{figure}
    \begin{center}
    \includegraphics[width=\columnwidth]{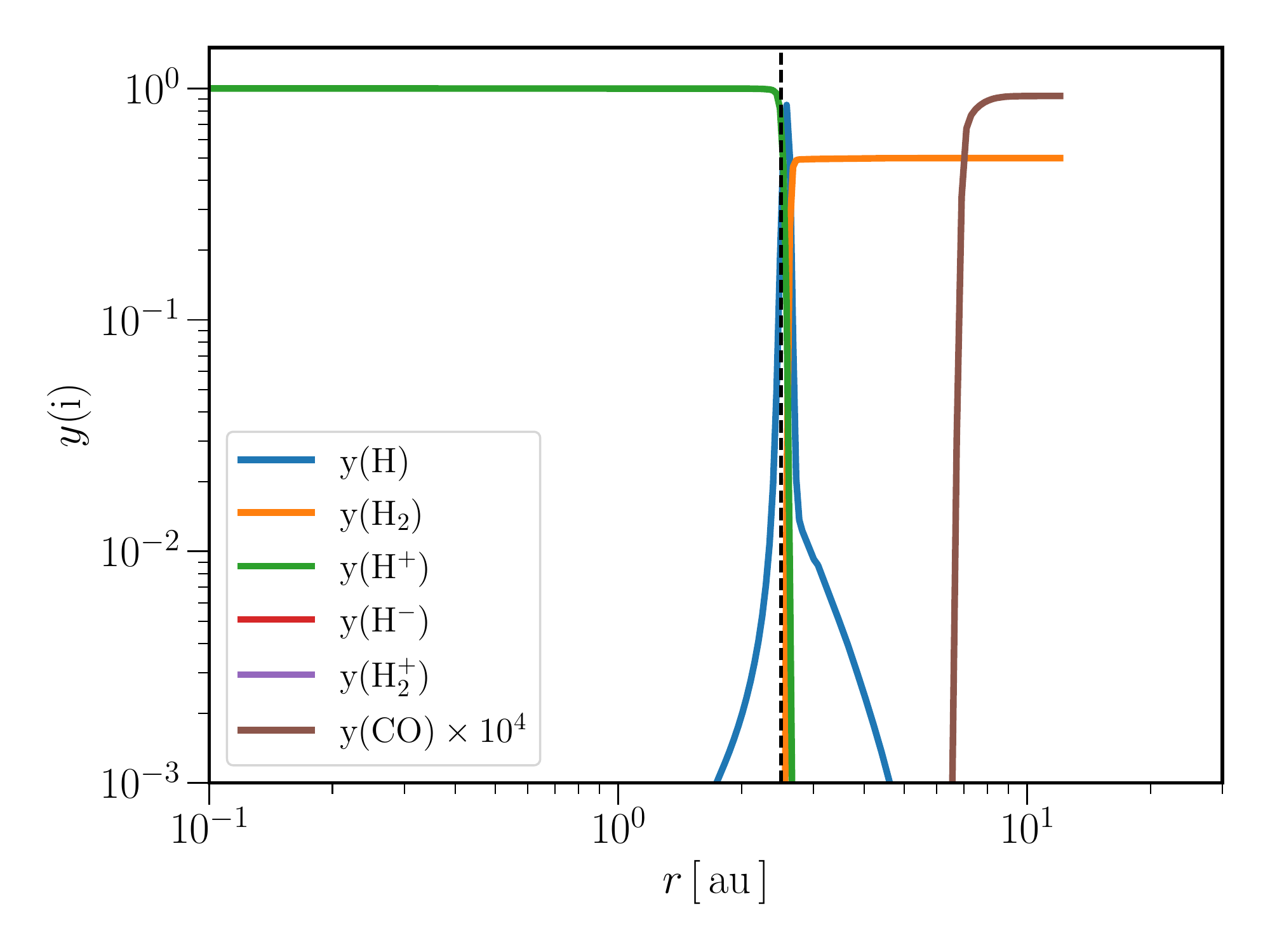}
    \end{center}
    \caption{
    The radial distributions of the chemical compositions, $\rm H$, $\rm H_2$, $\rm H^+$, $\rm H^-$, $\rm H_2^{+}$, and $\rm CO$.
    Same as Figure \ref{zu_THII}, the vertical dashed line shows the positions of the Str\"omgren radius. 
   }
    \label{zu_ychem1D}
\end{figure}
We first perform a simple test of H{\sc ii} region expansion in a static medium.
In the equilibrium state, the ionization front reaches the Str\"omgren radius as 
\begin{align}
  R_{\rm st} &= \left( \frac{3 f_{\rm ion} S_{\rm EUV}}{4 \pi \alpha_{\rm B} n_{\rm H}^2} \right)^{1/3} \nonumber \\ 
  &= 2.5 ~{\rm pc} \left( \frac{S_{\rm EUV}}{1.3 \times 10^{49} ~{\rm s^{-1}}} \right)^{1/3} \left( \frac{f_{\rm ion}}{0.45} \right)^{1/3} \left( \frac{n_{\rm H}}{10^2~{\rm cm^{-3}}} \right)^{-2/3}, \label{Rst}
\end{align}
where $S_{\rm EUV}$ and $f_{\rm ion}$ are the emissivity of ionizing photons and the absorption rate of neutral hydrogen.
Here, we consider the dust absorption in the H{\sc ii} regions. 
We use the H{\sc i} absorption rate in dusty medium derived by \citet{1972ApJ...177L..69P}.
Figure \ref{zu_ypHII} shows the ionization structure at $t=0.2~{\rm Myr}$.
At this time, the equilibrium state has already been realized. 
The spherical H{\sc ii} region appears around the radiation source, and its radius is almost the same as the Str\"omgren radius given by equation \eqref{Rst}. 
Temperature structures of gas and dust grains are shown in Figure \ref{zu_THII}.
In the H{\sc ii} region, the gas temperature increases to $T_{\rm g} \sim 8 \times 10^3~{\rm K}$ due to the photoionization heating.
The dust is also heated by photon absorption, and its temperature is higher than $10~{\rm K}$.
Figure \ref{zu_ychem1D} shows the radial distribution of the chemical compositions.
The photodissociation fronts of $\rm H_{2}$ and $\rm CO$ exist at $r \sim 2.5~{\rm pc}$ and $7~{\rm pc}$ outside the H{\sc ii} region.

\subsubsection{Dynamical expansion of H{\sc ii} region}

\begin{figure}
    \begin{center}
    \includegraphics[width=\columnwidth]{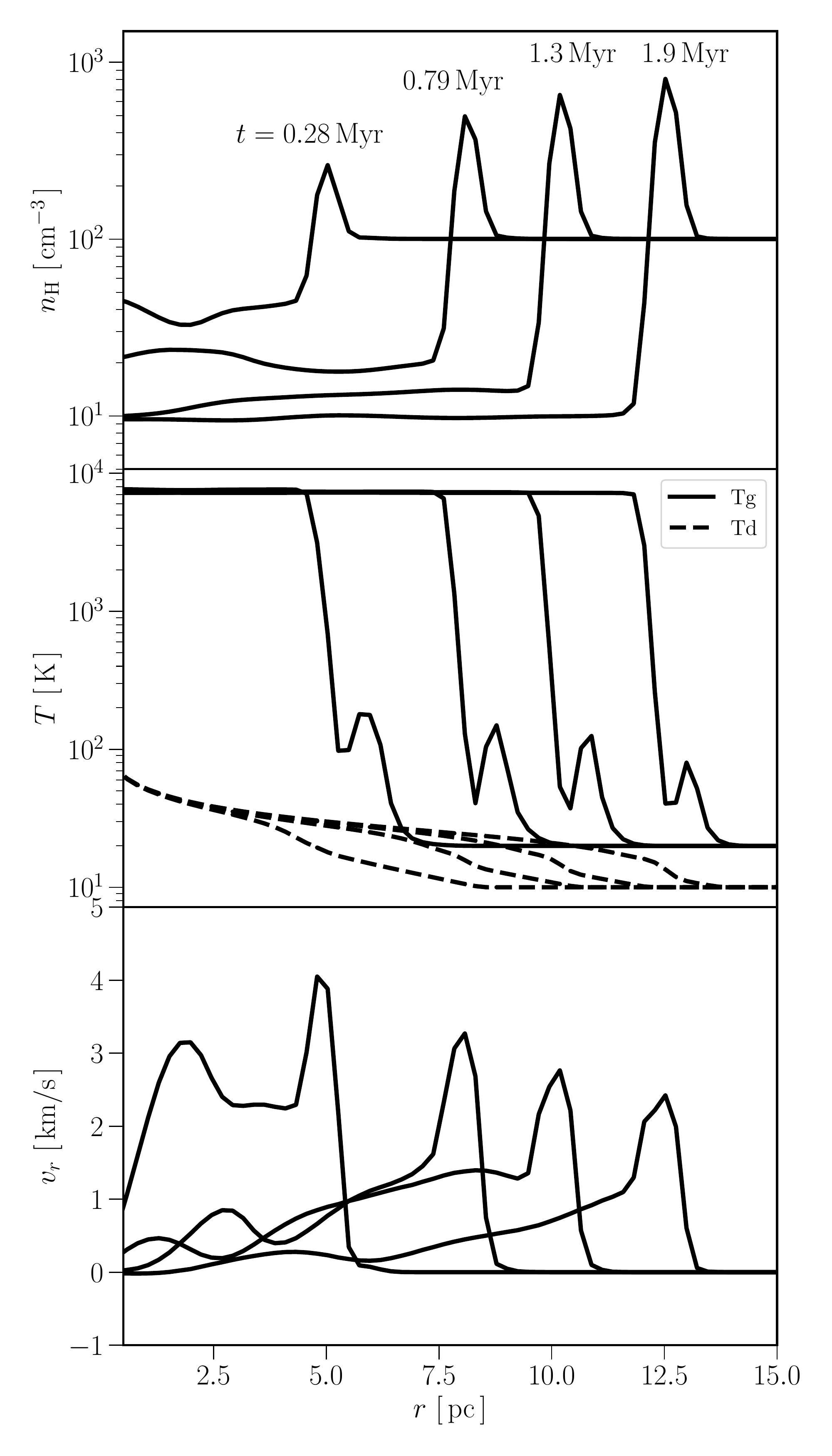}
    \end{center}
    \caption{The spherically averaged profiles of the dynamical expansion shell around the H{\sc ii} regions.
    Top, middle, and bottom panels show the radial profiles of the hydrogen number density ($n_{\rm H}$), the gas and the dust temperature ($T_{\rm g}$ and $T_{\rm d}$), and the radial velocity of gas $v_{\rm r}$.
    In each panel, the four snapshots are taken from $t=0.28~{\rm Myr}$, $0.79~{\rm Myr}$, $1.3~{\rm Myr}$, and $1.9~{\rm Myr}$.
    }
    \label{zu_HIIevolv}
\end{figure}
\begin{figure}
    \begin{center}
    \includegraphics[width=\columnwidth]{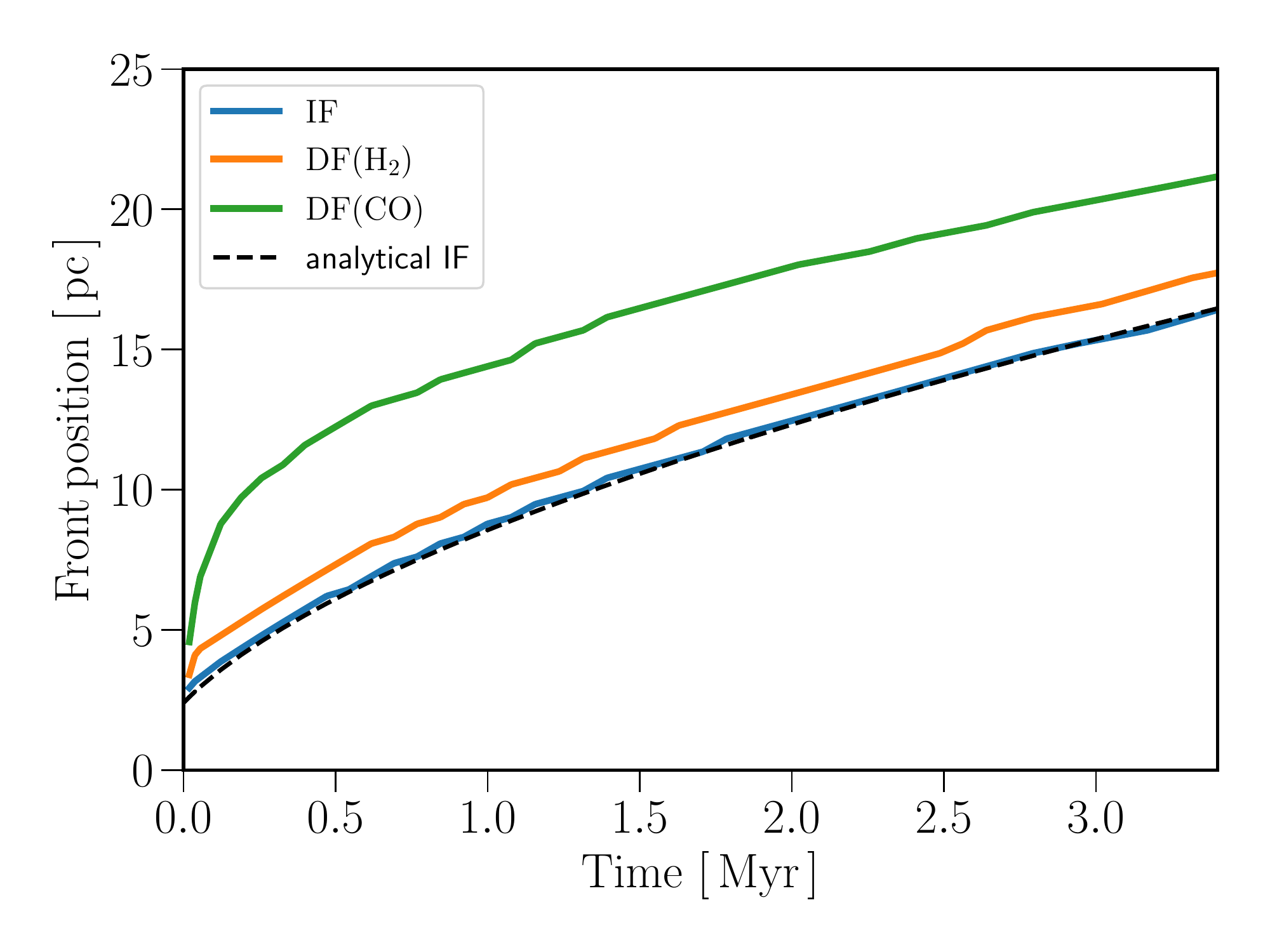}
    \end{center}
    \caption{The time evolution of the ionization front (IF, blue), the dissociation fronts of $\rm H_{2}$ and $\rm CO$ (DF, orange and green), and the ionization front analytically estimated by Eq. \eqref{solution_IF} (black).
    We set $t=0$ when the ionization front reaches the Str\"omgren radius $R_{\rm st}$.
    }
    \label{zu_HIIevolv2}
\end{figure}

We calculate the dynamical expansion of the H{\sc ii} regions. 
The initial setup is the same as in the previous subsection.
Figure \ref{zu_HIIevolv} shows the radial profiles of gas density, temperature, and radial velocity.
The thermal pressure from the H{\sc ii} region makes the shell structure of which the density is about 100 times higher than the inner region.
The gas temperature in the H{\sc ii} region is $ \gtrsim 8 \times 10^3 ~{\rm K}$ that is determined by the energy balance between the photoionization heating and the metal line cooling of O{\sc ii} and O{\sc iii}. 
Dust grains in the H{\sc ii} regions are also heated up to $30~{\rm K}$ due to the direct light from the central star.
The shell's radial velocities are $3-5~{\rm km/s}$, and gradually decrease.

Figure \ref{zu_HIIevolv2} shows the positions of the ionization front, photodissociation fronts of $\rm H_{2}$ and $\rm CO$ molecules as a function of time.  
The analytical solution for the ionization front is given by \citep{2006ApJ...646..240H}
\begin{align}
  R_{\rm IF} (t) = R_{\rm St} \left( 1 + \frac{7}{4} \sqrt{\frac{4}{3}} \frac{c_{\rm HII} t}{R_{\rm St}} \right)^{4/7}, \label{solution_IF}
\end{align}
where $R_{\rm St}$ and $c_{\rm HII}$ are the Str\"omgren radius and the sound speed of ionized gas.
The position of the ionization front nicely matches the analytical solution.
The photodissociation front propagates beyond the ionization front.
The self-shielding of $\rm H_2$ molecules is more efficient than that of CO molecules.
Therefore, the radius of the CO photodissociation region is larger than that of $\rm H_2$, and the region between the two photodissociation radii corresponds to the "CO-dark" molecular cloud
\citep[e.g.,][]{1988ApJ...334..771V, 2010ApJ...716.1191W,2020MNRAS.497.5061I}.
\citet{2006ApJ...646..240H} performed the dynamical expansion with almost the same setup (LD-S41 in their model).
They calculated the frequency-dependent radiation transport of photodissociation photons to estimate each $\rm H_2$ line absorption accurately. 
According to their result, the positions of the ionization front and the photodissociation front of $\rm H_2$ are almost the same.
In our simulations, the position of the photodissociation front is slightly outside the ionization front, but we reproduce their results well.

\section{Sink evolution of fiducial models}\label{appendix_sink_evolv}

\begin{figure}
    \begin{center}
    \includegraphics[width=\columnwidth]{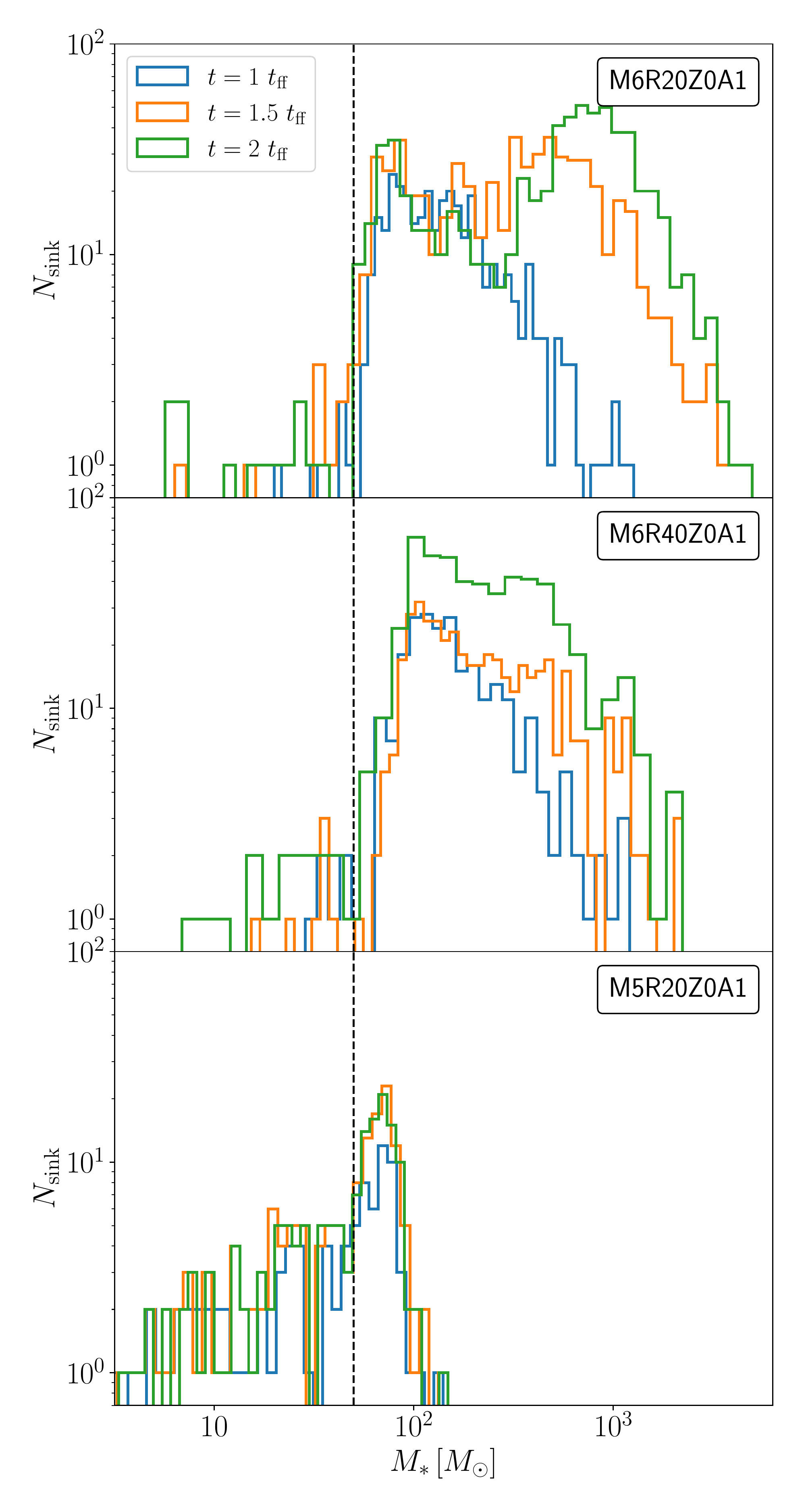}
    \end{center}
    \caption{ The mass distributions of sink particles at $t=1$ (blue), $1.5$ (orange), and $2~t_{\rm ff}$ (green). 
    Each panel shows the models of M6R20Z0A1 (top), M6R40Z0A1 (middle), and M5R20Z0A1 (bottom).
    }
    \label{zu_hist_pop}
\end{figure}
\begin{figure}
    \begin{center}
    \includegraphics[width=\columnwidth]{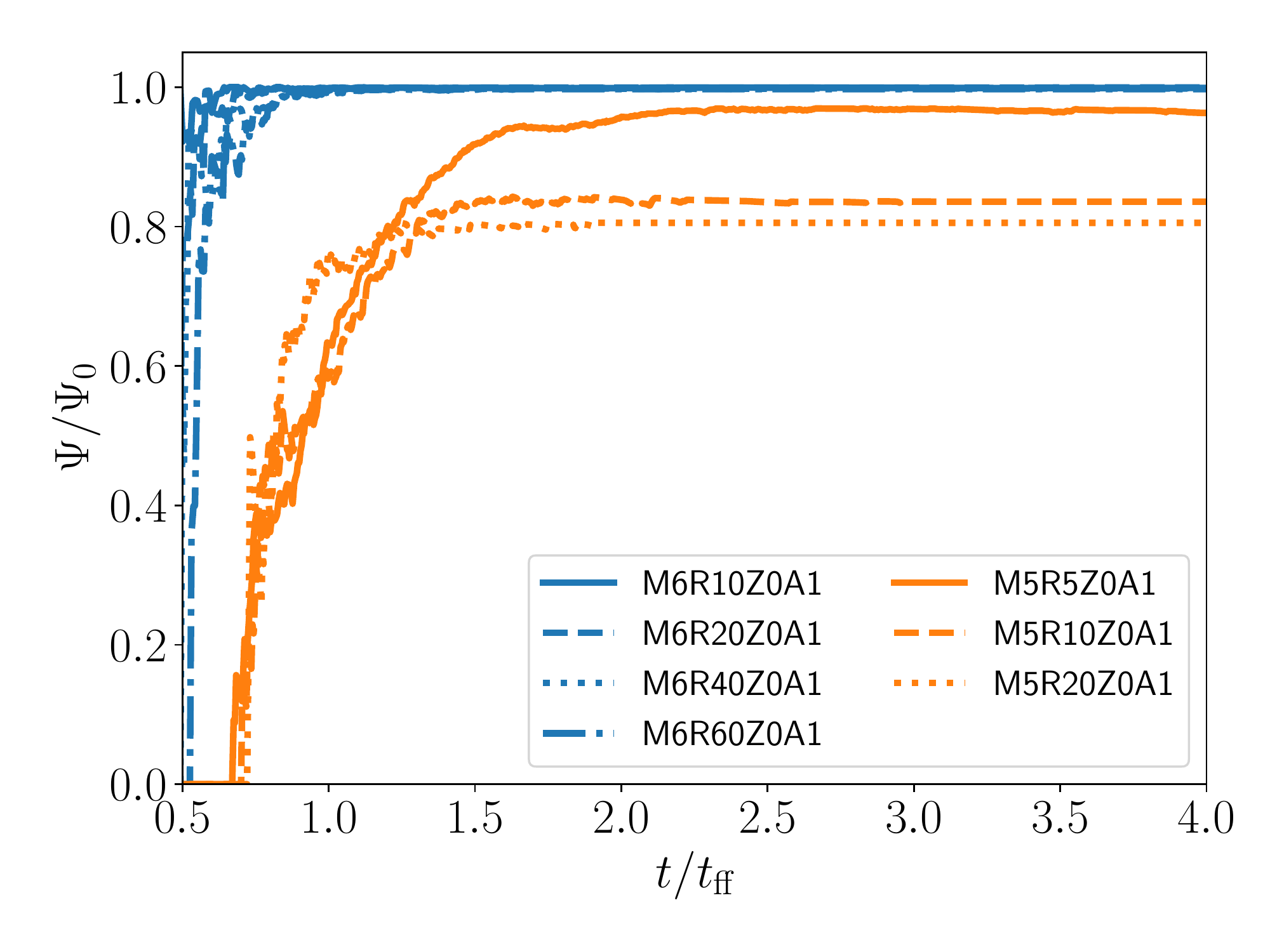}
    \end{center}
    \caption{The time evolution of mass-to-luminosity ratios ($\Phi$) in the cases with $(Z, \alpha_0) = (Z_{\odot}, 1)$.
    Each line is normalized by the values converting the total stellar mass to emissivity based on the IMF averaged one. 
    }
    \label{zu_mass_lumi}
\end{figure}

Figure \ref{zu_hist_pop} shows the histograms of the sink masses in the models of M6R20Z0A1, M6R40Z0A1, and M5R20Z0A1.
In the case of M6R20Z0A1, the sink mass distribution has a peak at $\sim 10^2~M_{\odot}$ at when the elapsed time is $1~t_{\rm ff}$.
After that, a stellar core forms, and
gas accretion onto sink particles continues until the end of the star formation.
Finally, the peak shifts to a higher mass around $\sim 10^3~M_{\odot}$ at $t \sim 2~t_{\rm ff}$.
In the case of M6R40Z0A1, the shapes of stellar mass distributions do not change significantly, and its peaks are around $\sim 10^2~M_{\odot}$ as shown in the middle panel of Figure \ref{zu_hist_pop}.
In both cases, most sink particles exceed $50~M_{\odot}$ 
that is the critical mass to be recognized as UV radiation sources.
In the case of M5R20Z0A1, on the other hand, the peak approaches $50~M_{\odot}$, and some sink particles are smaller than this threshold value.

To estimate the luminosity of a sink particle, we apply a pre-calculated emissivity per unit stellar mass averaged over the IMF. 
Figure \ref{zu_mass_lumi} shows the time evolution of the mass-to-luminosity ratio ($\Phi$) normalized by the values multiplying the emissivity times total stellar mass ($\phi_0$), i.e., the case assuming that all sink particles emit radiation. 
As shown in Figure \ref{zu_hist_pop}, most sink particles exceed $50~M_{\odot}$. Thus the normalized mass-to-luminosity ratio rapidly reaches unity in all cases with the clouds of $10^6~M_{\odot}$.
On the other hand, in the cases of $10^5~M_{\odot}$ clouds, the ratios increase more slowly and reach 0.8 at $t=t_{\rm ff}$.
In the diffuse cloud models of M5R20Z0A1 and M5R10Z0A1, this ratio remains constant at $\sim 0.8$, while it reaches $\gtrsim 0.9$ in the compact cloud model of M5R5Z0A1 in which the core formation occurs.

\section{Resolution study}\label{appendix_resolution_study}

\begin{figure}
    \begin{center}
    \includegraphics[width=\columnwidth]{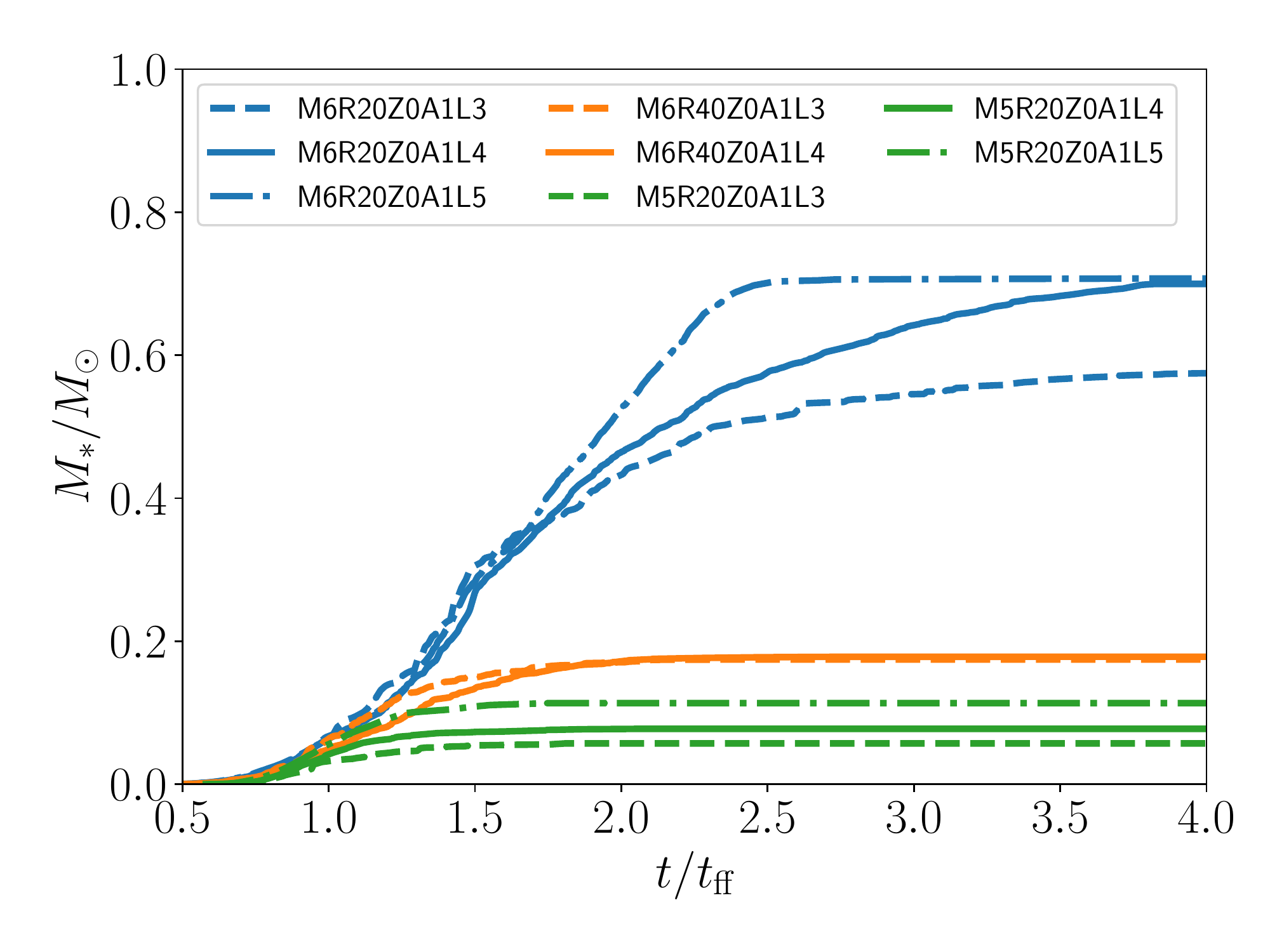}
    \end{center}
    \caption{
    The time evolution of the stellar mass in the cases of M6R20Z0A1, M6R40Z0A1, and M5R20Z0A1 with the maximum refinement level at $\l_{\rm max} = 3$, $4$, and $5$.
    In each model name, the labels of L3-L5 mean the maximum refinement levels 3 - 5.}
    \label{zu_stellar_mass_hist_reso}
\end{figure}
\begin{figure}
    \begin{center}
    \includegraphics[width=\columnwidth]{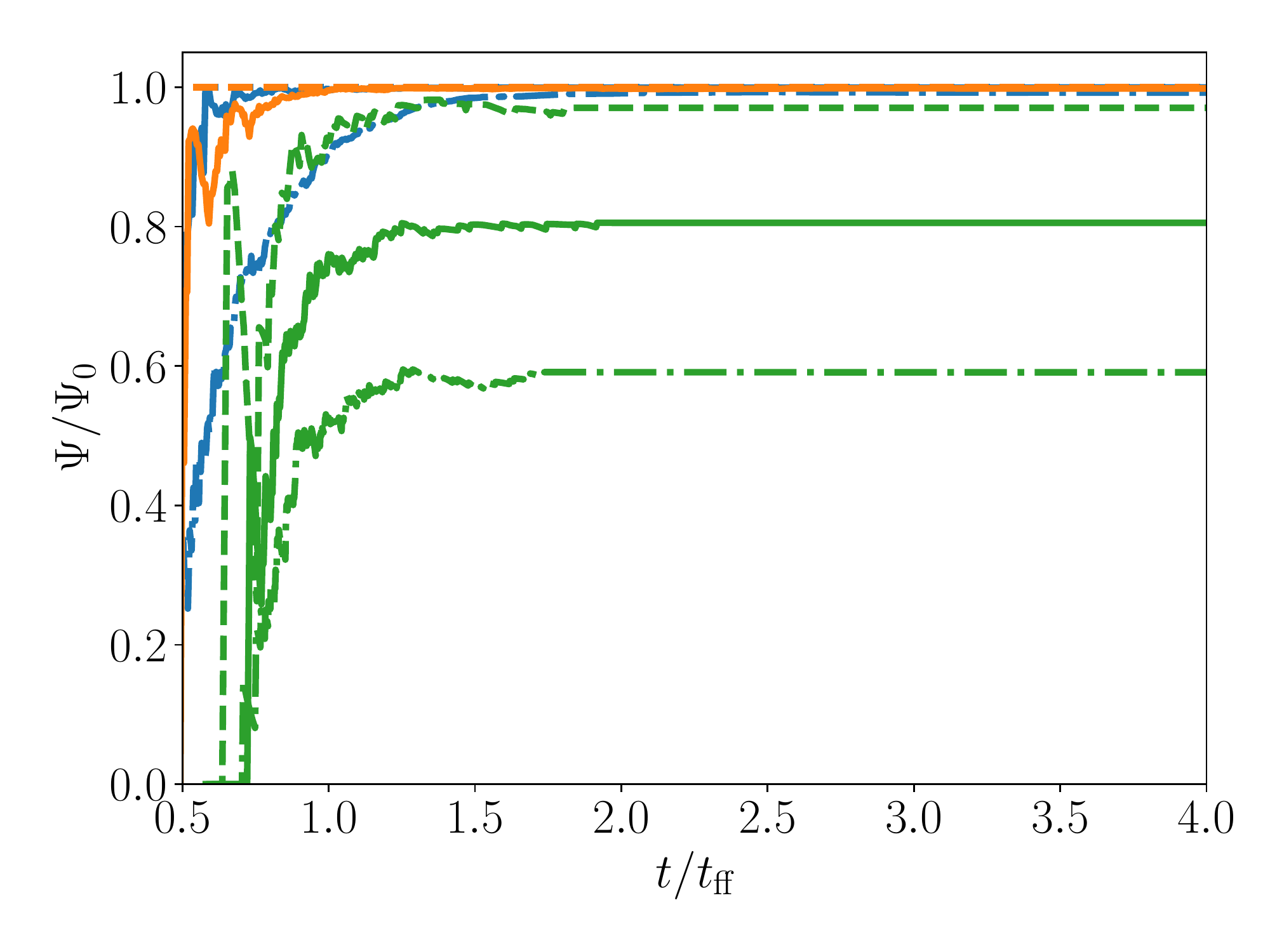}
    \end{center}
    \caption{
    The time evolution of the mass-to-luminosity ratios in the models of M6R20Z0A1, M6R40Z0A1, and M5R20Z0A1 at each maximum refinement level.
    The colors and styles are same as Figure \ref{zu_stellar_mass_hist_reso}.}
    \label{zu_masslumi_reso}
\end{figure}
\begin{figure}
    \begin{center}
    \includegraphics[width=\columnwidth]{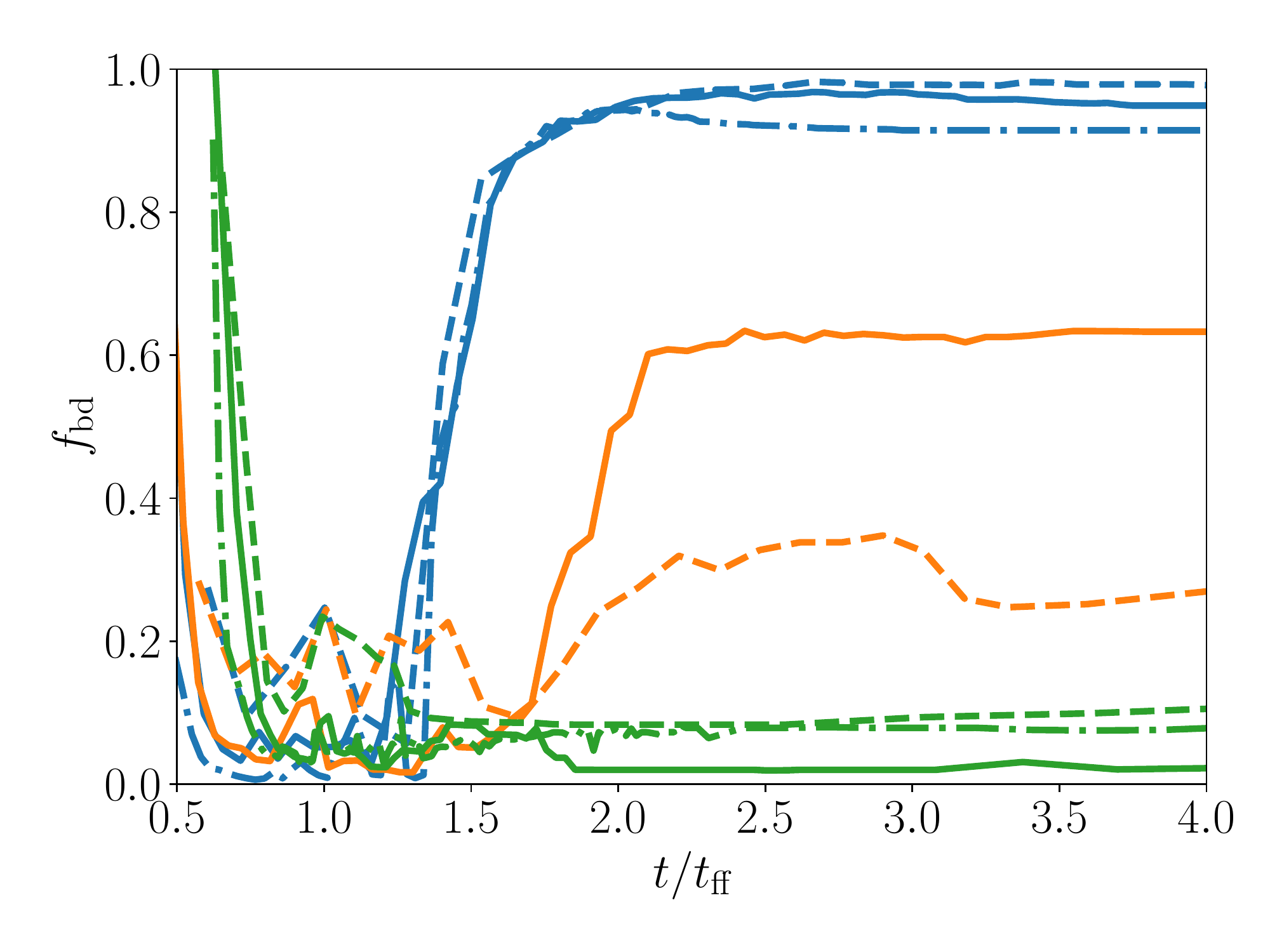}
    \end{center}
    \caption{
     The time evolution of bound fractions in the models of M6R20Z0A1, M6R40Z0A1, and M5R20Z0A1 at each maximum refinement level.
    The colors and styles are the same as Figure \ref{zu_stellar_mass_hist_reso}.}
    \label{zu_bdfrac_reso}
\end{figure}

Even in the current state-of-the-art simulations, we suffer from resolving individual stars in massive star forming clouds. 
Therefore, we consider the sink particle technique assuming a star cluster with the IMF as a sub-grid model.  
Here, we assume that only sink particles with the mass larger than $50~M_{\odot}$ become UV radiation sources. 
If the typical sink mass depends on the resolutions of the simulations, the total UV luminosity and star formation efficiency can also change. 
To investigate the resolution dependence, we perform the additional simulations with the maximum refinement level $l_{\rm max} = 3$ for the cases of M6R20Z0A1, M6R40Z0A1, and M5R20Z0A1, and $l_{\rm max} = 5$ for the cases of M6R20Z0A1, M6R40Z0A1, and M5R20Z0A1 (fiducial simulations use $l_{\rm max} = 4$).
We label the additional simulations as M6R20Z0A1L3 and M6R20Z0A1L5, where L3 and L5 represent the maximum level of refinement.

Figure \ref{zu_stellar_mass_hist_reso} shows the time evolution of the total stellar mass in each model.
In the cases for M6R20Z0A1, the results are converged until the elapsed time of 
$t \sim 1.7~t_{\rm ff}$.
At $t \sim 1.7~t_{\rm ff}$, stellar cores form, and the ambient gas accumulates around them. 
Then, as the stellar mass increases, the radiation pressure evacuates the gas and quenches the star formation.  
Note that, however, the radiation pressure can sensitively depend on the resolution of the simulations.
\citet{2018MNRAS.480.3468K} pointed out that the dust destruction front should be resolved to estimate the radiation pressure from the stellar light correctly.
In practice, we find that the star formation changes with the resolution. 
In the case with $l_{\rm max} = 3$, the SFR decreases after $t\sim 1.7~t_{\rm ff}$, and the final SFE is 0.58 that is 17\% lower than the cases with the higher maximum levels.
At $l_{\rm max} = 5$, the reduction of the SFR does not occurs until the star formation is completely quenched.
In the case with $l_{\rm max} = 4$ (fiducial resolution), the SFR decreases, but the final SFE is almost the same as that of $l_{\rm max} = 5$.
This indicates that the star formation becomes slower due to the radiation pressure, but most of gas around the stellar core is finally converted into stars.

In the models of M6R40Z0A1, stellar cores do not form. The gas clouds are rapidly disrupted by the photo-ionization feedback.
Therefore, the resolution dependence of radiation feedback does not appear, and the results are converged well.

In the cases of M5R20Z0A1, the SFEs change from 6\% to 11\% as the maximum refinement level increases from $l_{\rm max} = 3$ to $5$.
This decrease of the SEFs is related to the change in the mass-to-luminosity ratios as shown in Figure \ref{zu_masslumi_reso}.
The mass-to-luminosity ratio decreases as the resolution increases, and more stellar mass is needed to disrupt the cloud.
In particular, the total emissivity is almost the same in all resolutions.
It means that the model of radiation sources causes the increase of the SFEs in higher numerical resolutions.
We need the more suitable models of the radiation sources for the lower mass clouds with $\lesssim 10^5~M_{\odot}$.
On the other hand, the mass-to-luminosity ratios are almost unity in the massive clouds with $10^6~M_{\odot}$. The properties of radiation sources are independent of the numerical resolution in these cases. 

Figure \ref{zu_bdfrac_reso} shows the time evolution of bound fractions.
In the cases with M6R20Z0A1, the bound fractions converge in all cases.
As mentioned in Section \ref{prop_star_cluster}, the bound fractions are fluctuated when the SFE is $\epsilon_* \lesssim 0.1$ \citep{2020arXiv200804453G}.
The number of sink particles and their masses is altered with the different resolutions.
They amplify the bound fraction fluctuations when the SFEs are $\epsilon_* \lesssim 0.1$.
Thus, the bound fractions are scatted with the different resolutions in the models of M6R40Z0A1 and M5R20Z0A1.

In this study, we focus on the massive star cluster formation.
The final SFEs and the bound fractions of star clusters are converged at $l_{\rm max} \gtrsim 4$.
We conclude that the massive star cluster formation results obtained in this study are reasonable values and converged.


\bsp	
\label{lastpage}
\end{document}